\documentclass[aip,jcp,amsmath,amssymb,floatfix,citeautoscript,reprint,longbibliography]{revtex4-2}

\usepackage{graphicx}
\usepackage{dcolumn}
\usepackage{bm}

\usepackage{braket}
\usepackage[usenames,dvipsnames]{color}

\usepackage{ragged2e}
\usepackage{times}
\usepackage{txfonts}

\usepackage{xr}
\makeatletter
\newcommand*{\addFileDependency}[1]{
  \typeout{(#1)}
  \@addtofilelist{#1}
  \IfFileExists{#1}{}{\typeout{No file #1.}}
}
\makeatother
\newcommand*{\myexternaldocument}[1]{
    \externaldocument{#1}
    \addFileDependency{#1.tex}
    \addFileDependency{#1.aux}
}
\myexternaldocument{build/si}

\begin{document}

\title{Machine learning potentials from transfer learning of periodic correlated electronic structure methods: Application to liquid water with AFQMC, CCSD, and CCSD(T)}

\author{Michael S. Chen}
\affiliation{Department of Chemistry, Stanford University, Stanford, California, 94305, USA}

\author{Joonho Lee}
\affiliation{Department of Chemistry, Columbia University, New York, New York 10027, USA}

\author{Hong-Zhou Ye}
\affiliation{Department of Chemistry, Columbia University, New York, New York 10027, USA}

\author{Timothy C. Berkelbach}
\email{t.berkelbach@columbia.edu}
\affiliation{Department of Chemistry, Columbia University, New York, New York 10027, USA}
\affiliation{Center for Computational Quantum Physics, Flatiron Institute, New York, New York 10010, USA}

\author{David R. Reichman}
\email{drr2103@columbia.edu}
\affiliation{Department of Chemistry, Columbia University, New York, New York 10027, USA}

\author{Thomas E. Markland}
\email{tmarkland@stanford.edu}
\affiliation{Department of Chemistry, Stanford University, Stanford, California, 94305, USA}

\date{\today}

\begin{abstract}
Obtaining the atomistic structure and dynamics of disordered condensed phase systems from first principles remains one of the forefront challenges of chemical theory. Here we exploit recent advances in periodic electronic structure to show that, by leveraging transfer learning starting from lower tier electronic structure methods, one can obtain machine learned potential energy surfaces for liquid water from the higher tier AFQMC, CCSD, and CCSD(T) approaches using $\le$200 energies. By performing both classical and path integral molecular dynamics simulations on these machine learned potential energy surfaces we uncover the interplay of dynamical electron correlation and nuclear quantum effects across the entire liquid range of water while providing a general strategy for efficiently utilizing periodic correlated electronic structure methods to explore disordered condensed phase systems.
\end{abstract}

\maketitle
\normalsize

\section{Introduction}

{\it Ab initio} molecular dynamics (AIMD) simulations, where the forces and energies are generated at each time-step by performing an electronic structure calculation, provide an appealing route to simulate reactive chemical dynamics. However, for disordered condensed phase systems an accurate description typically requires many 100's of atoms to obtain a chemically reasonable description of bulk systems (e.g., water) and this grows into the 1000's for more heterogeneous systems (e.g., those with interfaces). Since AIMD simulations require an electronic structure calculation to be performed at each time-step, to statistically converge even simple thermodynamic properties necessitates many tens of thousands of {\it ab initio} calculations (10's of picosecond timescale at a $\sim$1~fs time step) and for slower converging properties many millions are needed (nanosecond timescale). The computational expense of these simulations is further compounded if one wants to incorporate nuclear quantum effects (NQE) via \textit{ab initio} path integral molecular dynamics simulations (PIMD). Due to its reasonable compromise between accuracy and efficiency, density functional theory (DFT) is currently the most frequently employed electronic structure method in condensed phase AIMD studies. However, the results  depend---sometimes sensitively---on the choice of the exchange-correlation functional and the inclusion of dispersion corrections\cite{Distasio2014,Gillan2016}. This issue motivates the use of beyond-DFT electronic structure theories, such as those based on the many-electron wavefunction. For example, work performed almost a decade ago demonstrated the AIMD simulations of liquid water using second-order M\" oller-Plesset perturbation theory~(MP2)~\cite{DelBen2013}. However, the high cost of more accurate methods precludes their direct use in AIMD.

Machine learned potentials (MLPs) have emerged as an extremely promising approach to accurately model \textit{ab initio} potential energy surfaces of condensed phase systems while being orders of magnitude more computationally efficient to evaluate. For liquid water, MLPs have been successfully developed at various levels of electronic structure ranging from different levels of DFT\cite{Morawietz2016,Morawietz2018,Zhang2018,Cheng2019,Schran2020,Yao2020,Zhang2021} to more recently using the random phase approximation (RPA)\cite{Yao2021} and MP2\cite{Lan2021,Liu2022}. The modeling of liquid water and other molecular systems with more accurate electronic structure methods, such as coupled-cluster theory or quantum Monte Carlo, has been so far limited to training on finite clusters of molecules\cite{Bukowski2007,Babin2013,Babin2014,Medders2014,Reddy2016,Schran2020a,Nandi2021,Daru2022,Yu2022,Archibald2018,Ryczko2022,Huang2022}. Recent advances have opened the door to efficiently calculating the properties of periodic condensed phase systems using higher-level methods like coupled cluster singles and doubles without and with perturbative triples (CCSD and CCSD(T))\cite{Purvis1982,Raghavachari1989} and phaseless auxiliary-field quantum Monte Carlo (AFQMC)\cite{Zhang2003Apr}. However, while these advances allow the energies of many 100's of condensed phase configurations to be evaluated, this is still considerably less than what would be typically required to train an accurate MLP.

Here, we demonstrate that by using an approach based on transfer learning starting from a variety of lower tier electronic structure methods one can generate a highly data-efficient approach to training MLPs using high level electronic structure methods. With this approach, we show that only 200 high-quality energies obtained from small periodic boxes containing 16 water molecules provide sufficient data for training our MLPs. Specifically, we train MLPs to periodic electronic structure calculations performed with AFQMC, CCSD, and CCSD(T). The MLPs are then used to perform AIMD and PIMD simulations of larger water boxes for the long times necessary to statistically converge static and dynamic properties using both classical and quantum mechanical treatment of nuclei. This allows us to achieve a careful comparison of the quality of the underlying electronic structure theories in describing water across its entire liquid temperature range and uncover the changes in the properties of liquid water as the dynamical electron correlation captured is increased. In addition, we provide a set of MLPs and a curated training set that can form the basis of future studies of water and aqueous systems. Our data-efficient approach provides a route to accurately obtain the properties of other disordered condensed-phase systems by combining high-level electronic structure theory and machine learning.

\section{Methods}

\subsection{Machine learning}

To develop a highly data efficient strategy that only requires a minimal number of energies from periodic high level electronic structure methods, such as CCSD, CCSD(T), and AFQMC, and produce accurate MLPs, we exploited a combination of approaches including an active learning procedure for curating a small but comprehensive training set, the training of MLPs to energies for small periodic boxes (16 waters) and showing that they reproduce properties when used for larger simulations (64 waters) and a transfer learning approach that leverages the transferability of physics from lower tier electronic structure methods. 

\subsubsection{Developing an efficient training set by leveraging a committee of machine learning potentials}
\label{sec:methods_active_learning}
We first applied an active learning procedure to curate a data efficient set of configurations to train our MLPs. We invoked the commonly employed Query-by-Committee (QbC)\cite{Seung1992,Krogh1994,Schran2020} approach to iteratively add configurations to the training set. At every iteration, the current dataset of configuration energies was used to train a committee of 8 MLPs, each a Behler-Parrinello neural network potential\cite{Behler2007,Morawietz2016} with a different random initialization of weights and random train-validation (90-10) splits of the dataset. To mitigate overfitting, we applied early stopping to each of these MLPs in the committee, monitoring the energy prediction error over a validation set. The mean potential energy surface obtained for this committee MLP was used to run a short MD simulation (SI Sec.~\ref{sec:si-md-details}) that was terminated either when the system becomes unphysical (i.e., reaching a temperature greater than 400K) or when 50~ps of simulation trajectory was generated. We then selected the 10 configurations where the committee MLP had the largest standard deviation in its potential energy prediction to recalculate at the target level of electronic structure theory and add to the training set for the next iteration of this procedure. To prevent selected configurations from being overly correlated with one another, no two selected configurations were closer than 100~fs apart. The initial dataset of 50 configurations used to initialize this procedure was also selected via an iterative QbC procedure, as applied to a single 100~ps SCC-DFTB\cite{Elstner1998} (SI Sec.~\ref{sec:si-md-details}) trajectory, that started with 10 randomly selected configurations and added 10 additional configurations for each of 4 subsequent iterations.

We generated 5 different 200 configuration datasets by running 5 instances of our QbC active learning scheme where the target level of electronic structure theory was DFT using the revPBE0-D3\cite{Perdew1996,Zhang1998,Adamo1999} functional (SI Sec.~\ref{sec:si-dft-hf-dftb}) which was chosen due to being computationally efficient compared to the high level electronic structure methods and since it has previously been shown to produce the properties of liquid water accurately when combined with path integral simulations\cite{Marsalek2017}. To select the generated dataset on which the higher tier electronic structure methods would be used we then used each of the 5 datasets to train a committee MLP using revPBE0-D3. We then selected the one that gave the lowest force prediction error (RMSE) when evaluated on a test set of 1000 water configurations (64 water molecules) drawn from previously published AIMD simulations\cite{Marsalek2017}. To check the transferability of the selected 200 configuration dataset, we recalculated the energies and forces for the generated datasets and the test set using the BLYP\cite{Becke1988,Lee1988} functional (SI Sec.~\ref{sec:si-dft-hf-dftb}). After training and evaluating a new set of BLYP trained MLPs, the same dataset that resulted in the lowest error revPBE0-D3 trained MLP also gave the lowest error BLYP trained MLP (SI Fig.~\ref{fig:trainingset_force_lc}). Given the transferability of the relative utility of this dataset, we used this same 200 configuration dataset to train our CCSD, CCSD(T), and AFQMC MLPs.

\subsubsection{Training machine learned potentials on configurations of small systems}

\begin{figure*}[]
    \begin{center}
        \includegraphics[width=0.95\textwidth]{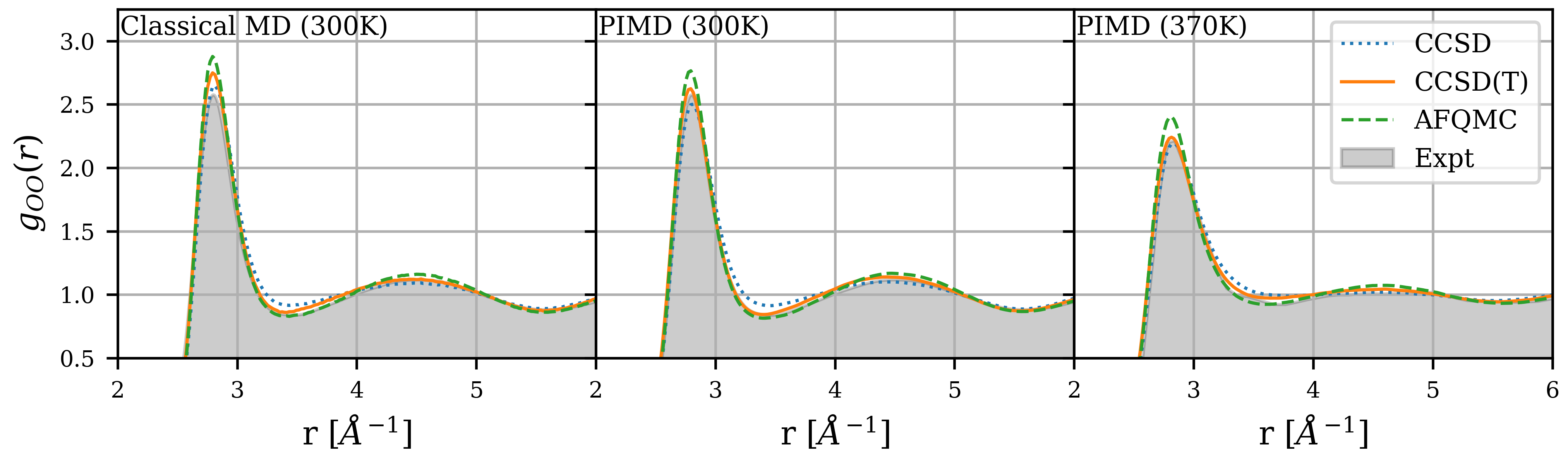}
    \end{center}
    \caption{Oxygen-oxygen RDFs for liquid water when running classical and PIMD simulations using NNPs fitted to either CCSD, CCSD(T), and AFQMC energies at both 300K and 370K. The PIMD oxygen-oxygen RDFs when using the CCSD(T) and AFQMC correspond closely with the experimental results shaded in grey at both 295K and 366K.}
	\label{fig:rdfs}
\end{figure*}

Given the significant scaling of CCSD, CCSD(T), and AFQMC's computational cost with system size, we sought to reduce the system size of the configurations used in our training set. We investigated the feasibility of training our committee MLP to the energies of a set configurations for small periodic boxes to accurately predict for properties associated with a larger simulation box. For liquid water running periodic molecular dynamics simulations of small water boxes (32 water molecules or fewer) leads to significant artifacts even in simple properties such as radial distribution functions (RDFs) when compared to larger system sizes\cite{Kuhne2009}. However, as demonstrated in SI Sec.~\ref{sec:si-revpbe0-refit-tests} for the revPBE0-D3 functional, if one trains a MLP on energies of small (16 molecule) periodic water configurations and then uses the resulting model to perform dynamics of a larger system (64 molecule), then the results obtained are in excellent agreement with those obtained from performing a AIMD simulation at the larger system size. Based on this demonstration we therefore trained the MLPs for the higher level methods on periodic configurations of 16 water molecules using the transfer learning approach described in the following section and then report the properties obtained in Sec.~\ref{sec:results_and_discussion} by performing MD and PIMD simulations using 64 water molecules. 

\subsubsection{Transfer learning approach to train the MLPs}
\label{sec:methods_transfer_learning}

With only hundreds of energies from the higher tier periodic electronic structure methods available, to make efficient use of this data we employed a transfer learning approach\cite{Pan2010}. To achieve this we first trained a committee MLP at a lower level of electronic structure theory, DFT or Hartree Fock (HF) on 531 configurations using both energies and gradients to improve the fitting. The parameters obtained for these MLPs were then used as the starting point for the fits to the higher level methods. This strategy exploits the idea that while the lower level methods may not reach the levels of chemical accuracy required for some applications they do contain fundamental physical information (e.g., about the fact that O and H when in close proximity form a high frequency chemical bond) that can be used to structure the neural network that underlies the MLP. Hence while the high level data is used to tune the accuracy of the MLP, it is leveraging the physics learnt by the initial training to the lower level method.

While using transfer learning to make efficient use of very small amounts of high level electronic structure data has distinct advantages over starting from a randomly initialized MLP, one must be mindful of the risk of hysteresis in the final MLP, i.e., that by biasing the weights by taking them from a model trained on a low level electronic structure method, that the final MLP will incorrectly contain remnants of the failures of the low level method. Hence to assess our transfer learning approach's ability to produce a final model that accurately reproduces the properties of the target high level electronic structure method, we initialized the procedure starting from a range of different low level methods that each give very different properties of liquid water. By comparing the structural and dynamical properties of liquid water obtained by performing MD and PIMD simulations on the final models obtained from these different starting points, one can thus evaluate which properties are obtained universally across the models initialized from different low level methods and thus accurately reflect the high level electronic structure approach. 

In practice, following the convergence of the lower level fits to both energy and gradient data (see SI Sec.~\ref{sec:si-ml-details}), we retrained the committee MLP (8 separate MLPs) to the energies of the higher level method employing the extended Kalman filter optimizer as implemented in the n2p2\cite{Singraber2019} package and using a 90-10 train-validation split in order to monitor the prediction error over the validation set for early-stopping each individual fit to prevent overfitting. Before applying this transfer learning approach to the high level CCSD, CCSD(T) and AFQMC methods for which AIMD is not possible for the timescales required, in SI Sec.~\ref{sec:si-revpbe0-refit-tests} we tested it by training an MLP with revPBE0-D3 as the higher level method and BLYP and HF as the lower level methods from which the transfer learning was performed. We chose DFT with the revPBE0-D3 exchange-correlation functional as the higher level in this benchmark due to its accurate description of the properties of liquid water and since we can compare the results of the transfer learned MLPs to AIMD and AI-PIMD trajectories that we have previously obtained \cite{Marsalek2017}. We chose BLYP and HF as the lower level methods since the former gives an incredibly overstructured and dynamically sluggish description of water while the latter gives the opposite.

As shown in SI Sec.~\ref{sec:si-revpbe0-refit-tests}, MD simulations using the final transfer learned MLP models of revPBE0-D3 using only 200 energies starting from either BLYP or HF correctly capture the target RDFs and VDOS for liquid water at 300K obtained from AIMD simulations, which are both markedly different from those given by low level methods themselves (BLYP water has a more structured oxygen-oxygen RDF and higher wavenumber hydrogen VDOS bend and stretch peak positions than revPBE0-D3 water, and vice-versa for HF water). The agreement with the reference revPBE0-D3 results for both models is particularly strong for the RDFs. For the VDOS, the BLYP-initialized model outperforms the HF-initialized model in capturing the high frequency O-H stretch peak. Both models also accurately reproduce the RDFs obtained from AI-PIMD simulations of revPBE0-D3 for liquid water\cite{Marsalek2017} at 300K (SI Fig.~\ref{fig:revpbe0-D3_refits_rdfs_pimd_300K}) with the only discrepancy again being in the VDOS (SI Fig.~\ref{fig:revpbe0-D3_refits_vdos_pimd_300K_370K}) where the BLYP-initialized MLP captures the full spectrum whereas HF shows an overstructured and blue shifted OH stretch region. 

Additionally, we compared our transfer learning approach to two common alternative machine learning approaches: directly training a committee MLP on the high level energies starting from randomly initialized weights and training a committee delta-learning model that corrects from the lower to higher level method.  To allow for a fair comparison, all three types of models were trained to the same 200 configuration training set of energies (revPBE0-D3) for liquid water using the same MLP architecture and optimization settings. The randomly initialized model resulted in a potential that was unstable for the purposes of running MD for a periodic simulation box of 64 water molecules, with the instantaneous temperature drifting severely after the first simulation step. For the delta-learning model, we trained a committee MLP to capture the energy difference between revPBE0-D3 and another committee MLP trained to BLYP. The delta-learning model was similarly as unstable as the randomly initialized model. Hence for liquid water with 200 energies at the higher level we found that the transfer learning approach we have detailed above provides the most accurate results.

Given the demonstrated efficacy of our transfer learning procedure, in Sec.~\ref{sec:results_and_discussion} we applied this approach to train committee MLPs to CCSD, CCSD(T), and AFQMC energies for liquid water using three different sets of MLP initializations: HF, BLYP, and revPBE0-D3. 

\subsection{Correlated electronic structure methods}
We perform correlated electronic structure calculations of liquid water with periodic boundary conditions at the $\Gamma$-point using AFQMC, CCSD, and CCSD(T). Periodic CCSD and CCSD(T) calculations were performed using PySCF~\cite{sun2020recent,McClain2017} where electron-repulsion integrals are handled using the range-separated density fitting method~\cite{Ye21JCP1,Ye21JCP2}, and AFQMC calculations were performed using QMCPACK\cite{kim2018qmcpack,Kent2020May} and ipie\cite{Malone2022Sep}. These calculations all began with a periodic spin-restricted HF calculation also perfomed using PySCF. We provide brief details here, and further information can be found in SI Secs.~\ref{sec:si-ccsd-ccsdt} and \ref{sec:si-afqmc}.

AFQMC is a projector Monte Carlo method where the ground state of a given Hamiltonian is obtained via imaginary time evolution. Without any approximations, it scales exponentially as the system size grows due to the fermionic sign problem. We use the phaseless approximation~\cite{Zhang2003Apr} to obtain an algorithm that scales with the system size $N$ as $O(N^3)-O(N^4)$ for each sample at the expense of introducing errors in the final ground state energy estimate. The phaseless approximation sets a boundary condition in imaginary time evolution using a priori chosen wavefunction called the trial wavefunction. The bias from this approximation becomes smaller as one improves the quality of trial wavefunctions. In this work we employ the simplest trial wavefunction, the spin-restricted Hartree-Fock determinant. A recent benchmark study examined the accuracy of AFQMC with Hartree-Fock trial wavefunctions over 1004 data points~\cite{Lee2022Oct}, and based on these results at this level of approximation we expect the accuracy of AFQMC to be between CCSD and CCSD(T) for the problems considered here.

\newcommand{\me}{\mathrm{e}}
\newcommand{\tr}[1]{\textrm{#1}}
Coupled cluster (CC) parameterizes the electronic wavefunction using an exponential ansatz, $\ket{\Psi_{\tr{CC}}} = \me^{\hat{T}} \ket{\Phi_{\tr{HF}}}$, where $\hat{T}$ creates all possible particle-hole excitations from the HF reference and is determined by iteratively solving a set of coupled non-linear equations\cite{Cizek1966}. CCSD approximates the full CC ansatz by truncating the $T$-operator at single and double-excitation levels, ~$\hat{T} = \hat{T}_1 + \hat{T}_2$\cite{Purvis1982}. CCSD(T) improves upon CCSD by further including the contribution from triple excitations in a perturbative (non-iterative) manner\cite{Raghavachari1989} and is often referred to as the ``gold standard" in the quantum chemistry of simple molecules. The computational cost of CCSD and CCSD(T) scales more steeply than that of AFQMC, as $O(N^6)$ and $O(N^7)$, respectively. In this work, we avoid the high computational cost of full CCSD and full CCSD(T) by systematically compressing the virtual space using the frozen natural orbital (FNO) approximation \cite{Taube08JCP,Lange20MP}, which we confirmed to introduce a negligible error (SI Sec.~\ref{sec:si-ccsd-ccsdt}).

\section{Results and Discussion}
\label{sec:results_and_discussion}
\begin{figure*}[]
    \begin{center}
        \includegraphics[width=0.95\textwidth]{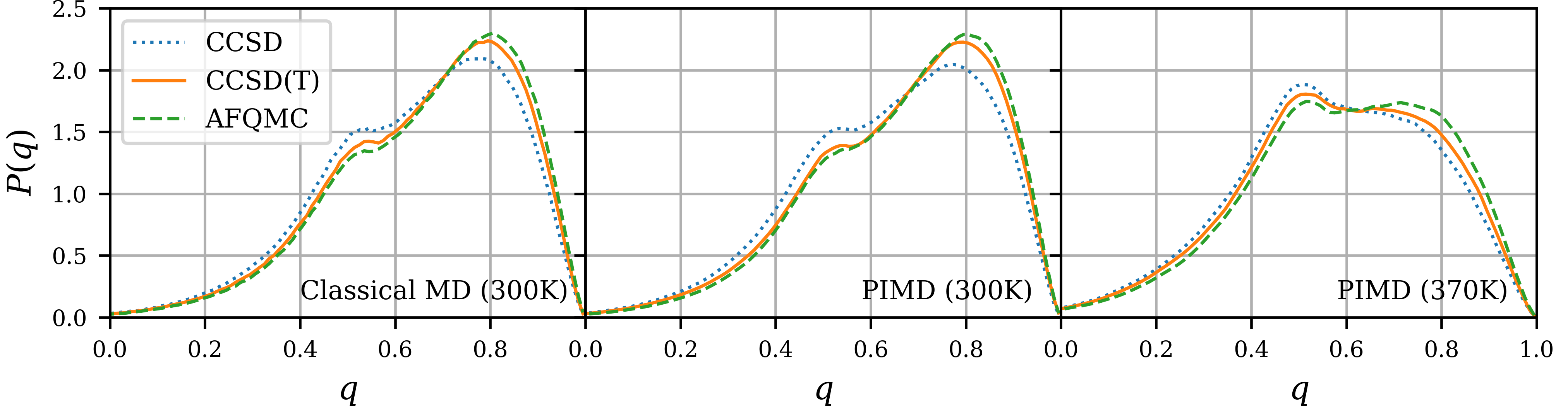}
    \end{center}
    \caption{Tetrahedral order parameter $q$ distributions as sampled via classical and PIMD simulations via NNPs fitted to either CCSD, CCSD(T), and AFQMC energies at both 300K and 370K.}
	\label{fig:tetraorder}
\end{figure*}

Having established the applicability of our data efficient approach to training MLPs that accurately reproduce the target potential energy surfaces on DFT, we now apply it to obtain the static and dynamic properties of liquid water using three correlated methods: CCSD, CCSD(T), and AFQMC. We trained a committee MLP, which as described in Sec.~\ref{sec:methods_active_learning} is formed of the mean of 8 independently trained MLPs, for each correlated electronic structure method. By performing MD and PIMD simulations in the NVT ensemble at 300~K and 370~K using the committee MLP for each electronic structure method, here we evaluate how different treatments of dynamical electron correlation affect the selected properties when the nuclei are treated classically or quantum mechanically. 

To provide an assessment of the accuracy of our transfer learned committee MLPs on different structural and dynamical properties as discussed in Sec.~\ref{sec:methods_transfer_learning} we can compare the consistency of the results obtained from models that have been initially trained to different lower level electronic structure methods. Hence, for each of the correlated methods we trained three transfer learned committee MLPs starting from HF, BLYP, and revPBE0-D3 as our lower level methods which are known to understructure, overstructure, and accurately reproduce the properties of water respectively. Performing classical MD at 300K from MLPs starting from these three different methods SI Fig.~\ref{fig:afqmc-rdfs-vdos-classical} shows that regardless of the lower level that the transfer learning was initialized with, the properties obtained at the high level, AFQMC in this case, coincide closely although the O-H stretch for the HF-initialized model is blue-shifted slightly. This demonstrates that our training set of 200 energies of periodic boxes consisting of 16 water molecules is large and diverse enough to train an accurate model of liquid water under these conditions using our transfer learning protocol. When nuclear quantum effects are included by peforming PIMD simulations using the committee MLPs, the oxygen-oxygen RDF and VDOS for the revPBE0-D3 and BLYP initialized transfer learning models remain consistent with one another (SI Fig.~\ref{fig:afqmc-rdfs-vdos-rpmd}).

In this section we show the results from the committee MLP generated using transfer learning from revPBE0-D3 for each of the correlated methods since, as shown in SI Fig.~\ref{tab:si_electronic_structure_correlations}, revPBE0-D3 shows the strongest correlation with the training set energies of CCSD, CCSD(T), and AFQMC out of the three initialization methods we used.

For our AFQMC results, it is important to note that the AFQMC energies contain stochastic error, with each of our N=200 training set AFQMC energies having a corresponding estimate for the standard error that ranges from 1-2~mH depending on the specific training set configuration. To evaluate how this level of error might affect our reported properties, we employed a test where we sample new training sets where the same N=200 configuration are used but a random value is added to each AFQMC energy to reflect the uncertainty of our AFQMC energies (see SI Sec.~\ref{sec:si-energy-noise-tests} for details). In total, 12 training sets were sampled and a transfer learned committee MLP was fit to each. SI Figs.~\ref{fig:afqmc_noise_rdf} and \ref{fig:afqmc_noise_vdos} serve to quantify the variations in the oxygen-oxygen RDF and hydrogen VDOS obtained from the 12 separate training sets due to the stochastic error in the AFQMC energies, with the grey shading representing the standard deviations. From this test we found that the stochastic error in our AFQMC calculations does introduce noticeable variations in the OH stretch peak of the hydrogen VDOS, particularly around the top of the peak, but the RDFs and VDOS are otherwise consistent for the different training sets.

\subsection{Static properties of water from correlated electronic structure methods}
\label{sec:results_static}
We first compare the static equilibrium properties for liquid water at 300~K and 370~K obtained via classical MD and PIMD simulations using the committee MLP for each correlated electronic structure method. For these properties PIMD exactly includes the NQEs for distinguishable particles, which is a highly reliable assumption for nuclei at this temperature. Figure~\ref{fig:rdfs} shows the oxygen-oxygen RDFs for each of CCSD, CCSD(T), and AFQMC as compared to the experimental results at 295~K\cite{Skinner2013} and 366~K\cite{Mariedahl2018}. At 300~K, classical MD CCSD(T) and AFQMC both give a first peak that is slightly higher than observed experimentally suggesting the liquid is overstructured. However, once NQEs are accounted for in the PIMD simulations both RDFs become slightly less structured and show better agreement with experiment, with that of CCSD(T) coinciding quantitatively. CCSD gives good agreement with experiment when used in classical MD simulations but is understructured when NQEs are included which is consistent with this electronic structure approach starting from a HF reference which gives a severely understructured liquid with the additional dynamical correlation added through the tiers of CC theory progressively structuring the liquid. At 370~K when used in PIMD simulations all methods give good agreement with the experiment with AFQMC again exhibiting a first peak that is higher than CCSD, CCSD(T) and experiment. SI Fig.~\ref{fig:mbpol_comp_rdfs_classical_300K} and Fig.~\ref{fig:mbpol_comp_rdfs_pimd_300K} show the hydrogen-hydrogen and oxygen-hydrogen RDFs at 300~K sampled via classical MD and PIMD, respectively, where all three electronic structure methods give similar results but AFQMC again exhibits a more structured hydrogen bond network with the first intermolecular OH peak at $\sim$1.85~$\AA$, which corresponds to hydrogen bonds, being slightly higher than the other methods.

The tetrahedral order parameter provides a measure of higher order structural correlations within water's hydrogen bond network beyond the purely radial information encoded in the RDFs. The tetrahedral order parameter $q$ is defined for a given water molecule as\cite{Errington2001},
\begin{equation}
    q = 1 - \frac{3}{8} \sum_{j=1}^{3}\sum_{k=j+1}^{4} \left( \cos{\theta_{jk}} + \frac{1}{3}\right),
\end{equation}
where $\theta_{jk}$ is the angle that a given water molecule's oxygen atom makes with two neighboring oxygen atoms $j$ and $k$. The tetrahedral order parameter thus ranges from 0 to 1 with higher values indicating that the hydrogen bond network possesses angles closer to that of a perfect tetrahedral arrangement of the four nearest neighbour oxygen atoms around a central water molecule. As shown in Fig.~\ref{fig:tetraorder} at 300~K for both classical MD and PIMD this property further highlights the understructured hydrogen bond network of CCSD compared to the more accurate correlated methods, CCSD(T) and AFQMC, that are in close agreement. At 370~K the distribution of the tetrahedral order parameter for all three methods shifts to lower values. 

The comparison of these static equilibrium properties suggests that the inclusion of higher order electron correlation contributions in methods like CCSD(T) and AFQMC, as compared to CCSD or HF, results in a greater degree of structuring in liquid water at 300~K and 370~K. Given the directional nature of this additional structuring, as seen from the greater probability density at higher $q$ in Figure~\ref{fig:tetraorder}, this arises from slightly stronger hydrogen bonds being formed when using the two higher level methods. Our comparisons of the oxygen-oxygen RDFs obtained from classical and PIMD simulations at 300~K also show that for these correlated methods the inclusion of NQEs works to slightly destructure liquid water. The relatively subtle overall effect of NQEs on liquid water around 300~K is known to arise from the close balance of competing quantum effects in this system\cite{Habershon2009}.

\subsection{Dynamical properties of water from correlated electronic structure methods}

\begin{table}[]
    \vspace{4mm}
    \centering
    \begin{tabular}{ | m{0.14\textwidth} || m{0.09\textwidth} | m{0.09\textwidth}| m{0.09\textwidth} | } 
        \hline
         & AFQMC & CCSD(T) & CCSD \\ \hline\hline
        Classical T=300K ($10^{-9} $ m$^2$/s) & 2.09 (0.05) & 2.21 (0.06) & 2.60 (0.08) \\ \hline
        TRPMD T=300K ($10^{-9} $ m$^2$/s) & 2.16 (0.09) & 2.30 (0.08) & 2.80 (0.11) \\ \hline
        TRPMD T=370K ($10^{-9} $ m$^2$/s) & 7.09 (0.11) & 8.16 (0.14) & 8.29 (0.14) \\ \hline
    \end{tabular}
    \caption{Diffusion coefficients for liquid water when running classical and TRPMD simulations using NNPs fitted to either CCSD, CCSD(T), and AFQMC energies at both 300K and 370K. Mean diffusion coefficients and standard errors of the mean are reported and are calculated using 20~ps length trajectories taken from 1~ns classical MD or 500~ps TRPMD trajectories. The experimental diffusion coefficient for water at 300~K and 370~K are 2.41$\pm$0.05\cite{Holz2000} and 8.26$\pm$0.02 ($10^{-9} $ m$^2$/s)\cite{Dietrich}, respectively.}
    \label{tab:diffusion-coeffs}
\end{table}

We now turn our attention to how the different correlated electronic structure methods behave when used to compute dynamical properties of liquid water, namely the self diffusion coefficient and VDOS. For these properties we compare the results obtained from classical MD and TRPMD. For these properties, since real time quantum dynamics is intractable for such a large atomistic condensed phase system for the timescales required to compute these properties, we use TRPMD to approximate the role of NQEs. TRPMD has previously been shown to be an accurate way for treating NQEs in the dynamics of condensed phase systems, however it is known to spuriously broaden high-frequency vibrational modes\cite{Rossi2014,Marsalek2017}. 

The diffusion coefficients obtained for the three correlated methods are shown in Table~\ref{tab:diffusion-coeffs} at 300~K and 370~K. Unlike the other properties we report, diffusion coefficients exhibit a notable scaling with system size that must be corrected for to make comparisons with experiment. Hence, as described in SI Sec.~\ref{sec:si-diffusion-coefficients}, the diffusion coefficients were computed using simulations of 64 water molecules and then extrapolated to the infinite system size limit using the previously derived system size scaling relation\cite{Yeh2004} and the experimental viscosity of water\cite{Kestin1978}. At 300~K, where the experimentally measured diffusion coefficient is 2.41$\pm$0.05$\times 10^{-9} $ m$^2$/s~\cite{Holz2000}, when classical MD is used AFQMC and CCSD(T) yield smaller diffusion coefficients than observed experimentally, 2.09$\pm$0.05 and 2.21$\pm$0.06 $\times 10^{-9} $ m$^2$/s respectively, while CCSD gives a larger value 2.60$\pm$0.08 $\times 10^{-9} $ m$^2$/s. This behavior is consistent with the trends observed for the electronic structure approaches in the structural properties, where CCSD formed a slightly understructured hydrogen bond network compared to the more accurate correlated methods, which here leads to faster dynamics. Upon including NQEs using TRPMD, the diffusion coefficients for all three electronic methods increase, consistent with the disruption of the hydrogen bond network upon including zero-point energy, which brings the CCSD(T) result (2.30$\pm$0.08 $\times 10^{-9} $ m$^2$/s) to within the statistical error bar of the experimentally observed value. Even with NQEs included the AFQMC diffusion coefficient is lower (2.16$\pm$0.09 $\times 10^{-9} $ m$^2$/s) than experiment, consistent with it forming a more structured liquid. However, it should be noted that the discrepancy in the diffusion coefficient is very small; to change water's diffusion coefficient from that observed at 300~K via classical MD using our CCSD(T) model to the value obtained by performing TRPMD dynamics would require less than a 2~K change in the temperature of the liquid\cite{Dietrich}. In addition, for dynamical properties TRPMD only approximately includes NQEs and hence the better agreement of CCSD(T) with the experimental value could be changed if a different approach was used to treat the quantum dynamics of the nuclei. At 370~K, when TRPMD is used CCSD and CCSD(T) give similar results (8.29$\pm$0.14 and 8.16$\pm$0.14 $\times 10^{-9} $ m$^2$/s, respectively), both of which are close to the experimental value of 8.26$\pm$0.02 $\times 10^{-9} $ m$^2$/s\cite{Dietrich}. The diffusion of AFQMC water is again considerably slower, which is consistent with the relatively greater degree of structure we saw in the latter's respective oxygen-oxygen RDF and $q$ distribution.

\begin{figure}
    \vspace{2mm}
    \begin{center}
        \includegraphics[width=0.475\textwidth]{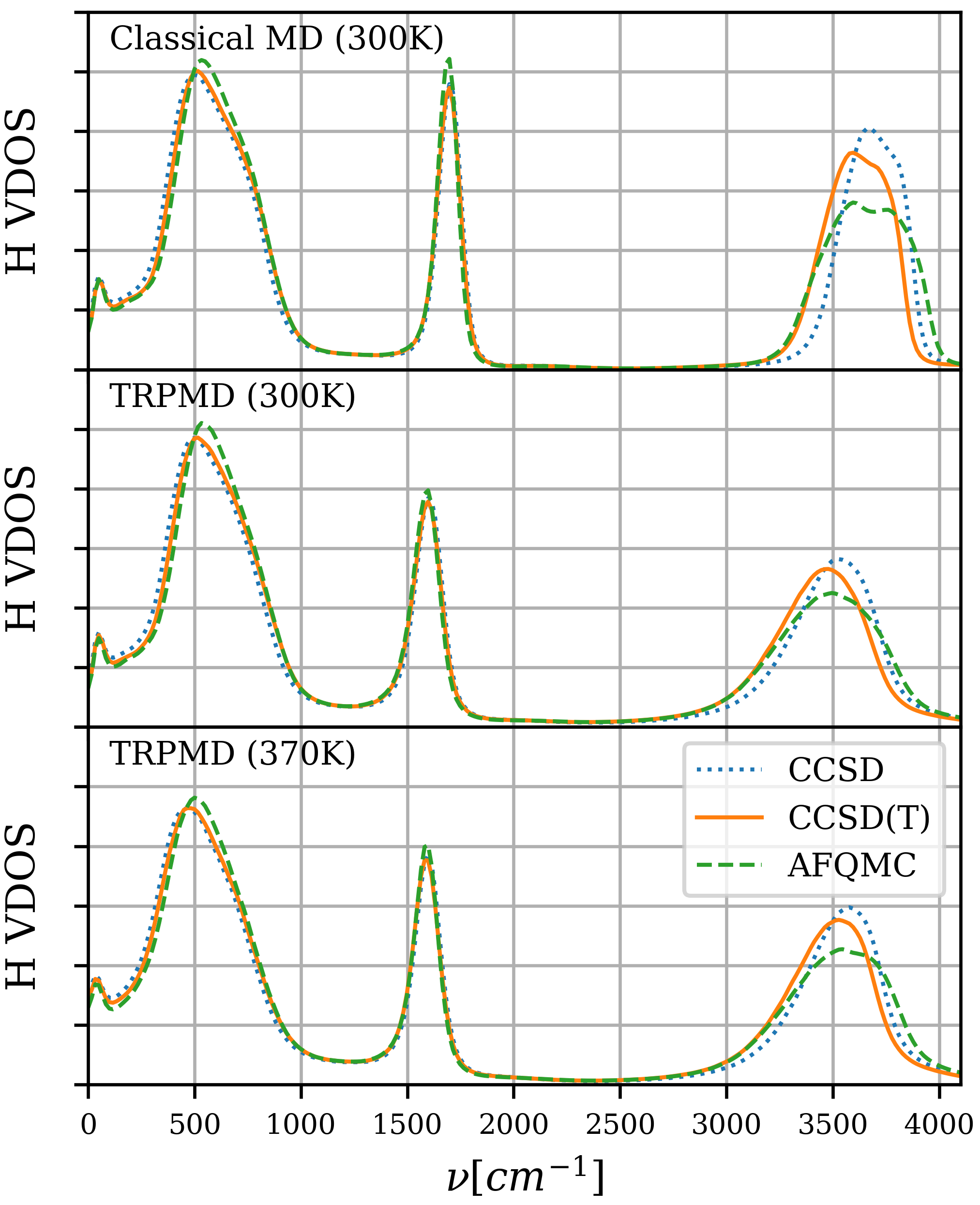}
    \end{center}
    \caption{The hydrogen VDOS for liquid water when running classical and TRPMD simulations using NNPs fitted to either CCSD, CCSD(T), and AFQMC energies at both 300K and 370K.}
    \label{fig:vdos}
\end{figure}

The VDOS in Figure~\ref{fig:vdos} provides more information on the frequency dependence of the dynamics of water for the three electronic structure methods, since the diffusion coefficient is simply proportional to its zero frequency limit. All three methods give qualitatively similar VDOS around the lower frequency librational band ($\sim$500~cm$^{-1}$) and peak associated with the H-O-H bending mode ($\sim$1600~cm$^{-1}$). At 300~K using classical MD, the main qualitative difference lies in the OH stretch peak (3000-4000 cm$^{-1}$), with CCSD(T) giving a peak that is redshifted by $\sim$80~cm$^{-1}$ with respect to the CCSD peak, while the AFQMC peak is slightly broader than the others and is centered closer to the CCSD result. At the low-frequency end of the O-H stretch peak, the VDOS for CCSD(T) and AFQMC coincide with one another. Low frequency O-H stretches are typically associated with stronger hydrogen bonds and the frequency of the O-H stretch peak has previously been demonstrated to be inversely correlated with the tetrahedrality parameter\cite{Morawietz2018}. Hence, the consistency between CCSD(T) and AFQMC at the low frequency part of their respective O-H stretch peaks is consistent with the structural evidence in Sec.~\ref{sec:results_static} showing that these two methods similarly structure water via slightly stronger hydrogen bonds as compared to CCSD. The TRPMD simulations at 300~K, which include NQEs via TRPMD simulations, lead to a $\sim$120~cm$^{-1}$ red-shift and broadening of the stretch peak and a $\sim$100~cm$^{-1}$ shift in the bend for all three methods, which is consistent with observations from previous studies\cite{Rossi2014,Marsalek2017}. Although some of the broadening likely arises from physical effects, TRPMD is known to introduce spurious broadening of high-frequency vibrational modes\cite{Rossi2014,Marsalek2017}. At 370~K the most significant difference is the shift in the zero frequency intensity, which we expect since this is directly related to the self-diffusion coefficient and otherwise the VDOS is qualitatively similar with respect to the 300~K results. This is expected since for the high frequency modes the zero point energy greatly exceeds the thermal energy in the the mode, i.e. $k_{B}T<<\hbar \omega/2$ where $k_{B}$ is the Boltzmann constant, $T$ is the temperature and $\omega$ is the frequency of the mode. 

Overall, the trends we observe in the dynamical properties largely mirror the evidence we presented for the structural properties showing that CCSD results in an understructured description of liquid water at 300~K, as compared to CCSD(T) and AFQMC, while at 300~K and 370~K AFQMC overstructures water. The differences in the diffusion coefficients in Table~\ref{tab:diffusion-coeffs} reflect this trend, with CCSD overestimating the experimental diffusion coefficient at 300~K with its understructured description of water and AFQMC underpredicting the diffusion coefficient at 370~K. With the VDOS the main differences between the three correlated methods manifest in the O-H stretch peak positions and breadth, with peaks given by MD at 300~K for both CCSD(T) and AFQMC skewed more to lower frequencies that are associated with stronger hydrogen bonds.

\section{Conclusion}

In summary, we leveraged developments in high-level periodic electronic structure theory and exploited methods to improve the data efficiency of fitting MLPs to investigate the static and dynamical properties of liquid water at the level of CCSD, CCSD(T), and AFQMC. We devised a data efficient protocol for training MLPs that uses small periodic boxes of water (16 molecules) sampled judiciously via an iterative QbC active learning procedure. To make the most out of the few configuration energies we can afford to compute, we also employed a transfer learning approach that leverages the transferability of physics between lower level electronic structure methods (e.g. DFT with the BLYP functional, revPBE0-D3 functional, or HF) and our target higher-level methods, using MLPs fit to the former to initialize a fine-tuning transfer learning fit to the latter. Using this approach we showed that we can train stable MLPs with as few as 50 configuration energies, capture the RDFs with 100 (SI Fig.~\ref{fig:revpbe0-D3_refits_rdfs_classical_300K}), and with 200 configuration energies obtain both accurate static and dynamical properties such as the diffusion constant and VDOS (SI Fig.~\ref{fig:revpbe0-D3_refits_vdos_classical_300K}). In contrast, with these same 200 energies we were unable to train stable models using delta learning or using random initialization of the model.

We used our MLPs trained to CCSD, CCSD(T), and AFQMC to examine how different static and dynamical properties of liquid water are affected by the level of dynamic electron correlation accounted for and the inclusion of NQEs. Our results indicate that CCSD tends to understructure liquid water and overpredict the diffusion coefficient, as compared to experiment. On the other hand, both CCSD(T) and AFQMC give oxygen-oxygen RDFs and diffusion coefficients that are more consistent with experimental values at 300~K, suggesting that the more accurate treatment of dynamical correlation present in these methods is sufficient for describing liquid water. The inclusion of NQEs for our 300~K simulations bring the CCSD(T) and AFQMC results in even closer agreement with experiment and seems to generally manifest as a slight destructuring of liquid water for all three electronic structure descriptions. This small destructuring upon including NQEs for these correlated methods is in contrast to some DFT exchange correlation functionals where due to the overprediction of the anharmonicity of the O-H coordinate the inclusion of NQEs works to structure the liquid phase\cite{Marsalek2017}.

Ultimately, we envision that the configurations and energies that form the training dataset, the resulting MLPs, and the protocols we employed here will be useful in their own separate respects for future work in modeling potential energy surfaces for condensed phase systems.

\section*{Acknowledgments}
T.E.M and M.S.C were supported by the National Science Foundation under Grant No.~CHE-2154291. This research also used resources of the National Energy Research Scientific Computing Center (NERSC), a U.S.~Department of Energy Office of Science User Facility operated under Contract No.~DE-AC02-05CH11231. This work was also supported by the US Air Force Office of Scientific Research under Grant No. FA9550-21-1-0400 (H.-Z.Y. and T.C.B.). We acknowledge the computing resources provided by Columbia University’s Shared Research Computing Facility project, which is supported by NIH Research Facility Improvement Grant No. 1G20RR030893-01, and associated funds from the New York State Empire State Development, Division of Science Technology and Innovation (NYSTAR) Contract No. C090171, both awarded April 15, 2010. The Flatiron Institute is a division of the Simons Foundation.

\bibliography{references}	

\begin{thebibliography}{61}%
\makeatletter
\providecommand \@ifxundefined [1]{%
 \@ifx{#1\undefined}
}%
\providecommand \@ifnum [1]{%
 \ifnum #1\expandafter \@firstoftwo
 \else \expandafter \@secondoftwo
 \fi
}%
\providecommand \@ifx [1]{%
 \ifx #1\expandafter \@firstoftwo
 \else \expandafter \@secondoftwo
 \fi
}%
\providecommand \natexlab [1]{#1}%
\providecommand \enquote  [1]{``#1''}%
\providecommand \bibnamefont  [1]{#1}%
\providecommand \bibfnamefont [1]{#1}%
\providecommand \citenamefont [1]{#1}%
\providecommand \href@noop [0]{\@secondoftwo}%
\providecommand \href [0]{\begingroup \@sanitize@url \@href}%
\providecommand \@href[1]{\@@startlink{#1}\@@href}%
\providecommand \@@href[1]{\endgroup#1\@@endlink}%
\providecommand \@sanitize@url [0]{\catcode `\\12\catcode `\$12\catcode
  `\&12\catcode `\#12\catcode `\^12\catcode `\_12\catcode `\%12\relax}%
\providecommand \@@startlink[1]{}%
\providecommand \@@endlink[0]{}%
\providecommand \url  [0]{\begingroup\@sanitize@url \@url }%
\providecommand \@url [1]{\endgroup\@href {#1}{\urlprefix }}%
\providecommand \urlprefix  [0]{URL }%
\providecommand \Eprint [0]{\href }%
\providecommand \doibase [0]{https://doi.org/}%
\providecommand \selectlanguage [0]{\@gobble}%
\providecommand \bibinfo  [0]{\@secondoftwo}%
\providecommand \bibfield  [0]{\@secondoftwo}%
\providecommand \translation [1]{[#1]}%
\providecommand \BibitemOpen [0]{}%
\providecommand \bibitemStop [0]{}%
\providecommand \bibitemNoStop [0]{.\EOS\space}%
\providecommand \EOS [0]{\spacefactor3000\relax}%
\providecommand \BibitemShut  [1]{\csname bibitem#1\endcsname}%
\let\auto@bib@innerbib\@empty
\bibitem [{\citenamefont {DiStasio}\ \emph {et~al.}(2014)\citenamefont
  {DiStasio}, \citenamefont {Santra}, \citenamefont {Li}, \citenamefont {Wu},\
  and\ \citenamefont {Car}}]{Distasio2014}%
  \BibitemOpen
  \bibfield  {author} {\bibinfo {author} {\bibfnamefont {R.~A.}\ \bibnamefont
  {DiStasio}}, \bibinfo {author} {\bibfnamefont {B.}~\bibnamefont {Santra}},
  \bibinfo {author} {\bibfnamefont {Z.}~\bibnamefont {Li}}, \bibinfo {author}
  {\bibfnamefont {X.}~\bibnamefont {Wu}},\ and\ \bibinfo {author}
  {\bibfnamefont {R.}~\bibnamefont {Car}},\ }\bibfield  {title} {\enquote
  {\bibinfo {title} {The individual and collective effects of exact exchange
  and dispersion interactions on the ab initio structure of liquid water},}\
  }\href {https://doi.org/10.1063/1.4893377} {\bibfield  {journal} {\bibinfo
  {journal} {The Journal of Chemical Physics}\ }\textbf {\bibinfo {volume}
  {141}},\ \bibinfo {pages} {084502} (\bibinfo {year} {2014})},\ \Eprint
  {https://arxiv.org/abs/https://doi.org/10.1063/1.4893377}
  {https://doi.org/10.1063/1.4893377} \BibitemShut {NoStop}%
\bibitem [{\citenamefont {Gillan}, \citenamefont {Alf{\`{e}}},\ and\
  \citenamefont {Michaelides}(2016)}]{Gillan2016}%
  \BibitemOpen
  \bibfield  {author} {\bibinfo {author} {\bibfnamefont {M.~J.}\ \bibnamefont
  {Gillan}}, \bibinfo {author} {\bibfnamefont {D.}~\bibnamefont {Alf{\`{e}}}},\
  and\ \bibinfo {author} {\bibfnamefont {A.}~\bibnamefont {Michaelides}},\
  }\bibfield  {title} {\enquote {\bibinfo {title} {{Perspective: How good is
  DFT for water?}}}\ }\href {https://doi.org/10.1063/1.4944633} {\bibfield
  {journal} {\bibinfo  {journal} {Journal of Chemical Physics}\ }\textbf
  {\bibinfo {volume} {144}} (\bibinfo {year} {2016}),\
  10.1063/1.4944633}\BibitemShut {NoStop}%
\bibitem [{\citenamefont {{Del Ben}}\ \emph {et~al.}(2013)\citenamefont {{Del
  Ben}}, \citenamefont {Sch{\"{o}}nherr}, \citenamefont {Hutter},\ and\
  \citenamefont {Vandevondele}}]{DelBen2013}%
  \BibitemOpen
  \bibfield  {author} {\bibinfo {author} {\bibfnamefont {M.}~\bibnamefont {{Del
  Ben}}}, \bibinfo {author} {\bibfnamefont {M.}~\bibnamefont
  {Sch{\"{o}}nherr}}, \bibinfo {author} {\bibfnamefont {J.}~\bibnamefont
  {Hutter}},\ and\ \bibinfo {author} {\bibfnamefont {J.}~\bibnamefont
  {Vandevondele}},\ }\bibfield  {title} {\enquote {\bibinfo {title} {{Bulk
  liquid water at ambient temperature and pressure from mp2 theory}},}\ }\href
  {https://doi.org/10.1021/jz401931f} {\bibfield  {journal} {\bibinfo
  {journal} {Journal of Physical Chemistry Letters}\ }\textbf {\bibinfo
  {volume} {4}},\ \bibinfo {pages} {3753--3759} (\bibinfo {year}
  {2013})}\BibitemShut {NoStop}%
\bibitem [{\citenamefont {Morawietz}\ \emph {et~al.}(2016)\citenamefont
  {Morawietz}, \citenamefont {Singraber}, \citenamefont {Dellago},\ and\
  \citenamefont {Behler}}]{Morawietz2016}%
  \BibitemOpen
  \bibfield  {author} {\bibinfo {author} {\bibfnamefont {T.}~\bibnamefont
  {Morawietz}}, \bibinfo {author} {\bibfnamefont {A.}~\bibnamefont
  {Singraber}}, \bibinfo {author} {\bibfnamefont {C.}~\bibnamefont {Dellago}},\
  and\ \bibinfo {author} {\bibfnamefont {J.}~\bibnamefont {Behler}},\
  }\bibfield  {title} {\enquote {\bibinfo {title} {{How van der Waals
  interactions determine the unique properties of water}},}\ }\href
  {https://doi.org/10.1073/pnas.1602375113} {\bibfield  {journal} {\bibinfo
  {journal} {Proceedings of the National Academy of Sciences}\ }\textbf
  {\bibinfo {volume} {113}},\ \bibinfo {pages} {8368--8373} (\bibinfo {year}
  {2016})},\ \Eprint {https://arxiv.org/abs/1606.07775} {arXiv:1606.07775}
  \BibitemShut {NoStop}%
\bibitem [{\citenamefont {Morawietz}\ \emph {et~al.}(2018)\citenamefont
  {Morawietz}, \citenamefont {Marsalek}, \citenamefont {Pattenaude},
  \citenamefont {Streacker}, \citenamefont {Ben-Amotz},\ and\ \citenamefont
  {Markland}}]{Morawietz2018}%
  \BibitemOpen
  \bibfield  {author} {\bibinfo {author} {\bibfnamefont {T.}~\bibnamefont
  {Morawietz}}, \bibinfo {author} {\bibfnamefont {O.}~\bibnamefont {Marsalek}},
  \bibinfo {author} {\bibfnamefont {S.~R.}\ \bibnamefont {Pattenaude}},
  \bibinfo {author} {\bibfnamefont {L.~M.}\ \bibnamefont {Streacker}}, \bibinfo
  {author} {\bibfnamefont {D.}~\bibnamefont {Ben-Amotz}},\ and\ \bibinfo
  {author} {\bibfnamefont {T.~E.}\ \bibnamefont {Markland}},\ }\bibfield
  {title} {\enquote {\bibinfo {title} {{The Interplay of Structure and Dynamics
  in the Raman Spectrum of Liquid Water over the Full Frequency and Temperature
  Range}},}\ }\href {https://doi.org/10.1021/acs.jpclett.8b00133} {\bibfield
  {journal} {\bibinfo  {journal} {Journal of Physical Chemistry Letters}\
  }\textbf {\bibinfo {volume} {9}},\ \bibinfo {pages} {851--857} (\bibinfo
  {year} {2018})}\BibitemShut {NoStop}%
\bibitem [{\citenamefont {Zhang}\ \emph {et~al.}(2018)\citenamefont {Zhang},
  \citenamefont {Han}, \citenamefont {Wang}, \citenamefont {Car},\ and\
  \citenamefont {E}}]{Zhang2018}%
  \BibitemOpen
  \bibfield  {author} {\bibinfo {author} {\bibfnamefont {L.}~\bibnamefont
  {Zhang}}, \bibinfo {author} {\bibfnamefont {J.}~\bibnamefont {Han}}, \bibinfo
  {author} {\bibfnamefont {H.}~\bibnamefont {Wang}}, \bibinfo {author}
  {\bibfnamefont {R.}~\bibnamefont {Car}},\ and\ \bibinfo {author}
  {\bibfnamefont {W.}~\bibnamefont {E}},\ }\bibfield  {title} {\enquote
  {\bibinfo {title} {Deep potential molecular dynamics: A scalable model with
  the accuracy of quantum mechanics},}\ }\href
  {https://doi.org/10.1103/PhysRevLett.120.143001} {\bibfield  {journal}
  {\bibinfo  {journal} {Phys. Rev. Lett.}\ }\textbf {\bibinfo {volume} {120}},\
  \bibinfo {pages} {143001} (\bibinfo {year} {2018})}\BibitemShut {NoStop}%
\bibitem [{\citenamefont {Cheng}\ \emph {et~al.}(2019)\citenamefont {Cheng},
  \citenamefont {Engel}, \citenamefont {Behler}, \citenamefont {Dellago},\ and\
  \citenamefont {Ceriotti}}]{Cheng2019}%
  \BibitemOpen
  \bibfield  {author} {\bibinfo {author} {\bibfnamefont {B.}~\bibnamefont
  {Cheng}}, \bibinfo {author} {\bibfnamefont {E.~A.}\ \bibnamefont {Engel}},
  \bibinfo {author} {\bibfnamefont {J.}~\bibnamefont {Behler}}, \bibinfo
  {author} {\bibfnamefont {C.}~\bibnamefont {Dellago}},\ and\ \bibinfo {author}
  {\bibfnamefont {M.}~\bibnamefont {Ceriotti}},\ }\bibfield  {title} {\enquote
  {\bibinfo {title} {Ab initio thermodynamics of liquid and solid water},}\
  }\href {https://doi.org/10.1073/pnas.1815117116} {\bibfield  {journal}
  {\bibinfo  {journal} {Proceedings of the National Academy of Sciences}\
  }\textbf {\bibinfo {volume} {116}},\ \bibinfo {pages} {1110--1115} (\bibinfo
  {year} {2019})},\ \Eprint
  {https://arxiv.org/abs/https://www.pnas.org/doi/pdf/10.1073/pnas.1815117116}
  {https://www.pnas.org/doi/pdf/10.1073/pnas.1815117116} \BibitemShut {NoStop}%
\bibitem [{\citenamefont {Schran}, \citenamefont {Brezina},\ and\ \citenamefont
  {Marsalek}(2020)}]{Schran2020}%
  \BibitemOpen
  \bibfield  {author} {\bibinfo {author} {\bibfnamefont {C.}~\bibnamefont
  {Schran}}, \bibinfo {author} {\bibfnamefont {K.}~\bibnamefont {Brezina}},\
  and\ \bibinfo {author} {\bibfnamefont {O.}~\bibnamefont {Marsalek}},\
  }\bibfield  {title} {\enquote {\bibinfo {title} {Committee neural network
  potentials control generalization errors and enable active learning},}\
  }\href {https://doi.org/10.1063/5.0016004} {\bibfield  {journal} {\bibinfo
  {journal} {The Journal of Chemical Physics}\ }\textbf {\bibinfo {volume}
  {153}},\ \bibinfo {pages} {104105} (\bibinfo {year} {2020})}\BibitemShut
  {NoStop}%
\bibitem [{\citenamefont {Yao}\ and\ \citenamefont {Kanai}(2020)}]{Yao2020}%
  \BibitemOpen
  \bibfield  {author} {\bibinfo {author} {\bibfnamefont {Y.}~\bibnamefont
  {Yao}}\ and\ \bibinfo {author} {\bibfnamefont {Y.}~\bibnamefont {Kanai}},\
  }\bibfield  {title} {\enquote {\bibinfo {title} {Temperature dependence of
  nuclear quantum effects on liquid water via artificial neural network model
  based on scan meta-gga functional},}\ }\href
  {https://doi.org/10.1063/5.0012815} {\bibfield  {journal} {\bibinfo
  {journal} {The Journal of Chemical Physics}\ }\textbf {\bibinfo {volume}
  {153}},\ \bibinfo {pages} {044114} (\bibinfo {year} {2020})},\ \Eprint
  {https://arxiv.org/abs/https://doi.org/10.1063/5.0012815}
  {https://doi.org/10.1063/5.0012815} \BibitemShut {NoStop}%
\bibitem [{\citenamefont {Zhang}\ \emph {et~al.}(2021)\citenamefont {Zhang},
  \citenamefont {Tang}, \citenamefont {Chen}, \citenamefont {Xu}, \citenamefont
  {Zhang}, \citenamefont {Qiu}, \citenamefont {Perdew}, \citenamefont {Klein},\
  and\ \citenamefont {Wu}}]{Zhang2021}%
  \BibitemOpen
  \bibfield  {author} {\bibinfo {author} {\bibfnamefont {C.}~\bibnamefont
  {Zhang}}, \bibinfo {author} {\bibfnamefont {F.}~\bibnamefont {Tang}},
  \bibinfo {author} {\bibfnamefont {M.}~\bibnamefont {Chen}}, \bibinfo {author}
  {\bibfnamefont {J.}~\bibnamefont {Xu}}, \bibinfo {author} {\bibfnamefont
  {L.}~\bibnamefont {Zhang}}, \bibinfo {author} {\bibfnamefont {D.~Y.}\
  \bibnamefont {Qiu}}, \bibinfo {author} {\bibfnamefont {J.~P.}\ \bibnamefont
  {Perdew}}, \bibinfo {author} {\bibfnamefont {M.~L.}\ \bibnamefont {Klein}},\
  and\ \bibinfo {author} {\bibfnamefont {X.}~\bibnamefont {Wu}},\ }\bibfield
  {title} {\enquote {\bibinfo {title} {Modeling liquid water by climbing up
  jacob’s ladder in density functional theory facilitated by using deep
  neural network potentials},}\ }\href
  {https://doi.org/10.1021/acs.jpcb.1c03884} {\bibfield  {journal} {\bibinfo
  {journal} {The Journal of Physical Chemistry B}\ }\textbf {\bibinfo {volume}
  {125}},\ \bibinfo {pages} {11444--11456} (\bibinfo {year} {2021})},\ \bibinfo
  {note} {pMID: 34533960},\ \Eprint
  {https://arxiv.org/abs/https://doi.org/10.1021/acs.jpcb.1c03884}
  {https://doi.org/10.1021/acs.jpcb.1c03884} \BibitemShut {NoStop}%
\bibitem [{\citenamefont {Yao}\ and\ \citenamefont {Kanai}(2021)}]{Yao2021}%
  \BibitemOpen
  \bibfield  {author} {\bibinfo {author} {\bibfnamefont {Y.}~\bibnamefont
  {Yao}}\ and\ \bibinfo {author} {\bibfnamefont {Y.}~\bibnamefont {Kanai}},\
  }\bibfield  {title} {\enquote {\bibinfo {title} {{Nuclear Quantum Effect and
  Its Temperature Dependence in Liquid Water from Random Phase Approximation
  via Artificial Neural Network}},}\ }\href
  {https://doi.org/10.1021/acs.jpclett.1c01566} {\bibfield  {journal} {\bibinfo
   {journal} {Journal of Physical Chemistry Letters}\ }\textbf {\bibinfo
  {volume} {12}},\ \bibinfo {pages} {6354--6362} (\bibinfo {year}
  {2021})}\BibitemShut {NoStop}%
\bibitem [{\citenamefont {Lan}\ \emph {et~al.}(2021)\citenamefont {Lan},
  \citenamefont {Wilkins}, \citenamefont {Rybkin}, \citenamefont {Iannuzzi},\
  and\ \citenamefont {Hutter}}]{Lan2021}%
  \BibitemOpen
  \bibfield  {author} {\bibinfo {author} {\bibfnamefont {J.}~\bibnamefont
  {Lan}}, \bibinfo {author} {\bibfnamefont {D.~M.}\ \bibnamefont {Wilkins}},
  \bibinfo {author} {\bibfnamefont {V.}~\bibnamefont {Rybkin}}, \bibinfo
  {author} {\bibfnamefont {M.}~\bibnamefont {Iannuzzi}},\ and\ \bibinfo
  {author} {\bibfnamefont {J.}~\bibnamefont {Hutter}},\ }\bibfield  {title}
  {\enquote {\bibinfo {title} {{Quantum Dynamics of Water from
  M{\o}ller-Plesset Perturbation Theory via a Neural Network Potential}},}\
  }\href {https://doi.org/10.26434/chemrxiv-2021-n32q8-v2} {\bibfield
  {journal} {\bibinfo  {journal} {chemRxiv}\ } (\bibinfo {year} {2021}),\
  10.26434/chemrxiv-2021-n32q8-v2}\BibitemShut {NoStop}%
\bibitem [{\citenamefont {Liu}, \citenamefont {Lan},\ and\ \citenamefont
  {He}(2022)}]{Liu2022}%
  \BibitemOpen
  \bibfield  {author} {\bibinfo {author} {\bibfnamefont {J.}~\bibnamefont
  {Liu}}, \bibinfo {author} {\bibfnamefont {J.}~\bibnamefont {Lan}},\ and\
  \bibinfo {author} {\bibfnamefont {X.}~\bibnamefont {He}},\ }\bibfield
  {title} {\enquote {\bibinfo {title} {Toward high-level machine learning
  potential for water based on quantum fragmentation and neural networks},}\
  }\href {https://doi.org/10.1021/acs.jpca.2c00601} {\bibfield  {journal}
  {\bibinfo  {journal} {The Journal of Physical Chemistry A}\ }\textbf
  {\bibinfo {volume} {126}},\ \bibinfo {pages} {3926--3936} (\bibinfo {year}
  {2022})},\ \bibinfo {note} {pMID: 35679610},\ \Eprint
  {https://arxiv.org/abs/https://doi.org/10.1021/acs.jpca.2c00601}
  {https://doi.org/10.1021/acs.jpca.2c00601} \BibitemShut {NoStop}%
\bibitem [{\citenamefont {Bukowski}\ \emph {et~al.}(2007)\citenamefont
  {Bukowski}, \citenamefont {Szalewicz}, \citenamefont {Groenenboom},\ and\
  \citenamefont {van~der Avoird}}]{Bukowski2007}%
  \BibitemOpen
  \bibfield  {author} {\bibinfo {author} {\bibfnamefont {R.}~\bibnamefont
  {Bukowski}}, \bibinfo {author} {\bibfnamefont {K.}~\bibnamefont {Szalewicz}},
  \bibinfo {author} {\bibfnamefont {G.~C.}\ \bibnamefont {Groenenboom}},\ and\
  \bibinfo {author} {\bibfnamefont {A.}~\bibnamefont {van~der Avoird}},\
  }\bibfield  {title} {\enquote {\bibinfo {title} {Predictions of the
  properties of water from first principles},}\ }\href
  {https://doi.org/10.1126/science.1136371} {\bibfield  {journal} {\bibinfo
  {journal} {Science}\ }\textbf {\bibinfo {volume} {315}},\ \bibinfo {pages}
  {1249--1252} (\bibinfo {year} {2007})},\ \Eprint
  {https://arxiv.org/abs/https://www.science.org/doi/pdf/10.1126/science.1136371}
  {https://www.science.org/doi/pdf/10.1126/science.1136371} \BibitemShut
  {NoStop}%
\bibitem [{\citenamefont {Babin}, \citenamefont {Leforestier},\ and\
  \citenamefont {Paesani}(2013)}]{Babin2013}%
  \BibitemOpen
  \bibfield  {author} {\bibinfo {author} {\bibfnamefont {V.}~\bibnamefont
  {Babin}}, \bibinfo {author} {\bibfnamefont {C.}~\bibnamefont {Leforestier}},\
  and\ \bibinfo {author} {\bibfnamefont {F.}~\bibnamefont {Paesani}},\
  }\bibfield  {title} {\enquote {\bibinfo {title} {Development of a “first
  principles” water potential with flexible monomers: Dimer potential energy
  surface, vrt spectrum, and second virial coefficient},}\ }\href
  {https://doi.org/10.1021/ct400863t} {\bibfield  {journal} {\bibinfo
  {journal} {Journal of Chemical Theory and Computation}\ }\textbf {\bibinfo
  {volume} {9}},\ \bibinfo {pages} {5395--5403} (\bibinfo {year} {2013})},\
  \bibinfo {note} {pMID: 26592277},\ \Eprint
  {https://arxiv.org/abs/https://doi.org/10.1021/ct400863t}
  {https://doi.org/10.1021/ct400863t} \BibitemShut {NoStop}%
\bibitem [{\citenamefont {Babin}, \citenamefont {Medders},\ and\ \citenamefont
  {Paesani}(2014)}]{Babin2014}%
  \BibitemOpen
  \bibfield  {author} {\bibinfo {author} {\bibfnamefont {V.}~\bibnamefont
  {Babin}}, \bibinfo {author} {\bibfnamefont {G.~R.}\ \bibnamefont {Medders}},\
  and\ \bibinfo {author} {\bibfnamefont {F.}~\bibnamefont {Paesani}},\
  }\bibfield  {title} {\enquote {\bibinfo {title} {Development of a “first
  principles” water potential with flexible monomers. ii: Trimer potential
  energy surface, third virial coefficient, and small clusters},}\ }\href
  {https://doi.org/10.1021/ct500079y} {\bibfield  {journal} {\bibinfo
  {journal} {Journal of Chemical Theory and Computation}\ }\textbf {\bibinfo
  {volume} {10}},\ \bibinfo {pages} {1599--1607} (\bibinfo {year} {2014})},\
  \bibinfo {note} {pMID: 26580372},\ \Eprint
  {https://arxiv.org/abs/https://doi.org/10.1021/ct500079y}
  {https://doi.org/10.1021/ct500079y} \BibitemShut {NoStop}%
\bibitem [{\citenamefont {Medders}, \citenamefont {Babin},\ and\ \citenamefont
  {Paesani}(2014)}]{Medders2014}%
  \BibitemOpen
  \bibfield  {author} {\bibinfo {author} {\bibfnamefont {G.~R.}\ \bibnamefont
  {Medders}}, \bibinfo {author} {\bibfnamefont {V.}~\bibnamefont {Babin}},\
  and\ \bibinfo {author} {\bibfnamefont {F.}~\bibnamefont {Paesani}},\
  }\bibfield  {title} {\enquote {\bibinfo {title} {Development of a
  “first-principles” water potential with flexible monomers. iii. liquid
  phase properties},}\ }\href {https://doi.org/10.1021/ct5004115} {\bibfield
  {journal} {\bibinfo  {journal} {Journal of Chemical Theory and Computation}\
  }\textbf {\bibinfo {volume} {10}},\ \bibinfo {pages} {2906--2910} (\bibinfo
  {year} {2014})},\ \bibinfo {note} {pMID: 26588266},\ \Eprint
  {https://arxiv.org/abs/https://doi.org/10.1021/ct5004115}
  {https://doi.org/10.1021/ct5004115} \BibitemShut {NoStop}%
\bibitem [{\citenamefont {Reddy}\ \emph {et~al.}(2016)\citenamefont {Reddy},
  \citenamefont {Straight}, \citenamefont {Bajaj}, \citenamefont {Huy~Pham},
  \citenamefont {Riera}, \citenamefont {Moberg}, \citenamefont {Morales},
  \citenamefont {Knight}, \citenamefont {Götz},\ and\ \citenamefont
  {Paesani}}]{Reddy2016}%
  \BibitemOpen
  \bibfield  {author} {\bibinfo {author} {\bibfnamefont {S.~K.}\ \bibnamefont
  {Reddy}}, \bibinfo {author} {\bibfnamefont {S.~C.}\ \bibnamefont {Straight}},
  \bibinfo {author} {\bibfnamefont {P.}~\bibnamefont {Bajaj}}, \bibinfo
  {author} {\bibfnamefont {C.}~\bibnamefont {Huy~Pham}}, \bibinfo {author}
  {\bibfnamefont {M.}~\bibnamefont {Riera}}, \bibinfo {author} {\bibfnamefont
  {D.~R.}\ \bibnamefont {Moberg}}, \bibinfo {author} {\bibfnamefont {M.~A.}\
  \bibnamefont {Morales}}, \bibinfo {author} {\bibfnamefont {C.}~\bibnamefont
  {Knight}}, \bibinfo {author} {\bibfnamefont {A.~W.}\ \bibnamefont {Götz}},\
  and\ \bibinfo {author} {\bibfnamefont {F.}~\bibnamefont {Paesani}},\
  }\bibfield  {title} {\enquote {\bibinfo {title} {On the accuracy of the
  mb-pol many-body potential for water: Interaction energies, vibrational
  frequencies, and classical thermodynamic and dynamical properties from
  clusters to liquid water and ice},}\ }\href
  {https://doi.org/10.1063/1.4967719} {\bibfield  {journal} {\bibinfo
  {journal} {The Journal of Chemical Physics}\ }\textbf {\bibinfo {volume}
  {145}},\ \bibinfo {pages} {194504} (\bibinfo {year} {2016})},\ \Eprint
  {https://arxiv.org/abs/https://doi.org/10.1063/1.4967719}
  {https://doi.org/10.1063/1.4967719} \BibitemShut {NoStop}%
\bibitem [{\citenamefont {Schran}, \citenamefont {Behler},\ and\ \citenamefont
  {Marx}(2020)}]{Schran2020a}%
  \BibitemOpen
  \bibfield  {author} {\bibinfo {author} {\bibfnamefont {C.}~\bibnamefont
  {Schran}}, \bibinfo {author} {\bibfnamefont {J.}~\bibnamefont {Behler}},\
  and\ \bibinfo {author} {\bibfnamefont {D.}~\bibnamefont {Marx}},\ }\bibfield
  {title} {\enquote {\bibinfo {title} {{Automated Fitting of Neural Network
  Potentials at Coupled Cluster Accuracy: Protonated Water Clusters as Testing
  Ground}},}\ }\href {https://doi.org/10.1021/acs.jctc.9b00805} {\bibfield
  {journal} {\bibinfo  {journal} {Journal of Chemical Theory and Computation}\
  }\textbf {\bibinfo {volume} {16}},\ \bibinfo {pages} {88--99} (\bibinfo
  {year} {2020})},\ \Eprint {https://arxiv.org/abs/1908.08734}
  {arXiv:1908.08734} \BibitemShut {NoStop}%
\bibitem [{\citenamefont {Nandi}\ \emph {et~al.}(2021)\citenamefont {Nandi},
  \citenamefont {Qu}, \citenamefont {Houston}, \citenamefont {Conte},
  \citenamefont {Yu},\ and\ \citenamefont {Bowman}}]{Nandi2021}%
  \BibitemOpen
  \bibfield  {author} {\bibinfo {author} {\bibfnamefont {A.}~\bibnamefont
  {Nandi}}, \bibinfo {author} {\bibfnamefont {C.}~\bibnamefont {Qu}}, \bibinfo
  {author} {\bibfnamefont {P.~L.}\ \bibnamefont {Houston}}, \bibinfo {author}
  {\bibfnamefont {R.}~\bibnamefont {Conte}}, \bibinfo {author} {\bibfnamefont
  {Q.}~\bibnamefont {Yu}},\ and\ \bibinfo {author} {\bibfnamefont {J.~M.}\
  \bibnamefont {Bowman}},\ }\bibfield  {title} {\enquote {\bibinfo {title} {{A
  CCSD(T)-Based 4-Body Potential for Water}},}\ }\href
  {https://doi.org/10.1021/acs.jpclett.1c03152} {\bibfield  {journal} {\bibinfo
   {journal} {Journal of Physical Chemistry Letters}\ }\textbf {\bibinfo
  {volume} {12}},\ \bibinfo {pages} {10318--10324} (\bibinfo {year}
  {2021})}\BibitemShut {NoStop}%
\bibitem [{\citenamefont {Daru}\ \emph {et~al.}(2022)\citenamefont {Daru},
  \citenamefont {Forbert}, \citenamefont {Behler},\ and\ \citenamefont
  {Marx}}]{Daru2022}%
  \BibitemOpen
  \bibfield  {author} {\bibinfo {author} {\bibfnamefont {J.}~\bibnamefont
  {Daru}}, \bibinfo {author} {\bibfnamefont {H.}~\bibnamefont {Forbert}},
  \bibinfo {author} {\bibfnamefont {J.}~\bibnamefont {Behler}},\ and\ \bibinfo
  {author} {\bibfnamefont {D.}~\bibnamefont {Marx}},\ }\bibfield  {title}
  {\enquote {\bibinfo {title} {Coupled cluster molecular dynamics of condensed
  phase systems enabled by machine learning potentials: Liquid water
  benchmark},}\ }\href {https://doi.org/10.1103/PhysRevLett.129.226001}
  {\bibfield  {journal} {\bibinfo  {journal} {Phys. Rev. Lett.}\ }\textbf
  {\bibinfo {volume} {129}},\ \bibinfo {pages} {226001} (\bibinfo {year}
  {2022})}\BibitemShut {NoStop}%
\bibitem [{\citenamefont {Yu}\ \emph {et~al.}(2022)\citenamefont {Yu},
  \citenamefont {Qu}, \citenamefont {Houston}, \citenamefont {Conte},
  \citenamefont {Nandi},\ and\ \citenamefont {Bowman}}]{Yu2022}%
  \BibitemOpen
  \bibfield  {author} {\bibinfo {author} {\bibfnamefont {Q.}~\bibnamefont
  {Yu}}, \bibinfo {author} {\bibfnamefont {C.}~\bibnamefont {Qu}}, \bibinfo
  {author} {\bibfnamefont {P.~L.}\ \bibnamefont {Houston}}, \bibinfo {author}
  {\bibfnamefont {R.}~\bibnamefont {Conte}}, \bibinfo {author} {\bibfnamefont
  {A.}~\bibnamefont {Nandi}},\ and\ \bibinfo {author} {\bibfnamefont {J.~M.}\
  \bibnamefont {Bowman}},\ }\bibfield  {title} {\enquote {\bibinfo {title}
  {q-aqua: A many-body ccsd(t) water potential, including four-body
  interactions, demonstrates the quantum nature of water from clusters to the
  liquid phase},}\ }\href {https://doi.org/10.1021/acs.jpclett.2c00966}
  {\bibfield  {journal} {\bibinfo  {journal} {The Journal of Physical Chemistry
  Letters}\ }\textbf {\bibinfo {volume} {13}},\ \bibinfo {pages} {5068--5074}
  (\bibinfo {year} {2022})},\ \bibinfo {note} {pMID: 35652912},\ \Eprint
  {https://arxiv.org/abs/https://doi.org/10.1021/acs.jpclett.2c00966}
  {https://doi.org/10.1021/acs.jpclett.2c00966} \BibitemShut {NoStop}%
\bibitem [{\citenamefont {Archibald}, \citenamefont {Krogel},\ and\
  \citenamefont {Kent}(2018)}]{Archibald2018}%
  \BibitemOpen
  \bibfield  {author} {\bibinfo {author} {\bibfnamefont {R.}~\bibnamefont
  {Archibald}}, \bibinfo {author} {\bibfnamefont {J.~T.}\ \bibnamefont
  {Krogel}},\ and\ \bibinfo {author} {\bibfnamefont {P.~R.}\ \bibnamefont
  {Kent}},\ }\bibfield  {title} {\enquote {\bibinfo {title} {{Gaussian process
  based optimization of molecular geometries using statistically sampled energy
  surfaces from quantum Monte Carlo}},}\ }\href
  {https://doi.org/10.1063/1.5040584} {\bibfield  {journal} {\bibinfo
  {journal} {Journal of Chemical Physics}\ }\textbf {\bibinfo {volume} {149}}
  (\bibinfo {year} {2018}),\ 10.1063/1.5040584}\BibitemShut {NoStop}%
\bibitem [{\citenamefont {Ryczko}, \citenamefont {Krogel},\ and\ \citenamefont
  {Tamblyn}(0)}]{Ryczko2022}%
  \BibitemOpen
  \bibfield  {author} {\bibinfo {author} {\bibfnamefont {K.}~\bibnamefont
  {Ryczko}}, \bibinfo {author} {\bibfnamefont {J.~T.}\ \bibnamefont {Krogel}},\
  and\ \bibinfo {author} {\bibfnamefont {I.}~\bibnamefont {Tamblyn}},\
  }\bibfield  {title} {\enquote {\bibinfo {title} {Machine learning diffusion
  monte carlo energies},}\ }\href {https://doi.org/10.1021/acs.jctc.2c00483}
  {\bibfield  {journal} {\bibinfo  {journal} {Journal of Chemical Theory and
  Computation}\ }\textbf {\bibinfo {volume} {0}},\ \bibinfo {pages} {null}
  (\bibinfo {year} {0})},\ \bibinfo {note} {pMID: 36317712},\ \Eprint
  {https://arxiv.org/abs/https://doi.org/10.1021/acs.jctc.2c00483}
  {https://doi.org/10.1021/acs.jctc.2c00483} \BibitemShut {NoStop}%
\bibitem [{\citenamefont {Huang}\ and\ \citenamefont
  {Rubenstein}(2022)}]{Huang2022}%
  \BibitemOpen
  \bibfield  {author} {\bibinfo {author} {\bibfnamefont {C.}~\bibnamefont
  {Huang}}\ and\ \bibinfo {author} {\bibfnamefont {B.~M.}\ \bibnamefont
  {Rubenstein}},\ }\bibfield  {title} {\enquote {\bibinfo {title} {Machine
  learning diffusion monte carlo forces},}\ }\href
  {https://doi.org/10.48550/arxiv.2211.07103} {\bibfield  {journal} {\bibinfo
  {journal} {arxiv}\ } (\bibinfo {year} {2022}),\
  10.48550/arxiv.2211.07103}\BibitemShut {NoStop}%
\bibitem [{\citenamefont {Purvis}\ and\ \citenamefont
  {Bartlett}(1982)}]{Purvis1982}%
  \BibitemOpen
  \bibfield  {author} {\bibinfo {author} {\bibfnamefont {G.~D.}\ \bibnamefont
  {Purvis}}\ and\ \bibinfo {author} {\bibfnamefont {R.~J.}\ \bibnamefont
  {Bartlett}},\ }\bibfield  {title} {\enquote {\bibinfo {title} {A full
  coupled‐cluster singles and doubles model: The inclusion of disconnected
  triples},}\ }\href {https://doi.org/10.1063/1.443164} {\bibfield  {journal}
  {\bibinfo  {journal} {The Journal of Chemical Physics}\ }\textbf {\bibinfo
  {volume} {76}},\ \bibinfo {pages} {1910--1918} (\bibinfo {year} {1982})},\
  \Eprint {https://arxiv.org/abs/https://doi.org/10.1063/1.443164}
  {https://doi.org/10.1063/1.443164} \BibitemShut {NoStop}%
\bibitem [{\citenamefont {Raghavachari}\ \emph {et~al.}(1989)\citenamefont
  {Raghavachari}, \citenamefont {Trucks}, \citenamefont {Pople},\ and\
  \citenamefont {Head-Gordon}}]{Raghavachari1989}%
  \BibitemOpen
  \bibfield  {author} {\bibinfo {author} {\bibfnamefont {K.}~\bibnamefont
  {Raghavachari}}, \bibinfo {author} {\bibfnamefont {G.~W.}\ \bibnamefont
  {Trucks}}, \bibinfo {author} {\bibfnamefont {J.~A.}\ \bibnamefont {Pople}},\
  and\ \bibinfo {author} {\bibfnamefont {M.}~\bibnamefont {Head-Gordon}},\
  }\bibfield  {title} {\enquote {\bibinfo {title} {A fifth-order perturbation
  comparison of electron correlation theories},}\ }\href
  {https://doi.org/https://doi.org/10.1016/S0009-2614(89)87395-6} {\bibfield
  {journal} {\bibinfo  {journal} {Chemical Physics Letters}\ }\textbf {\bibinfo
  {volume} {157}},\ \bibinfo {pages} {479--483} (\bibinfo {year}
  {1989})}\BibitemShut {NoStop}%
\bibitem [{\citenamefont {Zhang}\ and\ \citenamefont
  {Krakauer}(2003)}]{Zhang2003Apr}%
  \BibitemOpen
  \bibfield  {author} {\bibinfo {author} {\bibfnamefont {S.}~\bibnamefont
  {Zhang}}\ and\ \bibinfo {author} {\bibfnamefont {H.}~\bibnamefont
  {Krakauer}},\ }\bibfield  {title} {\enquote {\bibinfo {title} {{Quantum Monte
  Carlo Method using Phase-Free Random Walks with Slater Determinants}},}\
  }\href {https://doi.org/10.1103/PhysRevLett.90.136401} {\bibfield  {journal}
  {\bibinfo  {journal} {Phys. Rev. Lett.}\ }\textbf {\bibinfo {volume} {90}},\
  \bibinfo {pages} {136401} (\bibinfo {year} {2003})}\BibitemShut {NoStop}%
\bibitem [{\citenamefont {Seung}, \citenamefont {Oppert},\ and\ \citenamefont
  {Sompolinsky}(1992)}]{Seung1992}%
  \BibitemOpen
  \bibfield  {author} {\bibinfo {author} {\bibfnamefont {H.~S.}\ \bibnamefont
  {Seung}}, \bibinfo {author} {\bibfnamefont {M.}~\bibnamefont {Oppert}},\ and\
  \bibinfo {author} {\bibfnamefont {H.}~\bibnamefont {Sompolinsky}},\
  }\bibfield  {title} {\enquote {\bibinfo {title} {{Query by committee}},}\
  }in\ \href
  {http://delivery.acm.org/10.1145/140000/130417/p287-seung.pdf?ip=72.33.2.234{\&}id=130417{\&}acc=ACTIVE
  SERVICE{\&}key=066E7B0AFE2DCD37.066E7B0AFE2DCD37.4D4702B0C3E38B35.4D4702B0C3E38B35{\&}{\_}{\_}acm{\_}{\_}=1551281700{\_}3c9f5b70d3a5a3eec01803b438fdece6}
  {\emph {\bibinfo {booktitle} {Proceedings of the fifth annual workshop on
  Computational learning theory (COLT '92)}}}\ (\bibinfo  {publisher}
  {Association for Computing Machinery},\ \bibinfo {address} {New York},\
  \bibinfo {year} {1992})\ pp.\ \bibinfo {pages} {287--294}\BibitemShut
  {NoStop}%
\bibitem [{\citenamefont {Krogh}\ and\ \citenamefont
  {Vedelsby}(1994)}]{Krogh1994}%
  \BibitemOpen
  \bibfield  {author} {\bibinfo {author} {\bibfnamefont {A.}~\bibnamefont
  {Krogh}}\ and\ \bibinfo {author} {\bibfnamefont {J.}~\bibnamefont
  {Vedelsby}},\ }\bibfield  {title} {\enquote {\bibinfo {title} {{Neural
  Network Ensembles, Cross Validation, and Active Learning}},}\ }in\ \href
  {https://proceedings.neurips.cc/paper/1994/file/b8c37e33defde51cf91e1e03e51657da-Paper.pdf}
  {\emph {\bibinfo {booktitle} {Advances in Neural Information Processing
  Systems}}},\ Vol.~\bibinfo {volume} {7},\ \bibinfo {editor} {edited by\
  \bibinfo {editor} {\bibfnamefont {G.}~\bibnamefont {Tesauro}}, \bibinfo
  {editor} {\bibfnamefont {D.}~\bibnamefont {Touretzky}},\ and\ \bibinfo
  {editor} {\bibfnamefont {T.}~\bibnamefont {Leen}}}\ (\bibinfo  {publisher}
  {MIT Press},\ \bibinfo {year} {1994})\ pp.\ \bibinfo {pages}
  {231--238}\BibitemShut {NoStop}%
\bibitem [{\citenamefont {Behler}\ and\ \citenamefont
  {Parrinello}(2007)}]{Behler2007}%
  \BibitemOpen
  \bibfield  {author} {\bibinfo {author} {\bibfnamefont {J.}~\bibnamefont
  {Behler}}\ and\ \bibinfo {author} {\bibfnamefont {M.}~\bibnamefont
  {Parrinello}},\ }\bibfield  {title} {\enquote {\bibinfo {title} {{Generalized
  neural-network representation of high-dimensional potential-energy
  surfaces}},}\ }\href {https://doi.org/10.1103/PhysRevLett.98.146401}
  {\bibfield  {journal} {\bibinfo  {journal} {Physical Review Letters}\
  }\textbf {\bibinfo {volume} {98}},\ \bibinfo {pages} {146401} (\bibinfo
  {year} {2007})}\BibitemShut {NoStop}%
\bibitem [{\citenamefont {Elstner}\ \emph {et~al.}(1998)\citenamefont
  {Elstner}, \citenamefont {Porezag}, \citenamefont {Jungnickel}, \citenamefont
  {Elsner}, \citenamefont {Haugk}, \citenamefont {Frauenheim}, \citenamefont
  {Suhai},\ and\ \citenamefont {Seifert}}]{Elstner1998}%
  \BibitemOpen
  \bibfield  {author} {\bibinfo {author} {\bibfnamefont {M.}~\bibnamefont
  {Elstner}}, \bibinfo {author} {\bibfnamefont {D.}~\bibnamefont {Porezag}},
  \bibinfo {author} {\bibfnamefont {G.}~\bibnamefont {Jungnickel}}, \bibinfo
  {author} {\bibfnamefont {J.}~\bibnamefont {Elsner}}, \bibinfo {author}
  {\bibfnamefont {M.}~\bibnamefont {Haugk}}, \bibinfo {author} {\bibfnamefont
  {T.}~\bibnamefont {Frauenheim}}, \bibinfo {author} {\bibfnamefont
  {S.}~\bibnamefont {Suhai}},\ and\ \bibinfo {author} {\bibfnamefont
  {G.}~\bibnamefont {Seifert}},\ }\bibfield  {title} {\enquote {\bibinfo
  {title} {Self-consistent-charge density-functional tight-binding method for
  simulations of complex materials properties},}\ }\href
  {https://doi.org/10.1103/PhysRevB.58.7260} {\bibfield  {journal} {\bibinfo
  {journal} {Phys. Rev. B}\ }\textbf {\bibinfo {volume} {58}},\ \bibinfo
  {pages} {7260--7268} (\bibinfo {year} {1998})}\BibitemShut {NoStop}%
\bibitem [{\citenamefont {Perdew}, \citenamefont {Burke},\ and\ \citenamefont
  {Ernzerhof}(1996)}]{Perdew1996}%
  \BibitemOpen
  \bibfield  {author} {\bibinfo {author} {\bibfnamefont {J.~P.}\ \bibnamefont
  {Perdew}}, \bibinfo {author} {\bibfnamefont {K.}~\bibnamefont {Burke}},\ and\
  \bibinfo {author} {\bibfnamefont {M.}~\bibnamefont {Ernzerhof}},\ }\bibfield
  {title} {\enquote {\bibinfo {title} {Generalized gradient approximation made
  simple},}\ }\href {https://doi.org/10.1103/PhysRevLett.77.3865} {\bibfield
  {journal} {\bibinfo  {journal} {Phys. Rev. Lett.}\ }\textbf {\bibinfo
  {volume} {77}},\ \bibinfo {pages} {3865--3868} (\bibinfo {year}
  {1996})}\BibitemShut {NoStop}%
\bibitem [{\citenamefont {Zhang}\ and\ \citenamefont {Yang}(1998)}]{Zhang1998}%
  \BibitemOpen
  \bibfield  {author} {\bibinfo {author} {\bibfnamefont {Y.}~\bibnamefont
  {Zhang}}\ and\ \bibinfo {author} {\bibfnamefont {W.}~\bibnamefont {Yang}},\
  }\bibfield  {title} {\enquote {\bibinfo {title} {Comment on ``generalized
  gradient approximation made simple''},}\ }\href
  {https://doi.org/10.1103/PhysRevLett.80.890} {\bibfield  {journal} {\bibinfo
  {journal} {Phys. Rev. Lett.}\ }\textbf {\bibinfo {volume} {80}},\ \bibinfo
  {pages} {890--890} (\bibinfo {year} {1998})}\BibitemShut {NoStop}%
\bibitem [{\citenamefont {Adamo}\ and\ \citenamefont
  {Barone}(1999)}]{Adamo1999}%
  \BibitemOpen
  \bibfield  {author} {\bibinfo {author} {\bibfnamefont {C.}~\bibnamefont
  {Adamo}}\ and\ \bibinfo {author} {\bibfnamefont {V.}~\bibnamefont {Barone}},\
  }\bibfield  {title} {\enquote {\bibinfo {title} {Toward reliable density
  functional methods without adjustable parameters: The pbe0 model},}\ }\href
  {https://doi.org/10.1063/1.478522} {\bibfield  {journal} {\bibinfo  {journal}
  {The Journal of Chemical Physics}\ }\textbf {\bibinfo {volume} {110}},\
  \bibinfo {pages} {6158--6170} (\bibinfo {year} {1999})},\ \Eprint
  {https://arxiv.org/abs/https://doi.org/10.1063/1.478522}
  {https://doi.org/10.1063/1.478522} \BibitemShut {NoStop}%
\bibitem [{\citenamefont {Marsalek}\ and\ \citenamefont
  {Markland}(2017)}]{Marsalek2017}%
  \BibitemOpen
  \bibfield  {author} {\bibinfo {author} {\bibfnamefont {O.}~\bibnamefont
  {Marsalek}}\ and\ \bibinfo {author} {\bibfnamefont {T.~E.}\ \bibnamefont
  {Markland}},\ }\bibfield  {title} {\enquote {\bibinfo {title} {{Quantum
  Dynamics and Spectroscopy of Ab Initio Liquid Water: The Interplay of Nuclear
  and Electronic Quantum Effects}},}\ }\href
  {https://doi.org/10.1021/acs.jpclett.7b00391} {\bibfield  {journal} {\bibinfo
   {journal} {Journal of Physical Chemistry Letters}\ }\textbf {\bibinfo
  {volume} {8}},\ \bibinfo {pages} {1545--1551} (\bibinfo {year}
  {2017})}\BibitemShut {NoStop}%
\bibitem [{\citenamefont {Becke}(1988)}]{Becke1988}%
  \BibitemOpen
  \bibfield  {author} {\bibinfo {author} {\bibfnamefont {A.~D.}\ \bibnamefont
  {Becke}},\ }\bibfield  {title} {\enquote {\bibinfo {title}
  {Density-functional exchange-energy approximation with correct asymptotic
  behavior},}\ }\href {https://doi.org/10.1103/PhysRevA.38.3098} {\bibfield
  {journal} {\bibinfo  {journal} {Phys. Rev. A}\ }\textbf {\bibinfo {volume}
  {38}},\ \bibinfo {pages} {3098--3100} (\bibinfo {year} {1988})}\BibitemShut
  {NoStop}%
\bibitem [{\citenamefont {Lee}, \citenamefont {Yang},\ and\ \citenamefont
  {Parr}(1988)}]{Lee1988}%
  \BibitemOpen
  \bibfield  {author} {\bibinfo {author} {\bibfnamefont {C.}~\bibnamefont
  {Lee}}, \bibinfo {author} {\bibfnamefont {W.}~\bibnamefont {Yang}},\ and\
  \bibinfo {author} {\bibfnamefont {R.~G.}\ \bibnamefont {Parr}},\ }\bibfield
  {title} {\enquote {\bibinfo {title} {Development of the colle-salvetti
  correlation-energy formula into a functional of the electron density},}\
  }\href {https://doi.org/10.1103/PhysRevB.37.785} {\bibfield  {journal}
  {\bibinfo  {journal} {Phys. Rev. B}\ }\textbf {\bibinfo {volume} {37}},\
  \bibinfo {pages} {785--789} (\bibinfo {year} {1988})}\BibitemShut {NoStop}%
\bibitem [{\citenamefont {Kühne}, \citenamefont {Krack},\ and\ \citenamefont
  {Parrinello}(2009)}]{Kuhne2009}%
  \BibitemOpen
  \bibfield  {author} {\bibinfo {author} {\bibfnamefont {T.~D.}\ \bibnamefont
  {Kühne}}, \bibinfo {author} {\bibfnamefont {M.}~\bibnamefont {Krack}},\ and\
  \bibinfo {author} {\bibfnamefont {M.}~\bibnamefont {Parrinello}},\ }\bibfield
   {title} {\enquote {\bibinfo {title} {Static and dynamical properties of
  liquid water from first principles by a novel car-parrinello-like
  approach},}\ }\href {https://doi.org/10.1021/ct800417q} {\bibfield  {journal}
  {\bibinfo  {journal} {Journal of Chemical Theory and Computation}\ }\textbf
  {\bibinfo {volume} {5}},\ \bibinfo {pages} {235--241} (\bibinfo {year}
  {2009})},\ \bibinfo {note} {pMID: 26610101},\ \Eprint
  {https://arxiv.org/abs/https://doi.org/10.1021/ct800417q}
  {https://doi.org/10.1021/ct800417q} \BibitemShut {NoStop}%
\bibitem [{\citenamefont {Pan}\ and\ \citenamefont {Yang}(2010)}]{Pan2010}%
  \BibitemOpen
  \bibfield  {author} {\bibinfo {author} {\bibfnamefont {S.~J.}\ \bibnamefont
  {Pan}}\ and\ \bibinfo {author} {\bibfnamefont {Q.}~\bibnamefont {Yang}},\
  }\bibfield  {title} {\enquote {\bibinfo {title} {A survey on transfer
  learning},}\ }\href {https://doi.org/10.1109/TKDE.2009.191} {\bibfield
  {journal} {\bibinfo  {journal} {IEEE Transactions on Knowledge and Data
  Engineering}\ }\textbf {\bibinfo {volume} {22}},\ \bibinfo {pages}
  {1345--1359} (\bibinfo {year} {2010})}\BibitemShut {NoStop}%
\bibitem [{\citenamefont {Singraber}\ \emph {et~al.}(2019)\citenamefont
  {Singraber}, \citenamefont {Morawietz}, \citenamefont {Behler},\ and\
  \citenamefont {Dellago}}]{Singraber2019}%
  \BibitemOpen
  \bibfield  {author} {\bibinfo {author} {\bibfnamefont {A.}~\bibnamefont
  {Singraber}}, \bibinfo {author} {\bibfnamefont {T.}~\bibnamefont
  {Morawietz}}, \bibinfo {author} {\bibfnamefont {J.}~\bibnamefont {Behler}},\
  and\ \bibinfo {author} {\bibfnamefont {C.}~\bibnamefont {Dellago}},\
  }\bibfield  {title} {\enquote {\bibinfo {title} {Parallel multistream
  training of high-dimensional neural network potentials},}\ }\href
  {https://doi.org/10.1021/acs.jctc.8b01092} {\bibfield  {journal} {\bibinfo
  {journal} {Journal of Chemical Theory and Computation}\ }\textbf {\bibinfo
  {volume} {15}},\ \bibinfo {pages} {3075--3092} (\bibinfo {year}
  {2019})}\BibitemShut {NoStop}%
\bibitem [{\citenamefont {Sun}\ \emph {et~al.}(2020)\citenamefont {Sun},
  \citenamefont {Zhang}, \citenamefont {Banerjee}, \citenamefont {Bao},
  \citenamefont {Barbry}, \citenamefont {Blunt}, \citenamefont {Bogdanov},
  \citenamefont {Booth}, \citenamefont {Chen}, \citenamefont {Cui} \emph
  {et~al.}}]{sun2020recent}%
  \BibitemOpen
  \bibfield  {author} {\bibinfo {author} {\bibfnamefont {Q.}~\bibnamefont
  {Sun}}, \bibinfo {author} {\bibfnamefont {X.}~\bibnamefont {Zhang}}, \bibinfo
  {author} {\bibfnamefont {S.}~\bibnamefont {Banerjee}}, \bibinfo {author}
  {\bibfnamefont {P.}~\bibnamefont {Bao}}, \bibinfo {author} {\bibfnamefont
  {M.}~\bibnamefont {Barbry}}, \bibinfo {author} {\bibfnamefont {N.~S.}\
  \bibnamefont {Blunt}}, \bibinfo {author} {\bibfnamefont {N.~A.}\ \bibnamefont
  {Bogdanov}}, \bibinfo {author} {\bibfnamefont {G.~H.}\ \bibnamefont {Booth}},
  \bibinfo {author} {\bibfnamefont {J.}~\bibnamefont {Chen}}, \bibinfo {author}
  {\bibfnamefont {Z.-H.}\ \bibnamefont {Cui}}, \emph {et~al.},\ }\bibfield
  {title} {\enquote {\bibinfo {title} {Recent developments in the pyscf program
  package},}\ }\href@noop {} {\bibfield  {journal} {\bibinfo  {journal} {J.
  Chem. Phys.}\ }\textbf {\bibinfo {volume} {153}},\ \bibinfo {pages} {024109}
  (\bibinfo {year} {2020})}\BibitemShut {NoStop}%
\bibitem [{\citenamefont {McClain}\ \emph {et~al.}(2017)\citenamefont
  {McClain}, \citenamefont {Sun}, \citenamefont {Chan},\ and\ \citenamefont
  {Berkelbach}}]{McClain2017}%
  \BibitemOpen
  \bibfield  {author} {\bibinfo {author} {\bibfnamefont {J.}~\bibnamefont
  {McClain}}, \bibinfo {author} {\bibfnamefont {Q.}~\bibnamefont {Sun}},
  \bibinfo {author} {\bibfnamefont {G.~K.-L.}\ \bibnamefont {Chan}},\ and\
  \bibinfo {author} {\bibfnamefont {T.~C.}\ \bibnamefont {Berkelbach}},\
  }\bibfield  {title} {\enquote {\bibinfo {title} {Gaussian-based
  coupled-cluster theory for the ground-state and band structure of solids},}\
  }\href {https://doi.org/10.1021/acs.jctc.7b00049} {\bibfield  {journal}
  {\bibinfo  {journal} {Journal of Chemical Theory and Computation}\ }\textbf
  {\bibinfo {volume} {13}},\ \bibinfo {pages} {1209--1218} (\bibinfo {year}
  {2017})},\ \bibinfo {note} {pMID: 28218843},\ \Eprint
  {https://arxiv.org/abs/https://doi.org/10.1021/acs.jctc.7b00049}
  {https://doi.org/10.1021/acs.jctc.7b00049} \BibitemShut {NoStop}%
\bibitem [{\citenamefont {Ye}\ and\ \citenamefont
  {Berkelbach}(2021)}]{Ye21JCP1}%
  \BibitemOpen
  \bibfield  {author} {\bibinfo {author} {\bibfnamefont {H.-Z.}\ \bibnamefont
  {Ye}}\ and\ \bibinfo {author} {\bibfnamefont {T.~C.}\ \bibnamefont
  {Berkelbach}},\ }\bibfield  {title} {\enquote {\bibinfo {title} {Fast
  periodic gaussian density fitting by range separation},}\ }\href
  {https://doi.org/10.1063/5.0046617} {\bibfield  {journal} {\bibinfo
  {journal} {J. Chem. Phys.}\ }\textbf {\bibinfo {volume} {154}},\ \bibinfo
  {pages} {131104} (\bibinfo {year} {2021})}\BibitemShut {NoStop}%
\bibitem [{\citenamefont {Ye*}\ and\ \citenamefont
  {Berkelbach}(2021)}]{Ye21JCP2}%
  \BibitemOpen
  \bibfield  {author} {\bibinfo {author} {\bibfnamefont {H.-Z.}\ \bibnamefont
  {Ye*}}\ and\ \bibinfo {author} {\bibfnamefont {T.~C.}\ \bibnamefont
  {Berkelbach}},\ }\bibfield  {title} {\enquote {\bibinfo {title} {Tight
  distance-dependent estimators for screening two-center and three-center
  short-range coulomb integrals over gaussian basis functions},}\ }\href
  {https://doi.org/10.1063/5.0064151} {\bibfield  {journal} {\bibinfo
  {journal} {J. Chem. Phys.}\ }\textbf {\bibinfo {volume} {155}},\ \bibinfo
  {pages} {124106} (\bibinfo {year} {2021})}\BibitemShut {NoStop}%
\bibitem [{\citenamefont {Kim}\ \emph {et~al.}(2018)\citenamefont {Kim},
  \citenamefont {Baczewski}, \citenamefont {Beaudet}, \citenamefont {Benali},
  \citenamefont {Bennett}, \citenamefont {Berrill}, \citenamefont {Blunt},
  \citenamefont {Borda}, \citenamefont {Casula}, \citenamefont {Ceperley} \emph
  {et~al.}}]{kim2018qmcpack}%
  \BibitemOpen
  \bibfield  {author} {\bibinfo {author} {\bibfnamefont {J.}~\bibnamefont
  {Kim}}, \bibinfo {author} {\bibfnamefont {A.~D.}\ \bibnamefont {Baczewski}},
  \bibinfo {author} {\bibfnamefont {T.~D.}\ \bibnamefont {Beaudet}}, \bibinfo
  {author} {\bibfnamefont {A.}~\bibnamefont {Benali}}, \bibinfo {author}
  {\bibfnamefont {M.~C.}\ \bibnamefont {Bennett}}, \bibinfo {author}
  {\bibfnamefont {M.~A.}\ \bibnamefont {Berrill}}, \bibinfo {author}
  {\bibfnamefont {N.~S.}\ \bibnamefont {Blunt}}, \bibinfo {author}
  {\bibfnamefont {E.~J.~L.}\ \bibnamefont {Borda}}, \bibinfo {author}
  {\bibfnamefont {M.}~\bibnamefont {Casula}}, \bibinfo {author} {\bibfnamefont
  {D.~M.}\ \bibnamefont {Ceperley}}, \emph {et~al.},\ }\bibfield  {title}
  {\enquote {\bibinfo {title} {Qmcpack: an open source ab initio quantum monte
  carlo package for the electronic structure of atoms, molecules and solids},}\
  }\href@noop {} {\bibfield  {journal} {\bibinfo  {journal} {J. Phys. Cond.
  Mat.}\ }\textbf {\bibinfo {volume} {30}},\ \bibinfo {pages} {195901}
  (\bibinfo {year} {2018})}\BibitemShut {NoStop}%
\bibitem [{\citenamefont {Kent}\ \emph {et~al.}(2020)\citenamefont {Kent},
  \citenamefont {Annaberdiyev}, \citenamefont {Benali}, \citenamefont
  {Bennett}, \citenamefont {Landinez~Borda}, \citenamefont {Doak},
  \citenamefont {Hao}, \citenamefont {Jordan}, \citenamefont {Krogel},
  \citenamefont
  {Kyl{\ifmmode\ddot{a}\else\"{a}\fi}np{\ifmmode\ddot{a}\else\"{a}\fi}{\ifmmode\ddot{a}\else\"{a}\fi}},
  \citenamefont {Lee}, \citenamefont {Luo}, \citenamefont {Malone},
  \citenamefont {Melton}, \citenamefont {Mitas}, \citenamefont {Morales},
  \citenamefont {Neuscamman}, \citenamefont {Reboredo}, \citenamefont
  {Rubenstein}, \citenamefont {Saritas}, \citenamefont {Upadhyay},
  \citenamefont {Wang}, \citenamefont {Zhang},\ and\ \citenamefont
  {Zhao}}]{Kent2020May}%
  \BibitemOpen
  \bibfield  {author} {\bibinfo {author} {\bibfnamefont {P.~R.~C.}\
  \bibnamefont {Kent}}, \bibinfo {author} {\bibfnamefont {A.}~\bibnamefont
  {Annaberdiyev}}, \bibinfo {author} {\bibfnamefont {A.}~\bibnamefont
  {Benali}}, \bibinfo {author} {\bibfnamefont {M.~C.}\ \bibnamefont {Bennett}},
  \bibinfo {author} {\bibfnamefont {E.~J.}\ \bibnamefont {Landinez~Borda}},
  \bibinfo {author} {\bibfnamefont {P.}~\bibnamefont {Doak}}, \bibinfo {author}
  {\bibfnamefont {H.}~\bibnamefont {Hao}}, \bibinfo {author} {\bibfnamefont
  {K.~D.}\ \bibnamefont {Jordan}}, \bibinfo {author} {\bibfnamefont {J.~T.}\
  \bibnamefont {Krogel}}, \bibinfo {author} {\bibfnamefont {I.}~\bibnamefont
  {Kyl{\ifmmode\ddot{a}\else\"{a}\fi}np{\ifmmode\ddot{a}\else\"{a}\fi}{\ifmmode\ddot{a}\else\"{a}\fi}}},
  \bibinfo {author} {\bibfnamefont {J.}~\bibnamefont {Lee}}, \bibinfo {author}
  {\bibfnamefont {Y.}~\bibnamefont {Luo}}, \bibinfo {author} {\bibfnamefont
  {F.~D.}\ \bibnamefont {Malone}}, \bibinfo {author} {\bibfnamefont {C.~A.}\
  \bibnamefont {Melton}}, \bibinfo {author} {\bibfnamefont {L.}~\bibnamefont
  {Mitas}}, \bibinfo {author} {\bibfnamefont {M.~A.}\ \bibnamefont {Morales}},
  \bibinfo {author} {\bibfnamefont {E.}~\bibnamefont {Neuscamman}}, \bibinfo
  {author} {\bibfnamefont {F.~A.}\ \bibnamefont {Reboredo}}, \bibinfo {author}
  {\bibfnamefont {B.}~\bibnamefont {Rubenstein}}, \bibinfo {author}
  {\bibfnamefont {K.}~\bibnamefont {Saritas}}, \bibinfo {author} {\bibfnamefont
  {S.}~\bibnamefont {Upadhyay}}, \bibinfo {author} {\bibfnamefont
  {G.}~\bibnamefont {Wang}}, \bibinfo {author} {\bibfnamefont {S.}~\bibnamefont
  {Zhang}},\ and\ \bibinfo {author} {\bibfnamefont {L.}~\bibnamefont {Zhao}},\
  }\bibfield  {title} {\enquote {\bibinfo {title} {{QMCPACK: Advances in the
  development, efficiency, and application of auxiliary field and real-space
  variational and diffusion quantum Monte Carlo}},}\ }\href
  {https://doi.org/10.1063/5.0004860} {\bibfield  {journal} {\bibinfo
  {journal} {J. Chem. Phys.}\ }\textbf {\bibinfo {volume} {152}},\ \bibinfo
  {pages} {174105} (\bibinfo {year} {2020})}\BibitemShut {NoStop}%
\bibitem [{\citenamefont {Malone}\ \emph {et~al.}(2022)\citenamefont {Malone},
  \citenamefont {Mahajan}, \citenamefont {Spencer},\ and\ \citenamefont
  {Lee}}]{Malone2022Sep}%
  \BibitemOpen
  \bibfield  {author} {\bibinfo {author} {\bibfnamefont {F.~D.}\ \bibnamefont
  {Malone}}, \bibinfo {author} {\bibfnamefont {A.}~\bibnamefont {Mahajan}},
  \bibinfo {author} {\bibfnamefont {J.~S.}\ \bibnamefont {Spencer}},\ and\
  \bibinfo {author} {\bibfnamefont {J.}~\bibnamefont {Lee}},\ }\bibfield
  {title} {\enquote {\bibinfo {title} {{ipie: A Python-based Auxiliary-Field
  Quantum Monte Carlo Program with Flexibility and Efficiency on CPUs and
  GPUs}},}\ }\href {https://doi.org/10.48550/arXiv.2209.04015} {\bibfield
  {journal} {\bibinfo  {journal} {arXiv}\ } (\bibinfo {year} {2022}),\
  10.48550/arXiv.2209.04015},\ \Eprint {https://arxiv.org/abs/2209.04015}
  {2209.04015} \BibitemShut {NoStop}%
\bibitem [{\citenamefont {Lee}, \citenamefont {Pham},\ and\ \citenamefont
  {Reichman}(2022)}]{Lee2022Oct}%
  \BibitemOpen
  \bibfield  {author} {\bibinfo {author} {\bibfnamefont {J.}~\bibnamefont
  {Lee}}, \bibinfo {author} {\bibfnamefont {H.~Q.}\ \bibnamefont {Pham}},\ and\
  \bibinfo {author} {\bibfnamefont {D.~R.}\ \bibnamefont {Reichman}},\
  }\bibfield  {title} {\enquote {\bibinfo {title} {{Twenty Years of
  Auxiliary-Field Quantum Monte Carlo in Quantum Chemistry: An Overview and
  Assessment on Main Group Chemistry and Bond-Breaking}},}\ }\href
  {https://doi.org/10.1021/acs.jctc.2c00802} {\bibfield  {journal} {\bibinfo
  {journal} {J. Chem. Theory Comput.}\ }\textbf {\bibinfo {volume} {2022}}
  (\bibinfo {year} {2022}),\ 10.1021/acs.jctc.2c00802}\BibitemShut {NoStop}%
\bibitem [{\citenamefont {Čížek}(1966)}]{Cizek1966}%
  \BibitemOpen
  \bibfield  {author} {\bibinfo {author} {\bibfnamefont {J.}~\bibnamefont
  {Čížek}},\ }\bibfield  {title} {\enquote {\bibinfo {title} {On the
  correlation problem in atomic and molecular systems. calculation of
  wavefunction components in ursell‐type expansion using quantum‐field
  theoretical methods},}\ }\href {https://doi.org/10.1063/1.1727484} {\bibfield
   {journal} {\bibinfo  {journal} {The Journal of Chemical Physics}\ }\textbf
  {\bibinfo {volume} {45}},\ \bibinfo {pages} {4256--4266} (\bibinfo {year}
  {1966})},\ \Eprint {https://arxiv.org/abs/https://doi.org/10.1063/1.1727484}
  {https://doi.org/10.1063/1.1727484} \BibitemShut {NoStop}%
\bibitem [{\citenamefont {Taube}\ and\ \citenamefont
  {Bartlett}(2008)}]{Taube08JCP}%
  \BibitemOpen
  \bibfield  {author} {\bibinfo {author} {\bibfnamefont {A.~G.}\ \bibnamefont
  {Taube}}\ and\ \bibinfo {author} {\bibfnamefont {R.~J.}\ \bibnamefont
  {Bartlett}},\ }\bibfield  {title} {\enquote {\bibinfo {title} {Frozen natural
  orbital coupled-cluster theory: Forces and application to decomposition of
  nitroethane},}\ }\href {https://doi.org/10.1063/1.2902285} {\bibfield
  {journal} {\bibinfo  {journal} {J. Chem. Phys.}\ }\textbf {\bibinfo {volume}
  {128}},\ \bibinfo {pages} {164101} (\bibinfo {year} {2008})}\BibitemShut
  {NoStop}%
\bibitem [{\citenamefont {Lange}\ and\ \citenamefont
  {Berkelbach}(2020)}]{Lange20MP}%
  \BibitemOpen
  \bibfield  {author} {\bibinfo {author} {\bibfnamefont {M.~F.}\ \bibnamefont
  {Lange}}\ and\ \bibinfo {author} {\bibfnamefont {T.~C.}\ \bibnamefont
  {Berkelbach}},\ }\bibfield  {title} {\enquote {\bibinfo {title} {Active space
  approaches combining coupled-cluster and perturbation theory for ground
  states and excited states},}\ }\href
  {https://doi.org/10.1080/00268976.2020.1808726} {\bibfield  {journal}
  {\bibinfo  {journal} {Mol. Phys.}\ }\textbf {\bibinfo {volume} {118}},\
  \bibinfo {pages} {e1808726} (\bibinfo {year} {2020})}\BibitemShut {NoStop}%
\bibitem [{\citenamefont {Skinner}\ \emph {et~al.}(2013)\citenamefont
  {Skinner}, \citenamefont {Huang}, \citenamefont {Schlesinger}, \citenamefont
  {Pettersson}, \citenamefont {Nilsson},\ and\ \citenamefont
  {Benmore}}]{Skinner2013}%
  \BibitemOpen
  \bibfield  {author} {\bibinfo {author} {\bibfnamefont {L.~B.}\ \bibnamefont
  {Skinner}}, \bibinfo {author} {\bibfnamefont {C.}~\bibnamefont {Huang}},
  \bibinfo {author} {\bibfnamefont {D.}~\bibnamefont {Schlesinger}}, \bibinfo
  {author} {\bibfnamefont {L.~G.~M.}\ \bibnamefont {Pettersson}}, \bibinfo
  {author} {\bibfnamefont {A.}~\bibnamefont {Nilsson}},\ and\ \bibinfo {author}
  {\bibfnamefont {C.~J.}\ \bibnamefont {Benmore}},\ }\bibfield  {title}
  {\enquote {\bibinfo {title} {Benchmark oxygen-oxygen pair-distribution
  function of ambient water from x-ray diffraction measurements with a wide
  q-range},}\ }\href {https://doi.org/10.1063/1.4790861} {\bibfield  {journal}
  {\bibinfo  {journal} {The Journal of Chemical Physics}\ }\textbf {\bibinfo
  {volume} {138}},\ \bibinfo {pages} {074506} (\bibinfo {year}
  {2013})}\BibitemShut {NoStop}%
\bibitem [{\citenamefont {Mariedahl}\ \emph {et~al.}(2018)\citenamefont
  {Mariedahl}, \citenamefont {Perakis}, \citenamefont {Sp{\"{a}}h},
  \citenamefont {Pathak}, \citenamefont {Kim}, \citenamefont {Camisasca},
  \citenamefont {Schlesinger}, \citenamefont {Benmore}, \citenamefont
  {Pettersson}, \citenamefont {Nilsson},\ and\ \citenamefont
  {Amann-Winkel}}]{Mariedahl2018}%
  \BibitemOpen
  \bibfield  {author} {\bibinfo {author} {\bibfnamefont {D.}~\bibnamefont
  {Mariedahl}}, \bibinfo {author} {\bibfnamefont {F.}~\bibnamefont {Perakis}},
  \bibinfo {author} {\bibfnamefont {A.}~\bibnamefont {Sp{\"{a}}h}}, \bibinfo
  {author} {\bibfnamefont {H.}~\bibnamefont {Pathak}}, \bibinfo {author}
  {\bibfnamefont {K.~H.}\ \bibnamefont {Kim}}, \bibinfo {author} {\bibfnamefont
  {G.}~\bibnamefont {Camisasca}}, \bibinfo {author} {\bibfnamefont
  {D.}~\bibnamefont {Schlesinger}}, \bibinfo {author} {\bibfnamefont
  {C.}~\bibnamefont {Benmore}}, \bibinfo {author} {\bibfnamefont {L.~G.~M.}\
  \bibnamefont {Pettersson}}, \bibinfo {author} {\bibfnamefont
  {A.}~\bibnamefont {Nilsson}},\ and\ \bibinfo {author} {\bibfnamefont
  {K.}~\bibnamefont {Amann-Winkel}},\ }\bibfield  {title} {\enquote {\bibinfo
  {title} {{X-ray Scattering and O-O Pair-Distribution Functions of Amorphous
  Ices}},}\ }\href {https://doi.org/10.1021/acs.jpcb.8b04823} {\bibfield
  {journal} {\bibinfo  {journal} {Journal of Physical Chemistry B}\ }\textbf
  {\bibinfo {volume} {122}},\ \bibinfo {pages} {7616--7624} (\bibinfo {year}
  {2018})}\BibitemShut {NoStop}%
\bibitem [{\citenamefont {{Jeffrey R. Errington}}\ and\ \citenamefont {{Pablo
  G. Debenedetti}}(2001)}]{Errington2001}%
  \BibitemOpen
  \bibfield  {author} {\bibinfo {author} {\bibnamefont {{Jeffrey R.
  Errington}}}\ and\ \bibinfo {author} {\bibnamefont {{Pablo G.
  Debenedetti}}},\ }\bibfield  {title} {\enquote {\bibinfo {title}
  {{Relationship between structural order and the anomalies of liquid
  water}},}\ }\href
  {https://www-nature-com.ezp.sub.su.se/articles/35053024.pdf} {\bibfield
  {journal} {\bibinfo  {journal} {Nature}\ }\textbf {\bibinfo {volume} {409}},\
  \bibinfo {pages} {318} (\bibinfo {year} {2001})}\BibitemShut {NoStop}%
\bibitem [{\citenamefont {Habershon}, \citenamefont {Markland},\ and\
  \citenamefont {Manolopoulos}(2009)}]{Habershon2009}%
  \BibitemOpen
  \bibfield  {author} {\bibinfo {author} {\bibfnamefont {S.}~\bibnamefont
  {Habershon}}, \bibinfo {author} {\bibfnamefont {T.~E.}\ \bibnamefont
  {Markland}},\ and\ \bibinfo {author} {\bibfnamefont {D.~E.}\ \bibnamefont
  {Manolopoulos}},\ }\bibfield  {title} {\enquote {\bibinfo {title} {Competing
  quantum effects in the dynamics of a flexible water model},}\ }\href
  {https://doi.org/10.1063/1.3167790} {\bibfield  {journal} {\bibinfo
  {journal} {The Journal of Chemical Physics}\ }\textbf {\bibinfo {volume}
  {131}},\ \bibinfo {pages} {024501} (\bibinfo {year} {2009})}\BibitemShut
  {NoStop}%
\bibitem [{\citenamefont {Holz}, \citenamefont {Heil},\ and\ \citenamefont
  {Sacco}(2000)}]{Holz2000}%
  \BibitemOpen
  \bibfield  {author} {\bibinfo {author} {\bibfnamefont {M.}~\bibnamefont
  {Holz}}, \bibinfo {author} {\bibfnamefont {S.~R.}\ \bibnamefont {Heil}},\
  and\ \bibinfo {author} {\bibfnamefont {A.}~\bibnamefont {Sacco}},\ }\bibfield
   {title} {\enquote {\bibinfo {title} {Temperature-dependent self-diffusion
  coefficients of water and six selected molecular liquids for calibration in
  accurate 1h nmr pfg measurements},}\ }\href
  {https://doi.org/10.1039/B005319H} {\bibfield  {journal} {\bibinfo  {journal}
  {Phys. Chem. Chem. Phys.}\ }\textbf {\bibinfo {volume} {2}},\ \bibinfo
  {pages} {4740--4742} (\bibinfo {year} {2000})}\BibitemShut {NoStop}%
\bibitem [{\citenamefont {Dietrich}(2002)}]{Dietrich}%
  \BibitemOpen
  \bibfield  {author} {\bibinfo {author} {\bibfnamefont {O.}~\bibnamefont
  {Dietrich}},\ }\href@noop {} {\enquote {\bibinfo {title} {{Diffusion
  Coefficients of Water}},}\ }\bibinfo {howpublished}
  {\url{https://dtrx.de/od/diff/}} (\bibinfo {year} {2002}),\ \bibinfo {note}
  {[Accessed: 30-August-2022]}\BibitemShut {NoStop}%
\bibitem [{\citenamefont {Rossi}\ \emph {et~al.}(2014)\citenamefont {Rossi},
  \citenamefont {Liu}, \citenamefont {Paesani}, \citenamefont {Bowman},\ and\
  \citenamefont {Ceriotti}}]{Rossi2014}%
  \BibitemOpen
  \bibfield  {author} {\bibinfo {author} {\bibfnamefont {M.}~\bibnamefont
  {Rossi}}, \bibinfo {author} {\bibfnamefont {H.}~\bibnamefont {Liu}}, \bibinfo
  {author} {\bibfnamefont {F.}~\bibnamefont {Paesani}}, \bibinfo {author}
  {\bibfnamefont {J.}~\bibnamefont {Bowman}},\ and\ \bibinfo {author}
  {\bibfnamefont {M.}~\bibnamefont {Ceriotti}},\ }\bibfield  {title} {\enquote
  {\bibinfo {title} {Communication: On the consistency of approximate quantum
  dynamics simulation methods for vibrational spectra in the condensed
  phase},}\ }\href {https://doi.org/10.1063/1.4901214} {\bibfield  {journal}
  {\bibinfo  {journal} {The Journal of Chemical Physics}\ }\textbf {\bibinfo
  {volume} {141}},\ \bibinfo {pages} {181101} (\bibinfo {year}
  {2014})}\BibitemShut {NoStop}%
\bibitem [{\citenamefont {Yeh}\ and\ \citenamefont {Hummer}(2004)}]{Yeh2004}%
  \BibitemOpen
  \bibfield  {author} {\bibinfo {author} {\bibfnamefont {I.~C.}\ \bibnamefont
  {Yeh}}\ and\ \bibinfo {author} {\bibfnamefont {G.}~\bibnamefont {Hummer}},\
  }\bibfield  {title} {\enquote {\bibinfo {title} {{System-size dependence of
  diffusion coefficients and viscosities from molecular dynamics simulations
  with periodic boundary conditions}},}\ }\href
  {https://doi.org/10.1021/jp0477147} {\bibfield  {journal} {\bibinfo
  {journal} {Journal of Physical Chemistry B}\ }\textbf {\bibinfo {volume}
  {108}},\ \bibinfo {pages} {15873--15879} (\bibinfo {year}
  {2004})}\BibitemShut {NoStop}%
\bibitem [{\citenamefont {Kestin}, \citenamefont {Sokolov},\ and\ \citenamefont
  {Wakeham}(1978)}]{Kestin1978}%
  \BibitemOpen
  \bibfield  {author} {\bibinfo {author} {\bibfnamefont {J.}~\bibnamefont
  {Kestin}}, \bibinfo {author} {\bibfnamefont {M.}~\bibnamefont {Sokolov}},\
  and\ \bibinfo {author} {\bibfnamefont {W.~A.}\ \bibnamefont {Wakeham}},\
  }\bibfield  {title} {\enquote {\bibinfo {title} {Viscosity of liquid water in
  the range -8c to 150c},}\ }\href {https://doi.org/10.1063/1.555581}
  {\bibfield  {journal} {\bibinfo  {journal} {Journal of Physical and Chemical
  Reference Data}\ }\textbf {\bibinfo {volume} {7}},\ \bibinfo {pages}
  {941--948} (\bibinfo {year} {1978})}\BibitemShut {NoStop}%
\end{thebibliography}%


\begin{thebibliography}{47}%
\makeatletter
\providecommand \@ifxundefined [1]{%
 \@ifx{#1\undefined}
}%
\providecommand \@ifnum [1]{%
 \ifnum #1\expandafter \@firstoftwo
 \else \expandafter \@secondoftwo
 \fi
}%
\providecommand \@ifx [1]{%
 \ifx #1\expandafter \@firstoftwo
 \else \expandafter \@secondoftwo
 \fi
}%
\providecommand \natexlab [1]{#1}%
\providecommand \enquote  [1]{``#1''}%
\providecommand \bibnamefont  [1]{#1}%
\providecommand \bibfnamefont [1]{#1}%
\providecommand \citenamefont [1]{#1}%
\providecommand \href@noop [0]{\@secondoftwo}%
\providecommand \href [0]{\begingroup \@sanitize@url \@href}%
\providecommand \@href[1]{\@@startlink{#1}\@@href}%
\providecommand \@@href[1]{\endgroup#1\@@endlink}%
\providecommand \@sanitize@url [0]{\catcode `\\12\catcode `\$12\catcode
  `\&12\catcode `\#12\catcode `\^12\catcode `\_12\catcode `\%12\relax}%
\providecommand \@@startlink[1]{}%
\providecommand \@@endlink[0]{}%
\providecommand \url  [0]{\begingroup\@sanitize@url \@url }%
\providecommand \@url [1]{\endgroup\@href {#1}{\urlprefix }}%
\providecommand \urlprefix  [0]{URL }%
\providecommand \Eprint [0]{\href }%
\providecommand \doibase [0]{https://doi.org/}%
\providecommand \selectlanguage [0]{\@gobble}%
\providecommand \bibinfo  [0]{\@secondoftwo}%
\providecommand \bibfield  [0]{\@secondoftwo}%
\providecommand \translation [1]{[#1]}%
\providecommand \BibitemOpen [0]{}%
\providecommand \bibitemStop [0]{}%
\providecommand \bibitemNoStop [0]{.\EOS\space}%
\providecommand \EOS [0]{\spacefactor3000\relax}%
\providecommand \BibitemShut  [1]{\csname bibitem#1\endcsname}%
\let\auto@bib@innerbib\@empty
\bibitem [{\citenamefont {Hutter}\ \emph {et~al.}(2014)\citenamefont {Hutter},
  \citenamefont {Iannuzzi}, \citenamefont {Schiffmann},\ and\ \citenamefont
  {VandeVondele}}]{Hutter2014}%
  \BibitemOpen
  \bibfield  {author} {\bibinfo {author} {\bibfnamefont {J.}~\bibnamefont
  {Hutter}}, \bibinfo {author} {\bibfnamefont {M.}~\bibnamefont {Iannuzzi}},
  \bibinfo {author} {\bibfnamefont {F.}~\bibnamefont {Schiffmann}},\ and\
  \bibinfo {author} {\bibfnamefont {J.}~\bibnamefont {VandeVondele}},\
  }\bibfield  {title} {\bibinfo {title} {cp2k: atomistic simulations of
  condensed matter systems},\ }\href
  {https://doi.org/https://doi.org/10.1002/wcms.1159} {\bibfield  {journal}
  {\bibinfo  {journal} {WIREs Computational Molecular Science}\ }\textbf
  {\bibinfo {volume} {4}},\ \bibinfo {pages} {15} (\bibinfo {year} {2014})},\
  \Eprint
  {https://arxiv.org/abs/https://wires.onlinelibrary.wiley.com/doi/pdf/10.1002/wcms.1159}
  {https://wires.onlinelibrary.wiley.com/doi/pdf/10.1002/wcms.1159}
  \BibitemShut {NoStop}%
\bibitem [{\citenamefont {Becke}(1988)}]{Becke1988}%
  \BibitemOpen
  \bibfield  {author} {\bibinfo {author} {\bibfnamefont {A.~D.}\ \bibnamefont
  {Becke}},\ }\bibfield  {title} {\bibinfo {title} {Density-functional
  exchange-energy approximation with correct asymptotic behavior},\ }\href
  {https://doi.org/10.1103/PhysRevA.38.3098} {\bibfield  {journal} {\bibinfo
  {journal} {Phys. Rev. A}\ }\textbf {\bibinfo {volume} {38}},\ \bibinfo
  {pages} {3098} (\bibinfo {year} {1988})}\BibitemShut {NoStop}%
\bibitem [{\citenamefont {Lee}\ \emph {et~al.}(1988)\citenamefont {Lee},
  \citenamefont {Yang},\ and\ \citenamefont {Parr}}]{Lee1988}%
  \BibitemOpen
  \bibfield  {author} {\bibinfo {author} {\bibfnamefont {C.}~\bibnamefont
  {Lee}}, \bibinfo {author} {\bibfnamefont {W.}~\bibnamefont {Yang}},\ and\
  \bibinfo {author} {\bibfnamefont {R.~G.}\ \bibnamefont {Parr}},\ }\bibfield
  {title} {\bibinfo {title} {Development of the colle-salvetti
  correlation-energy formula into a functional of the electron density},\
  }\href {https://doi.org/10.1103/PhysRevB.37.785} {\bibfield  {journal}
  {\bibinfo  {journal} {Phys. Rev. B}\ }\textbf {\bibinfo {volume} {37}},\
  \bibinfo {pages} {785} (\bibinfo {year} {1988})}\BibitemShut {NoStop}%
\bibitem [{\citenamefont {Perdew}\ \emph {et~al.}(1996)\citenamefont {Perdew},
  \citenamefont {Burke},\ and\ \citenamefont {Ernzerhof}}]{Perdew1996}%
  \BibitemOpen
  \bibfield  {author} {\bibinfo {author} {\bibfnamefont {J.~P.}\ \bibnamefont
  {Perdew}}, \bibinfo {author} {\bibfnamefont {K.}~\bibnamefont {Burke}},\ and\
  \bibinfo {author} {\bibfnamefont {M.}~\bibnamefont {Ernzerhof}},\ }\bibfield
  {title} {\bibinfo {title} {Generalized gradient approximation made simple},\
  }\href {https://doi.org/10.1103/PhysRevLett.77.3865} {\bibfield  {journal}
  {\bibinfo  {journal} {Phys. Rev. Lett.}\ }\textbf {\bibinfo {volume} {77}},\
  \bibinfo {pages} {3865} (\bibinfo {year} {1996})}\BibitemShut {NoStop}%
\bibitem [{\citenamefont {Zhang}\ and\ \citenamefont {Yang}(1998)}]{Zhang1998}%
  \BibitemOpen
  \bibfield  {author} {\bibinfo {author} {\bibfnamefont {Y.}~\bibnamefont
  {Zhang}}\ and\ \bibinfo {author} {\bibfnamefont {W.}~\bibnamefont {Yang}},\
  }\bibfield  {title} {\bibinfo {title} {Comment on ``generalized gradient
  approximation made simple''},\ }\href
  {https://doi.org/10.1103/PhysRevLett.80.890} {\bibfield  {journal} {\bibinfo
  {journal} {Phys. Rev. Lett.}\ }\textbf {\bibinfo {volume} {80}},\ \bibinfo
  {pages} {890} (\bibinfo {year} {1998})}\BibitemShut {NoStop}%
\bibitem [{\citenamefont {Adamo}\ and\ \citenamefont
  {Barone}(1999)}]{Adamo1999}%
  \BibitemOpen
  \bibfield  {author} {\bibinfo {author} {\bibfnamefont {C.}~\bibnamefont
  {Adamo}}\ and\ \bibinfo {author} {\bibfnamefont {V.}~\bibnamefont {Barone}},\
  }\bibfield  {title} {\bibinfo {title} {Toward reliable density functional
  methods without adjustable parameters: The pbe0 model},\ }\href
  {https://doi.org/10.1063/1.478522} {\bibfield  {journal} {\bibinfo  {journal}
  {The Journal of Chemical Physics}\ }\textbf {\bibinfo {volume} {110}},\
  \bibinfo {pages} {6158} (\bibinfo {year} {1999})},\ \Eprint
  {https://arxiv.org/abs/https://doi.org/10.1063/1.478522}
  {https://doi.org/10.1063/1.478522} \BibitemShut {NoStop}%
\bibitem [{\citenamefont {Elstner}\ \emph {et~al.}(1998)\citenamefont
  {Elstner}, \citenamefont {Porezag}, \citenamefont {Jungnickel}, \citenamefont
  {Elsner}, \citenamefont {Haugk}, \citenamefont {Frauenheim}, \citenamefont
  {Suhai},\ and\ \citenamefont {Seifert}}]{Elstner1998}%
  \BibitemOpen
  \bibfield  {author} {\bibinfo {author} {\bibfnamefont {M.}~\bibnamefont
  {Elstner}}, \bibinfo {author} {\bibfnamefont {D.}~\bibnamefont {Porezag}},
  \bibinfo {author} {\bibfnamefont {G.}~\bibnamefont {Jungnickel}}, \bibinfo
  {author} {\bibfnamefont {J.}~\bibnamefont {Elsner}}, \bibinfo {author}
  {\bibfnamefont {M.}~\bibnamefont {Haugk}}, \bibinfo {author} {\bibfnamefont
  {T.}~\bibnamefont {Frauenheim}}, \bibinfo {author} {\bibfnamefont
  {S.}~\bibnamefont {Suhai}},\ and\ \bibinfo {author} {\bibfnamefont
  {G.}~\bibnamefont {Seifert}},\ }\bibfield  {title} {\bibinfo {title}
  {Self-consistent-charge density-functional tight-binding method for
  simulations of complex materials properties},\ }\href
  {https://doi.org/10.1103/PhysRevB.58.7260} {\bibfield  {journal} {\bibinfo
  {journal} {Phys. Rev. B}\ }\textbf {\bibinfo {volume} {58}},\ \bibinfo
  {pages} {7260} (\bibinfo {year} {1998})}\BibitemShut {NoStop}%
\bibitem [{\citenamefont {Goedecker}\ \emph {et~al.}(1996)\citenamefont
  {Goedecker}, \citenamefont {Teter},\ and\ \citenamefont
  {Hutter}}]{Goedecker1996Jul}%
  \BibitemOpen
  \bibfield  {author} {\bibinfo {author} {\bibfnamefont {S.}~\bibnamefont
  {Goedecker}}, \bibinfo {author} {\bibfnamefont {M.}~\bibnamefont {Teter}},\
  and\ \bibinfo {author} {\bibfnamefont {J.}~\bibnamefont {Hutter}},\
  }\bibfield  {title} {\bibinfo {title} {{Separable dual-space Gaussian
  pseudopotentials}},\ }\href {https://doi.org/10.1103/PhysRevB.54.1703}
  {\bibfield  {journal} {\bibinfo  {journal} {Phys. Rev. B}\ }\textbf {\bibinfo
  {volume} {54}},\ \bibinfo {pages} {1703} (\bibinfo {year}
  {1996})}\BibitemShut {NoStop}%
\bibitem [{\citenamefont {Lippert}\ \emph {et~al.}(1997)\citenamefont
  {Lippert}, \citenamefont {Hutter},\ and\ \citenamefont
  {Parrinello}}]{Lippert1997}%
  \BibitemOpen
  \bibfield  {author} {\bibinfo {author} {\bibfnamefont {G.}~\bibnamefont
  {Lippert}}, \bibinfo {author} {\bibfnamefont {J.}~\bibnamefont {Hutter}},\
  and\ \bibinfo {author} {\bibfnamefont {M.}~\bibnamefont {Parrinello}},\
  }\bibfield  {title} {\bibinfo {title} {A hybrid gaussian and plane wave
  density functional scheme},\ }\href {https://doi.org/10.1080/002689797170220}
  {\bibfield  {journal} {\bibinfo  {journal} {Molecular Physics}\ }\textbf
  {\bibinfo {volume} {92}},\ \bibinfo {pages} {477} (\bibinfo {year} {1997})},\
  \Eprint {https://arxiv.org/abs/https://doi.org/10.1080/002689797170220}
  {https://doi.org/10.1080/002689797170220} \BibitemShut {NoStop}%
\bibitem [{\citenamefont {VandeVondele}\ \emph {et~al.}(2005)\citenamefont
  {VandeVondele}, \citenamefont {Krack}, \citenamefont {Mohamed}, \citenamefont
  {Parrinello}, \citenamefont {Chassaing},\ and\ \citenamefont
  {Hutter}}]{VandeVondele2005}%
  \BibitemOpen
  \bibfield  {author} {\bibinfo {author} {\bibfnamefont {J.}~\bibnamefont
  {VandeVondele}}, \bibinfo {author} {\bibfnamefont {M.}~\bibnamefont {Krack}},
  \bibinfo {author} {\bibfnamefont {F.}~\bibnamefont {Mohamed}}, \bibinfo
  {author} {\bibfnamefont {M.}~\bibnamefont {Parrinello}}, \bibinfo {author}
  {\bibfnamefont {T.}~\bibnamefont {Chassaing}},\ and\ \bibinfo {author}
  {\bibfnamefont {J.}~\bibnamefont {Hutter}},\ }\bibfield  {title} {\bibinfo
  {title} {Quickstep: Fast and accurate density functional calculations using a
  mixed gaussian and plane waves approach},\ }\href
  {https://doi.org/https://doi.org/10.1016/j.cpc.2004.12.014} {\bibfield
  {journal} {\bibinfo  {journal} {Computer Physics Communications}\ }\textbf
  {\bibinfo {volume} {167}},\ \bibinfo {pages} {103} (\bibinfo {year}
  {2005})}\BibitemShut {NoStop}%
\bibitem [{\citenamefont {VandeVondele}\ and\ \citenamefont
  {Hutter}(2007)}]{VandeVondele2007Sep}%
  \BibitemOpen
  \bibfield  {author} {\bibinfo {author} {\bibfnamefont {J.}~\bibnamefont
  {VandeVondele}}\ and\ \bibinfo {author} {\bibfnamefont {J.}~\bibnamefont
  {Hutter}},\ }\bibfield  {title} {\bibinfo {title} {{Gaussian basis sets for
  accurate calculations on molecular systems in gas and condensed phases}},\
  }\href {https://doi.org/10.1063/1.2770708} {\bibfield  {journal} {\bibinfo
  {journal} {J. Chem. Phys.}\ }\textbf {\bibinfo {volume} {127}},\ \bibinfo
  {pages} {114105} (\bibinfo {year} {2007})}\BibitemShut {NoStop}%
\bibitem [{\citenamefont {Guidon}\ \emph {et~al.}(2009)\citenamefont {Guidon},
  \citenamefont {Hutter},\ and\ \citenamefont {VandeVondele}}]{Guidon2009}%
  \BibitemOpen
  \bibfield  {author} {\bibinfo {author} {\bibfnamefont {M.}~\bibnamefont
  {Guidon}}, \bibinfo {author} {\bibfnamefont {J.}~\bibnamefont {Hutter}},\
  and\ \bibinfo {author} {\bibfnamefont {J.}~\bibnamefont {VandeVondele}},\
  }\bibfield  {title} {\bibinfo {title} {Robust periodic hartree-fock exchange
  for large-scale simulations using gaussian basis sets},\ }\href
  {https://doi.org/10.1021/ct900494g} {\bibfield  {journal} {\bibinfo
  {journal} {Journal of Chemical Theory and Computation}\ }\textbf {\bibinfo
  {volume} {5}},\ \bibinfo {pages} {3010} (\bibinfo {year} {2009})},\ \bibinfo
  {note} {pMID: 26609981},\ \Eprint
  {https://arxiv.org/abs/https://doi.org/10.1021/ct900494g}
  {https://doi.org/10.1021/ct900494g} \BibitemShut {NoStop}%
\bibitem [{\citenamefont {Guidon}\ \emph {et~al.}(2010)\citenamefont {Guidon},
  \citenamefont {Hutter},\ and\ \citenamefont {VandeVondele}}]{Guidon2010}%
  \BibitemOpen
  \bibfield  {author} {\bibinfo {author} {\bibfnamefont {M.}~\bibnamefont
  {Guidon}}, \bibinfo {author} {\bibfnamefont {J.}~\bibnamefont {Hutter}},\
  and\ \bibinfo {author} {\bibfnamefont {J.}~\bibnamefont {VandeVondele}},\
  }\bibfield  {title} {\bibinfo {title} {Auxiliary density matrix methods for
  hartree-fock exchange calculations},\ }\href
  {https://doi.org/10.1021/ct1002225} {\bibfield  {journal} {\bibinfo
  {journal} {Journal of Chemical Theory and Computation}\ }\textbf {\bibinfo
  {volume} {6}},\ \bibinfo {pages} {2348} (\bibinfo {year} {2010})},\ \bibinfo
  {note} {pMID: 26613491},\ \Eprint
  {https://arxiv.org/abs/https://doi.org/10.1021/ct1002225}
  {https://doi.org/10.1021/ct1002225} \BibitemShut {NoStop}%
\bibitem [{\citenamefont {Grimme}\ \emph {et~al.}(2010)\citenamefont {Grimme},
  \citenamefont {Antony}, \citenamefont {Ehrlich},\ and\ \citenamefont
  {Krieg}}]{Grimme2010}%
  \BibitemOpen
  \bibfield  {author} {\bibinfo {author} {\bibfnamefont {S.}~\bibnamefont
  {Grimme}}, \bibinfo {author} {\bibfnamefont {J.}~\bibnamefont {Antony}},
  \bibinfo {author} {\bibfnamefont {S.}~\bibnamefont {Ehrlich}},\ and\ \bibinfo
  {author} {\bibfnamefont {H.}~\bibnamefont {Krieg}},\ }\bibfield  {title}
  {\bibinfo {title} {A consistent and accurate ab initio parametrization of
  density functional dispersion correction (dft-d) for the 94 elements h-pu},\
  }\href {https://doi.org/10.1063/1.3382344} {\bibfield  {journal} {\bibinfo
  {journal} {The Journal of Chemical Physics}\ }\textbf {\bibinfo {volume}
  {132}},\ \bibinfo {pages} {154104} (\bibinfo {year} {2010})},\ \Eprint
  {https://arxiv.org/abs/https://doi.org/10.1063/1.3382344}
  {https://doi.org/10.1063/1.3382344} \BibitemShut {NoStop}%
\bibitem [{\citenamefont {Ewald}(1921)}]{Ewald1921}%
  \BibitemOpen
  \bibfield  {author} {\bibinfo {author} {\bibfnamefont {P.~P.}\ \bibnamefont
  {Ewald}},\ }\bibfield  {title} {\bibinfo {title} {Die berechnung optischer
  und elektrostatischer gitterpotentiale},\ }\href
  {https://doi.org/https://doi.org/10.1002/andp.19213690304} {\bibfield
  {journal} {\bibinfo  {journal} {Annalen der Physik}\ }\textbf {\bibinfo
  {volume} {369}},\ \bibinfo {pages} {253} (\bibinfo {year} {1921})},\ \Eprint
  {https://arxiv.org/abs/https://onlinelibrary.wiley.com/doi/pdf/10.1002/andp.19213690304}
  {https://onlinelibrary.wiley.com/doi/pdf/10.1002/andp.19213690304}
  \BibitemShut {NoStop}%
\bibitem [{\citenamefont {Essmann}\ \emph {et~al.}(1995)\citenamefont
  {Essmann}, \citenamefont {Perera}, \citenamefont {Berkowitz}, \citenamefont
  {Darden}, \citenamefont {Lee},\ and\ \citenamefont {Pedersen}}]{Essman1995}%
  \BibitemOpen
  \bibfield  {author} {\bibinfo {author} {\bibfnamefont {U.}~\bibnamefont
  {Essmann}}, \bibinfo {author} {\bibfnamefont {L.}~\bibnamefont {Perera}},
  \bibinfo {author} {\bibfnamefont {M.~L.}\ \bibnamefont {Berkowitz}}, \bibinfo
  {author} {\bibfnamefont {T.}~\bibnamefont {Darden}}, \bibinfo {author}
  {\bibfnamefont {H.}~\bibnamefont {Lee}},\ and\ \bibinfo {author}
  {\bibfnamefont {L.~G.}\ \bibnamefont {Pedersen}},\ }\bibfield  {title}
  {\bibinfo {title} {A smooth particle mesh ewald method},\ }\href
  {https://doi.org/10.1063/1.470117} {\bibfield  {journal} {\bibinfo  {journal}
  {The Journal of Chemical Physics}\ }\textbf {\bibinfo {volume} {103}},\
  \bibinfo {pages} {8577} (\bibinfo {year} {1995})},\ \Eprint
  {https://arxiv.org/abs/https://doi.org/10.1063/1.470117}
  {https://doi.org/10.1063/1.470117} \BibitemShut {NoStop}%
\bibitem [{\citenamefont {Sun}\ \emph {et~al.}(2020)\citenamefont {Sun},
  \citenamefont {Zhang}, \citenamefont {Banerjee}, \citenamefont {Bao},
  \citenamefont {Barbry}, \citenamefont {Blunt}, \citenamefont {Bogdanov},
  \citenamefont {Booth}, \citenamefont {Chen}, \citenamefont {Cui} \emph
  {et~al.}}]{sun2020recent}%
  \BibitemOpen
  \bibfield  {author} {\bibinfo {author} {\bibfnamefont {Q.}~\bibnamefont
  {Sun}}, \bibinfo {author} {\bibfnamefont {X.}~\bibnamefont {Zhang}}, \bibinfo
  {author} {\bibfnamefont {S.}~\bibnamefont {Banerjee}}, \bibinfo {author}
  {\bibfnamefont {P.}~\bibnamefont {Bao}}, \bibinfo {author} {\bibfnamefont
  {M.}~\bibnamefont {Barbry}}, \bibinfo {author} {\bibfnamefont {N.~S.}\
  \bibnamefont {Blunt}}, \bibinfo {author} {\bibfnamefont {N.~A.}\ \bibnamefont
  {Bogdanov}}, \bibinfo {author} {\bibfnamefont {G.~H.}\ \bibnamefont {Booth}},
  \bibinfo {author} {\bibfnamefont {J.}~\bibnamefont {Chen}}, \bibinfo {author}
  {\bibfnamefont {Z.-H.}\ \bibnamefont {Cui}}, \emph {et~al.},\ }\bibfield
  {title} {\bibinfo {title} {Recent developments in the pyscf program
  package},\ }\href@noop {} {\bibfield  {journal} {\bibinfo  {journal} {J.
  Chem. Phys.}\ }\textbf {\bibinfo {volume} {153}},\ \bibinfo {pages} {024109}
  (\bibinfo {year} {2020})}\BibitemShut {NoStop}%
\bibitem [{\citenamefont {Taube}\ and\ \citenamefont
  {Bartlett}(2008)}]{Taube08JCP}%
  \BibitemOpen
  \bibfield  {author} {\bibinfo {author} {\bibfnamefont {A.~G.}\ \bibnamefont
  {Taube}}\ and\ \bibinfo {author} {\bibfnamefont {R.~J.}\ \bibnamefont
  {Bartlett}},\ }\bibfield  {title} {\bibinfo {title} {Frozen natural orbital
  coupled-cluster theory: Forces and application to decomposition of
  nitroethane},\ }\href {https://doi.org/10.1063/1.2902285} {\bibfield
  {journal} {\bibinfo  {journal} {J. Chem. Phys.}\ }\textbf {\bibinfo {volume}
  {128}},\ \bibinfo {pages} {164101} (\bibinfo {year} {2008})}\BibitemShut
  {NoStop}%
\bibitem [{\citenamefont {Lange}\ and\ \citenamefont
  {Berkelbach}(2020)}]{Lange20MP}%
  \BibitemOpen
  \bibfield  {author} {\bibinfo {author} {\bibfnamefont {M.~F.}\ \bibnamefont
  {Lange}}\ and\ \bibinfo {author} {\bibfnamefont {T.~C.}\ \bibnamefont
  {Berkelbach}},\ }\bibfield  {title} {\bibinfo {title} {Active space
  approaches combining coupled-cluster and perturbation theory for ground
  states and excited states},\ }\href
  {https://doi.org/10.1080/00268976.2020.1808726} {\bibfield  {journal}
  {\bibinfo  {journal} {Mol. Phys.}\ }\textbf {\bibinfo {volume} {118}},\
  \bibinfo {pages} {e1808726} (\bibinfo {year} {2020})}\BibitemShut {NoStop}%
\bibitem [{\citenamefont {Kim}\ \emph {et~al.}(2018)\citenamefont {Kim},
  \citenamefont {Baczewski}, \citenamefont {Beaudet}, \citenamefont {Benali},
  \citenamefont {Bennett}, \citenamefont {Berrill}, \citenamefont {Blunt},
  \citenamefont {Borda}, \citenamefont {Casula}, \citenamefont {Ceperley} \emph
  {et~al.}}]{kim2018qmcpack}%
  \BibitemOpen
  \bibfield  {author} {\bibinfo {author} {\bibfnamefont {J.}~\bibnamefont
  {Kim}}, \bibinfo {author} {\bibfnamefont {A.~D.}\ \bibnamefont {Baczewski}},
  \bibinfo {author} {\bibfnamefont {T.~D.}\ \bibnamefont {Beaudet}}, \bibinfo
  {author} {\bibfnamefont {A.}~\bibnamefont {Benali}}, \bibinfo {author}
  {\bibfnamefont {M.~C.}\ \bibnamefont {Bennett}}, \bibinfo {author}
  {\bibfnamefont {M.~A.}\ \bibnamefont {Berrill}}, \bibinfo {author}
  {\bibfnamefont {N.~S.}\ \bibnamefont {Blunt}}, \bibinfo {author}
  {\bibfnamefont {E.~J.~L.}\ \bibnamefont {Borda}}, \bibinfo {author}
  {\bibfnamefont {M.}~\bibnamefont {Casula}}, \bibinfo {author} {\bibfnamefont
  {D.~M.}\ \bibnamefont {Ceperley}}, \emph {et~al.},\ }\bibfield  {title}
  {\bibinfo {title} {Qmcpack: an open source ab initio quantum monte carlo
  package for the electronic structure of atoms, molecules and solids},\
  }\href@noop {} {\bibfield  {journal} {\bibinfo  {journal} {J. Phys. Cond.
  Mat.}\ }\textbf {\bibinfo {volume} {30}},\ \bibinfo {pages} {195901}
  (\bibinfo {year} {2018})}\BibitemShut {NoStop}%
\bibitem [{\citenamefont {Kent}\ \emph {et~al.}(2020)\citenamefont {Kent},
  \citenamefont {Annaberdiyev}, \citenamefont {Benali}, \citenamefont
  {Bennett}, \citenamefont {Landinez~Borda}, \citenamefont {Doak},
  \citenamefont {Hao}, \citenamefont {Jordan}, \citenamefont {Krogel},
  \citenamefont
  {Kyl{\ifmmode\ddot{a}\else\"{a}\fi}np{\ifmmode\ddot{a}\else\"{a}\fi}{\ifmmode\ddot{a}\else\"{a}\fi}},
  \citenamefont {Lee}, \citenamefont {Luo}, \citenamefont {Malone},
  \citenamefont {Melton}, \citenamefont {Mitas}, \citenamefont {Morales},
  \citenamefont {Neuscamman}, \citenamefont {Reboredo}, \citenamefont
  {Rubenstein}, \citenamefont {Saritas}, \citenamefont {Upadhyay},
  \citenamefont {Wang}, \citenamefont {Zhang},\ and\ \citenamefont
  {Zhao}}]{Kent2020May}%
  \BibitemOpen
  \bibfield  {author} {\bibinfo {author} {\bibfnamefont {P.~R.~C.}\
  \bibnamefont {Kent}}, \bibinfo {author} {\bibfnamefont {A.}~\bibnamefont
  {Annaberdiyev}}, \bibinfo {author} {\bibfnamefont {A.}~\bibnamefont
  {Benali}}, \bibinfo {author} {\bibfnamefont {M.~C.}\ \bibnamefont {Bennett}},
  \bibinfo {author} {\bibfnamefont {E.~J.}\ \bibnamefont {Landinez~Borda}},
  \bibinfo {author} {\bibfnamefont {P.}~\bibnamefont {Doak}}, \bibinfo {author}
  {\bibfnamefont {H.}~\bibnamefont {Hao}}, \bibinfo {author} {\bibfnamefont
  {K.~D.}\ \bibnamefont {Jordan}}, \bibinfo {author} {\bibfnamefont {J.~T.}\
  \bibnamefont {Krogel}}, \bibinfo {author} {\bibfnamefont {I.}~\bibnamefont
  {Kyl{\ifmmode\ddot{a}\else\"{a}\fi}np{\ifmmode\ddot{a}\else\"{a}\fi}{\ifmmode\ddot{a}\else\"{a}\fi}}},
  \bibinfo {author} {\bibfnamefont {J.}~\bibnamefont {Lee}}, \bibinfo {author}
  {\bibfnamefont {Y.}~\bibnamefont {Luo}}, \bibinfo {author} {\bibfnamefont
  {F.~D.}\ \bibnamefont {Malone}}, \bibinfo {author} {\bibfnamefont {C.~A.}\
  \bibnamefont {Melton}}, \bibinfo {author} {\bibfnamefont {L.}~\bibnamefont
  {Mitas}}, \bibinfo {author} {\bibfnamefont {M.~A.}\ \bibnamefont {Morales}},
  \bibinfo {author} {\bibfnamefont {E.}~\bibnamefont {Neuscamman}}, \bibinfo
  {author} {\bibfnamefont {F.~A.}\ \bibnamefont {Reboredo}}, \bibinfo {author}
  {\bibfnamefont {B.}~\bibnamefont {Rubenstein}}, \bibinfo {author}
  {\bibfnamefont {K.}~\bibnamefont {Saritas}}, \bibinfo {author} {\bibfnamefont
  {S.}~\bibnamefont {Upadhyay}}, \bibinfo {author} {\bibfnamefont
  {G.}~\bibnamefont {Wang}}, \bibinfo {author} {\bibfnamefont {S.}~\bibnamefont
  {Zhang}},\ and\ \bibinfo {author} {\bibfnamefont {L.}~\bibnamefont {Zhao}},\
  }\bibfield  {title} {\bibinfo {title} {{QMCPACK: Advances in the development,
  efficiency, and application of auxiliary field and real-space variational and
  diffusion quantum Monte Carlo}},\ }\href {https://doi.org/10.1063/5.0004860}
  {\bibfield  {journal} {\bibinfo  {journal} {J. Chem. Phys.}\ }\textbf
  {\bibinfo {volume} {152}},\ \bibinfo {pages} {174105} (\bibinfo {year}
  {2020})}\BibitemShut {NoStop}%
\bibitem [{\citenamefont {Malone}\ \emph {et~al.}(2022)\citenamefont {Malone},
  \citenamefont {Mahajan}, \citenamefont {Spencer},\ and\ \citenamefont
  {Lee}}]{Malone2022Sep}%
  \BibitemOpen
  \bibfield  {author} {\bibinfo {author} {\bibfnamefont {F.~D.}\ \bibnamefont
  {Malone}}, \bibinfo {author} {\bibfnamefont {A.}~\bibnamefont {Mahajan}},
  \bibinfo {author} {\bibfnamefont {J.~S.}\ \bibnamefont {Spencer}},\ and\
  \bibinfo {author} {\bibfnamefont {J.}~\bibnamefont {Lee}},\ }\bibfield
  {title} {\bibinfo {title} {{ipie: A Python-based Auxiliary-Field Quantum
  Monte Carlo Program with Flexibility and Efficiency on CPUs and GPUs}},\
  }\bibfield  {journal} {\bibinfo  {journal} {arXiv}\ }\href
  {https://doi.org/10.48550/arXiv.2209.04015} {10.48550/arXiv.2209.04015}
  (\bibinfo {year} {2022}),\ \Eprint {https://arxiv.org/abs/2209.04015}
  {2209.04015} \BibitemShut {NoStop}%
\bibitem [{\citenamefont {Hartwigsen}\ \emph {et~al.}(1998)\citenamefont
  {Hartwigsen}, \citenamefont {Goedecker},\ and\ \citenamefont
  {Hutter}}]{Hartwigsen1998Aug}%
  \BibitemOpen
  \bibfield  {author} {\bibinfo {author} {\bibfnamefont {C.}~\bibnamefont
  {Hartwigsen}}, \bibinfo {author} {\bibfnamefont {S.}~\bibnamefont
  {Goedecker}},\ and\ \bibinfo {author} {\bibfnamefont {J.}~\bibnamefont
  {Hutter}},\ }\bibfield  {title} {\bibinfo {title} {{Relativistic separable
  dual-space Gaussian pseudopotentials from H to Rn}},\ }\href
  {https://doi.org/10.1103/PhysRevB.58.3641} {\bibfield  {journal} {\bibinfo
  {journal} {Phys. Rev. B}\ }\textbf {\bibinfo {volume} {58}},\ \bibinfo
  {pages} {3641} (\bibinfo {year} {1998})}\BibitemShut {NoStop}%
\bibitem [{\citenamefont {Fraser}\ \emph {et~al.}(1814)\citenamefont {Fraser},
  \citenamefont {Foulkes}, \citenamefont {Rajagopal}, \citenamefont {Needs},
  \citenamefont {Kenny},\ and\ \citenamefont {Williamson}}]{Fraser1814Jan}%
  \BibitemOpen
  \bibfield  {author} {\bibinfo {author} {\bibfnamefont {L.~M.}\ \bibnamefont
  {Fraser}}, \bibinfo {author} {\bibfnamefont {W.~M.~C.}\ \bibnamefont
  {Foulkes}}, \bibinfo {author} {\bibfnamefont {G.}~\bibnamefont {Rajagopal}},
  \bibinfo {author} {\bibfnamefont {R.~J.}\ \bibnamefont {Needs}}, \bibinfo
  {author} {\bibfnamefont {S.~D.}\ \bibnamefont {Kenny}},\ and\ \bibinfo
  {author} {\bibfnamefont {A.~J.}\ \bibnamefont {Williamson}},\ }\bibfield
  {title} {\bibinfo {title} {{Finite-size effects and Coulomb interactions in
  quantum Monte Carlo calculations for homogeneous systems with periodic
  boundary conditions}},\ }\href {https://doi.org/10.1103/PhysRevB.53.1814}
  {\bibfield  {journal} {\bibinfo  {journal} {Phys. Rev. B}\ }\textbf {\bibinfo
  {volume} {53}},\ \bibinfo {pages} {1814} (\bibinfo {year}
  {1814})}\BibitemShut {NoStop}%
\bibitem [{\citenamefont {{Singraber, Andreas}}()}]{n2p2}%
  \BibitemOpen
  \bibfield  {author} {\bibinfo {author} {\bibnamefont {{Singraber,
  Andreas}}},\ }\href {https://compphysvienna.github.io/n2p2/} {\bibinfo
  {title} {n2p2 - the neural network potential package}}\BibitemShut {NoStop}%
\bibitem [{\citenamefont {Morawietz}\ \emph {et~al.}(2016)\citenamefont
  {Morawietz}, \citenamefont {Singraber}, \citenamefont {Dellago},\ and\
  \citenamefont {Behler}}]{Morawietz2016}%
  \BibitemOpen
  \bibfield  {author} {\bibinfo {author} {\bibfnamefont {T.}~\bibnamefont
  {Morawietz}}, \bibinfo {author} {\bibfnamefont {A.}~\bibnamefont
  {Singraber}}, \bibinfo {author} {\bibfnamefont {C.}~\bibnamefont {Dellago}},\
  and\ \bibinfo {author} {\bibfnamefont {J.}~\bibnamefont {Behler}},\
  }\bibfield  {title} {\bibinfo {title} {{How van der Waals interactions
  determine the unique properties of water}},\ }\href
  {https://doi.org/10.1073/pnas.1602375113} {\bibfield  {journal} {\bibinfo
  {journal} {Proceedings of the National Academy of Sciences}\ }\textbf
  {\bibinfo {volume} {113}},\ \bibinfo {pages} {8368} (\bibinfo {year}
  {2016})},\ \Eprint {https://arxiv.org/abs/1606.07775} {arXiv:1606.07775}
  \BibitemShut {NoStop}%
\bibitem [{\citenamefont {Behler}\ and\ \citenamefont
  {Parrinello}(2007)}]{Behler2007}%
  \BibitemOpen
  \bibfield  {author} {\bibinfo {author} {\bibfnamefont {J.}~\bibnamefont
  {Behler}}\ and\ \bibinfo {author} {\bibfnamefont {M.}~\bibnamefont
  {Parrinello}},\ }\bibfield  {title} {\bibinfo {title} {{Generalized
  neural-network representation of high-dimensional potential-energy
  surfaces}},\ }\href {https://doi.org/10.1103/PhysRevLett.98.146401}
  {\bibfield  {journal} {\bibinfo  {journal} {Physical Review Letters}\
  }\textbf {\bibinfo {volume} {98}},\ \bibinfo {pages} {146401} (\bibinfo
  {year} {2007})}\BibitemShut {NoStop}%
\bibitem [{\citenamefont {Nguyen}\ and\ \citenamefont
  {Widrow}(1990)}]{Nguyen1990}%
  \BibitemOpen
  \bibfield  {author} {\bibinfo {author} {\bibfnamefont {D.}~\bibnamefont
  {Nguyen}}\ and\ \bibinfo {author} {\bibfnamefont {B.}~\bibnamefont
  {Widrow}},\ }\bibfield  {title} {\bibinfo {title} {Improving the learning
  speed of 2-layer neural networks by choosing initial values of the adaptive
  weights},\ }in\ \href {https://doi.org/10.1109/IJCNN.1990.137819} {\emph
  {\bibinfo {booktitle} {1990 IJCNN International Joint Conference on Neural
  Networks}}}\ (\bibinfo {year} {1990})\ pp.\ \bibinfo {pages} {21--26
  vol.3}\BibitemShut {NoStop}%
\bibitem [{\citenamefont {Kalman}(1960)}]{Kalman1960}%
  \BibitemOpen
  \bibfield  {author} {\bibinfo {author} {\bibfnamefont {R.~E.}\ \bibnamefont
  {Kalman}},\ }\bibfield  {title} {\bibinfo {title} {{A New Approach to Linear
  Filtering and Prediction Problems}},\ }\href
  {https://doi.org/10.1115/1.3662552} {\bibfield  {journal} {\bibinfo
  {journal} {Journal of Basic Engineering}\ }\textbf {\bibinfo {volume} {82}},\
  \bibinfo {pages} {35} (\bibinfo {year} {1960})},\ \Eprint
  {https://arxiv.org/abs/https://asmedigitalcollection.asme.org/fluidsengineering/article-pdf/82/1/35/5518977/35\_1.pdf}
  {https://asmedigitalcollection.asme.org/fluidsengineering/article-pdf/82/1/35/5518977/35\_1.pdf}
  \BibitemShut {NoStop}%
\bibitem [{\citenamefont {Kalman}\ and\ \citenamefont
  {Bucy}(1961)}]{Kalman1961}%
  \BibitemOpen
  \bibfield  {author} {\bibinfo {author} {\bibfnamefont {R.~E.}\ \bibnamefont
  {Kalman}}\ and\ \bibinfo {author} {\bibfnamefont {R.~S.}\ \bibnamefont
  {Bucy}},\ }\bibfield  {title} {\bibinfo {title} {{New Results in Linear
  Filtering and Prediction Theory}},\ }\href
  {https://doi.org/10.1115/1.3658902} {\bibfield  {journal} {\bibinfo
  {journal} {Journal of Basic Engineering}\ }\textbf {\bibinfo {volume} {83}},\
  \bibinfo {pages} {95} (\bibinfo {year} {1961})},\ \Eprint
  {https://arxiv.org/abs/https://asmedigitalcollection.asme.org/fluidsengineering/article-pdf/83/1/95/5503549/95\_1.pdf}
  {https://asmedigitalcollection.asme.org/fluidsengineering/article-pdf/83/1/95/5503549/95\_1.pdf}
  \BibitemShut {NoStop}%
\bibitem [{\citenamefont {Smith}\ \emph {et~al.}(1962)\citenamefont {Smith},
  \citenamefont {Schmidt},\ and\ \citenamefont {McGee}}]{Smith1962}%
  \BibitemOpen
  \bibfield  {author} {\bibinfo {author} {\bibfnamefont {G.~L.}\ \bibnamefont
  {Smith}}, \bibinfo {author} {\bibfnamefont {S.~F.}\ \bibnamefont {Schmidt}},\
  and\ \bibinfo {author} {\bibfnamefont {L.~A.}\ \bibnamefont {McGee}},\
  }\href@noop {} {\emph {\bibinfo {title} {{Technical Report R-135: Application
  of statistical filter theory to the optimal estimation of position and
  velocity on board a circumlunar vehicle}}}},\ \bibinfo {type} {Tech. Rep.}\
  (\bibinfo  {institution} {National Aeronautics and Space Administration},\
  \bibinfo {year} {1962})\BibitemShut {NoStop}%
\bibitem [{\citenamefont {Singraber}\ \emph {et~al.}(2019)\citenamefont
  {Singraber}, \citenamefont {Morawietz}, \citenamefont {Behler},\ and\
  \citenamefont {Dellago}}]{Singraber2019}%
  \BibitemOpen
  \bibfield  {author} {\bibinfo {author} {\bibfnamefont {A.}~\bibnamefont
  {Singraber}}, \bibinfo {author} {\bibfnamefont {T.}~\bibnamefont
  {Morawietz}}, \bibinfo {author} {\bibfnamefont {J.}~\bibnamefont {Behler}},\
  and\ \bibinfo {author} {\bibfnamefont {C.}~\bibnamefont {Dellago}},\
  }\bibfield  {title} {\bibinfo {title} {Parallel multistream training of
  high-dimensional neural network potentials},\ }\href
  {https://doi.org/10.1021/acs.jctc.8b01092} {\bibfield  {journal} {\bibinfo
  {journal} {Journal of Chemical Theory and Computation}\ }\textbf {\bibinfo
  {volume} {15}},\ \bibinfo {pages} {3075} (\bibinfo {year}
  {2019})}\BibitemShut {NoStop}%
\bibitem [{\citenamefont {Schran}\ \emph {et~al.}(2020)\citenamefont {Schran},
  \citenamefont {Brezina},\ and\ \citenamefont {Marsalek}}]{Schran2020}%
  \BibitemOpen
  \bibfield  {author} {\bibinfo {author} {\bibfnamefont {C.}~\bibnamefont
  {Schran}}, \bibinfo {author} {\bibfnamefont {K.}~\bibnamefont {Brezina}},\
  and\ \bibinfo {author} {\bibfnamefont {O.}~\bibnamefont {Marsalek}},\
  }\bibfield  {title} {\bibinfo {title} {Committee neural network potentials
  control generalization errors and enable active learning},\ }\href
  {https://doi.org/10.1063/5.0016004} {\bibfield  {journal} {\bibinfo
  {journal} {The Journal of Chemical Physics}\ }\textbf {\bibinfo {volume}
  {153}},\ \bibinfo {pages} {104105} (\bibinfo {year} {2020})}\BibitemShut
  {NoStop}%
\bibitem [{\citenamefont {Ceriotti}\ \emph {et~al.}(2014)\citenamefont
  {Ceriotti}, \citenamefont {More},\ and\ \citenamefont
  {Manolopoulos}}]{Ceriotti2014}%
  \BibitemOpen
  \bibfield  {author} {\bibinfo {author} {\bibfnamefont {M.}~\bibnamefont
  {Ceriotti}}, \bibinfo {author} {\bibfnamefont {J.}~\bibnamefont {More}},\
  and\ \bibinfo {author} {\bibfnamefont {D.~E.}\ \bibnamefont {Manolopoulos}},\
  }\bibfield  {title} {\bibinfo {title} {i-pi: A python interface for ab initio
  path integral molecular dynamics simulations},\ }\href
  {https://doi.org/https://doi.org/10.1016/j.cpc.2013.10.027} {\bibfield
  {journal} {\bibinfo  {journal} {Computer Physics Communications}\ }\textbf
  {\bibinfo {volume} {185}},\ \bibinfo {pages} {1019} (\bibinfo {year}
  {2014})}\BibitemShut {NoStop}%
\bibitem [{\citenamefont {Kapil}\ \emph {et~al.}(2019)\citenamefont {Kapil},
  \citenamefont {Rossi}, \citenamefont {Marsalek}, \citenamefont {Petraglia},
  \citenamefont {Litman}, \citenamefont {Spura}, \citenamefont {Cheng},
  \citenamefont {Cuzzocrea}, \citenamefont {Meißner}, \citenamefont {Wilkins},
  \citenamefont {Helfrecht}, \citenamefont {Juda}, \citenamefont {Bienvenue},
  \citenamefont {Fang}, \citenamefont {Kessler}, \citenamefont {Poltavsky},
  \citenamefont {Vandenbrande}, \citenamefont {Wieme}, \citenamefont
  {Corminboeuf}, \citenamefont {Kühne}, \citenamefont {Manolopoulos},
  \citenamefont {Markland}, \citenamefont {Richardson}, \citenamefont
  {Tkatchenko}, \citenamefont {Tribello}, \citenamefont {{Van Speybroeck}},\
  and\ \citenamefont {Ceriotti}}]{Kapil2019}%
  \BibitemOpen
  \bibfield  {author} {\bibinfo {author} {\bibfnamefont {V.}~\bibnamefont
  {Kapil}}, \bibinfo {author} {\bibfnamefont {M.}~\bibnamefont {Rossi}},
  \bibinfo {author} {\bibfnamefont {O.}~\bibnamefont {Marsalek}}, \bibinfo
  {author} {\bibfnamefont {R.}~\bibnamefont {Petraglia}}, \bibinfo {author}
  {\bibfnamefont {Y.}~\bibnamefont {Litman}}, \bibinfo {author} {\bibfnamefont
  {T.}~\bibnamefont {Spura}}, \bibinfo {author} {\bibfnamefont
  {B.}~\bibnamefont {Cheng}}, \bibinfo {author} {\bibfnamefont
  {A.}~\bibnamefont {Cuzzocrea}}, \bibinfo {author} {\bibfnamefont {R.~H.}\
  \bibnamefont {Meißner}}, \bibinfo {author} {\bibfnamefont {D.~M.}\
  \bibnamefont {Wilkins}}, \bibinfo {author} {\bibfnamefont {B.~A.}\
  \bibnamefont {Helfrecht}}, \bibinfo {author} {\bibfnamefont {P.}~\bibnamefont
  {Juda}}, \bibinfo {author} {\bibfnamefont {S.~P.}\ \bibnamefont {Bienvenue}},
  \bibinfo {author} {\bibfnamefont {W.}~\bibnamefont {Fang}}, \bibinfo {author}
  {\bibfnamefont {J.}~\bibnamefont {Kessler}}, \bibinfo {author} {\bibfnamefont
  {I.}~\bibnamefont {Poltavsky}}, \bibinfo {author} {\bibfnamefont
  {S.}~\bibnamefont {Vandenbrande}}, \bibinfo {author} {\bibfnamefont
  {J.}~\bibnamefont {Wieme}}, \bibinfo {author} {\bibfnamefont
  {C.}~\bibnamefont {Corminboeuf}}, \bibinfo {author} {\bibfnamefont {T.~D.}\
  \bibnamefont {Kühne}}, \bibinfo {author} {\bibfnamefont {D.~E.}\
  \bibnamefont {Manolopoulos}}, \bibinfo {author} {\bibfnamefont {T.~E.}\
  \bibnamefont {Markland}}, \bibinfo {author} {\bibfnamefont {J.~O.}\
  \bibnamefont {Richardson}}, \bibinfo {author} {\bibfnamefont
  {A.}~\bibnamefont {Tkatchenko}}, \bibinfo {author} {\bibfnamefont {G.~A.}\
  \bibnamefont {Tribello}}, \bibinfo {author} {\bibfnamefont {V.}~\bibnamefont
  {{Van Speybroeck}}},\ and\ \bibinfo {author} {\bibfnamefont {M.}~\bibnamefont
  {Ceriotti}},\ }\bibfield  {title} {\bibinfo {title} {i-pi 2.0: A universal
  force engine for advanced molecular simulations},\ }\href
  {https://doi.org/https://doi.org/10.1016/j.cpc.2018.09.020} {\bibfield
  {journal} {\bibinfo  {journal} {Computer Physics Communications}\ }\textbf
  {\bibinfo {volume} {236}},\ \bibinfo {pages} {214} (\bibinfo {year}
  {2019})}\BibitemShut {NoStop}%
\bibitem [{\citenamefont {Thompson}\ \emph {et~al.}(2022)\citenamefont
  {Thompson}, \citenamefont {Aktulga}, \citenamefont {Berger}, \citenamefont
  {Bolintineanu}, \citenamefont {Brown}, \citenamefont {Crozier}, \citenamefont
  {in~'t Veld}, \citenamefont {Kohlmeyer}, \citenamefont {Moore}, \citenamefont
  {Nguyen}, \citenamefont {Shan}, \citenamefont {Stevens}, \citenamefont
  {Tranchida}, \citenamefont {Trott},\ and\ \citenamefont {Plimpton}}]{LAMMPS}%
  \BibitemOpen
  \bibfield  {author} {\bibinfo {author} {\bibfnamefont {A.~P.}\ \bibnamefont
  {Thompson}}, \bibinfo {author} {\bibfnamefont {H.~M.}\ \bibnamefont
  {Aktulga}}, \bibinfo {author} {\bibfnamefont {R.}~\bibnamefont {Berger}},
  \bibinfo {author} {\bibfnamefont {D.~S.}\ \bibnamefont {Bolintineanu}},
  \bibinfo {author} {\bibfnamefont {W.~M.}\ \bibnamefont {Brown}}, \bibinfo
  {author} {\bibfnamefont {P.~S.}\ \bibnamefont {Crozier}}, \bibinfo {author}
  {\bibfnamefont {P.~J.}\ \bibnamefont {in~'t Veld}}, \bibinfo {author}
  {\bibfnamefont {A.}~\bibnamefont {Kohlmeyer}}, \bibinfo {author}
  {\bibfnamefont {S.~G.}\ \bibnamefont {Moore}}, \bibinfo {author}
  {\bibfnamefont {T.~D.}\ \bibnamefont {Nguyen}}, \bibinfo {author}
  {\bibfnamefont {R.}~\bibnamefont {Shan}}, \bibinfo {author} {\bibfnamefont
  {M.~J.}\ \bibnamefont {Stevens}}, \bibinfo {author} {\bibfnamefont
  {J.}~\bibnamefont {Tranchida}}, \bibinfo {author} {\bibfnamefont
  {C.}~\bibnamefont {Trott}},\ and\ \bibinfo {author} {\bibfnamefont {S.~J.}\
  \bibnamefont {Plimpton}},\ }\bibfield  {title} {\bibinfo {title} {{LAMMPS} -
  a flexible simulation tool for particle-based materials modeling at the
  atomic, meso, and continuum scales},\ }\href
  {https://doi.org/10.1016/j.cpc.2021.108171} {\bibfield  {journal} {\bibinfo
  {journal} {Comp. Phys. Comm.}\ }\textbf {\bibinfo {volume} {271}},\ \bibinfo
  {pages} {108171} (\bibinfo {year} {2022})}\BibitemShut {NoStop}%
\bibitem [{\citenamefont {Bussi}\ \emph {et~al.}(2007)\citenamefont {Bussi},
  \citenamefont {Donadio},\ and\ \citenamefont {Parrinello}}]{Bussi2007}%
  \BibitemOpen
  \bibfield  {author} {\bibinfo {author} {\bibfnamefont {G.}~\bibnamefont
  {Bussi}}, \bibinfo {author} {\bibfnamefont {D.}~\bibnamefont {Donadio}},\
  and\ \bibinfo {author} {\bibfnamefont {M.}~\bibnamefont {Parrinello}},\
  }\bibfield  {title} {\bibinfo {title} {Canonical sampling through velocity
  rescaling},\ }\href {https://doi.org/10.1063/1.2408420} {\bibfield  {journal}
  {\bibinfo  {journal} {The Journal of Chemical Physics}\ }\textbf {\bibinfo
  {volume} {126}},\ \bibinfo {pages} {014101} (\bibinfo {year} {2007})},\
  \Eprint {https://arxiv.org/abs/https://doi.org/10.1063/1.2408420}
  {https://doi.org/10.1063/1.2408420} \BibitemShut {NoStop}%
\bibitem [{\citenamefont {Ceriotti}\ \emph {et~al.}(2010)\citenamefont
  {Ceriotti}, \citenamefont {Parrinello}, \citenamefont {Markland},\ and\
  \citenamefont {Manolopoulos}}]{Ceriotti2010}%
  \BibitemOpen
  \bibfield  {author} {\bibinfo {author} {\bibfnamefont {M.}~\bibnamefont
  {Ceriotti}}, \bibinfo {author} {\bibfnamefont {M.}~\bibnamefont
  {Parrinello}}, \bibinfo {author} {\bibfnamefont {T.~E.}\ \bibnamefont
  {Markland}},\ and\ \bibinfo {author} {\bibfnamefont {D.~E.}\ \bibnamefont
  {Manolopoulos}},\ }\bibfield  {title} {\bibinfo {title} {Efficient stochastic
  thermostatting of path integral molecular dynamics},\ }\href
  {https://doi.org/10.1063/1.3489925} {\bibfield  {journal} {\bibinfo
  {journal} {The Journal of Chemical Physics}\ }\textbf {\bibinfo {volume}
  {133}},\ \bibinfo {pages} {124104} (\bibinfo {year} {2010})},\ \Eprint
  {https://arxiv.org/abs/https://doi.org/10.1063/1.3489925}
  {https://doi.org/10.1063/1.3489925} \BibitemShut {NoStop}%
\bibitem [{\citenamefont {Vedamuthu}\ \emph {et~al.}(1996)\citenamefont
  {Vedamuthu}, \citenamefont {Singh},\ and\ \citenamefont
  {Robinson}}]{Vedamuthu1996}%
  \BibitemOpen
  \bibfield  {author} {\bibinfo {author} {\bibfnamefont {M.}~\bibnamefont
  {Vedamuthu}}, \bibinfo {author} {\bibfnamefont {S.}~\bibnamefont {Singh}},\
  and\ \bibinfo {author} {\bibfnamefont {G.~W.}\ \bibnamefont {Robinson}},\
  }\bibfield  {title} {\bibinfo {title} {Simple relationship between the
  properties of isotopic water},\ }\href {https://doi.org/10.1021/jp953268z}
  {\bibfield  {journal} {\bibinfo  {journal} {The Journal of Physical
  Chemistry}\ }\textbf {\bibinfo {volume} {100}},\ \bibinfo {pages} {3825}
  (\bibinfo {year} {1996})},\ \Eprint
  {https://arxiv.org/abs/https://doi.org/10.1021/jp953268z}
  {https://doi.org/10.1021/jp953268z} \BibitemShut {NoStop}%
\bibitem [{\citenamefont {Mallamace}\ \emph {et~al.}(2007)\citenamefont
  {Mallamace}, \citenamefont {Branca}, \citenamefont {Broccio}, \citenamefont
  {Corsaro}, \citenamefont {Mou},\ and\ \citenamefont {Chen}}]{Mallamace2007}%
  \BibitemOpen
  \bibfield  {author} {\bibinfo {author} {\bibfnamefont {F.}~\bibnamefont
  {Mallamace}}, \bibinfo {author} {\bibfnamefont {C.}~\bibnamefont {Branca}},
  \bibinfo {author} {\bibfnamefont {M.}~\bibnamefont {Broccio}}, \bibinfo
  {author} {\bibfnamefont {C.}~\bibnamefont {Corsaro}}, \bibinfo {author}
  {\bibfnamefont {C.-Y.}\ \bibnamefont {Mou}},\ and\ \bibinfo {author}
  {\bibfnamefont {S.-H.}\ \bibnamefont {Chen}},\ }\bibfield  {title} {\bibinfo
  {title} {The anomalous behavior of the density of water in the range 30 k
  \&lt; <i>t</i> \&lt; 373 k},\ }\href
  {https://doi.org/10.1073/pnas.0706504104} {\bibfield  {journal} {\bibinfo
  {journal} {Proceedings of the National Academy of Sciences}\ }\textbf
  {\bibinfo {volume} {104}},\ \bibinfo {pages} {18387} (\bibinfo {year}
  {2007})},\ \Eprint
  {https://arxiv.org/abs/https://www.pnas.org/doi/pdf/10.1073/pnas.0706504104}
  {https://www.pnas.org/doi/pdf/10.1073/pnas.0706504104} \BibitemShut {NoStop}%
\bibitem [{\citenamefont {Yeh}\ and\ \citenamefont {Hummer}(2004)}]{Yeh2004}%
  \BibitemOpen
  \bibfield  {author} {\bibinfo {author} {\bibfnamefont {I.~C.}\ \bibnamefont
  {Yeh}}\ and\ \bibinfo {author} {\bibfnamefont {G.}~\bibnamefont {Hummer}},\
  }\bibfield  {title} {\bibinfo {title} {{System-size dependence of diffusion
  coefficients and viscosities from molecular dynamics simulations with
  periodic boundary conditions}},\ }\href {https://doi.org/10.1021/jp0477147}
  {\bibfield  {journal} {\bibinfo  {journal} {Journal of Physical Chemistry B}\
  }\textbf {\bibinfo {volume} {108}},\ \bibinfo {pages} {15873} (\bibinfo
  {year} {2004})}\BibitemShut {NoStop}%
\bibitem [{\citenamefont {Kestin}\ \emph {et~al.}(1978)\citenamefont {Kestin},
  \citenamefont {Sokolov},\ and\ \citenamefont {Wakeham}}]{Kestin1978}%
  \BibitemOpen
  \bibfield  {author} {\bibinfo {author} {\bibfnamefont {J.}~\bibnamefont
  {Kestin}}, \bibinfo {author} {\bibfnamefont {M.}~\bibnamefont {Sokolov}},\
  and\ \bibinfo {author} {\bibfnamefont {W.~A.}\ \bibnamefont {Wakeham}},\
  }\bibfield  {title} {\bibinfo {title} {Viscosity of liquid water in the range
  -8c to 150c},\ }\href {https://doi.org/10.1063/1.555581} {\bibfield
  {journal} {\bibinfo  {journal} {Journal of Physical and Chemical Reference
  Data}\ }\textbf {\bibinfo {volume} {7}},\ \bibinfo {pages} {941} (\bibinfo
  {year} {1978})}\BibitemShut {NoStop}%
\bibitem [{\citenamefont {Marsalek}\ and\ \citenamefont
  {Markland}(2017)}]{Marsalek2017}%
  \BibitemOpen
  \bibfield  {author} {\bibinfo {author} {\bibfnamefont {O.}~\bibnamefont
  {Marsalek}}\ and\ \bibinfo {author} {\bibfnamefont {T.~E.}\ \bibnamefont
  {Markland}},\ }\bibfield  {title} {\bibinfo {title} {{Quantum Dynamics and
  Spectroscopy of Ab Initio Liquid Water: The Interplay of Nuclear and
  Electronic Quantum Effects}},\ }\href
  {https://doi.org/10.1021/acs.jpclett.7b00391} {\bibfield  {journal} {\bibinfo
   {journal} {Journal of Physical Chemistry Letters}\ }\textbf {\bibinfo
  {volume} {8}},\ \bibinfo {pages} {1545} (\bibinfo {year} {2017})}\BibitemShut
  {NoStop}%
\bibitem [{\citenamefont {Babin}\ \emph {et~al.}(2013)\citenamefont {Babin},
  \citenamefont {Leforestier},\ and\ \citenamefont {Paesani}}]{Babin2013}%
  \BibitemOpen
  \bibfield  {author} {\bibinfo {author} {\bibfnamefont {V.}~\bibnamefont
  {Babin}}, \bibinfo {author} {\bibfnamefont {C.}~\bibnamefont {Leforestier}},\
  and\ \bibinfo {author} {\bibfnamefont {F.}~\bibnamefont {Paesani}},\
  }\bibfield  {title} {\bibinfo {title} {Development of a “first
  principles” water potential with flexible monomers: Dimer potential energy
  surface, vrt spectrum, and second virial coefficient},\ }\href
  {https://doi.org/10.1021/ct400863t} {\bibfield  {journal} {\bibinfo
  {journal} {Journal of Chemical Theory and Computation}\ }\textbf {\bibinfo
  {volume} {9}},\ \bibinfo {pages} {5395} (\bibinfo {year} {2013})},\ \bibinfo
  {note} {pMID: 26592277},\ \Eprint
  {https://arxiv.org/abs/https://doi.org/10.1021/ct400863t}
  {https://doi.org/10.1021/ct400863t} \BibitemShut {NoStop}%
\bibitem [{\citenamefont {Babin}\ \emph {et~al.}(2014)\citenamefont {Babin},
  \citenamefont {Medders},\ and\ \citenamefont {Paesani}}]{Babin2014}%
  \BibitemOpen
  \bibfield  {author} {\bibinfo {author} {\bibfnamefont {V.}~\bibnamefont
  {Babin}}, \bibinfo {author} {\bibfnamefont {G.~R.}\ \bibnamefont {Medders}},\
  and\ \bibinfo {author} {\bibfnamefont {F.}~\bibnamefont {Paesani}},\
  }\bibfield  {title} {\bibinfo {title} {Development of a “first
  principles” water potential with flexible monomers. ii: Trimer potential
  energy surface, third virial coefficient, and small clusters},\ }\href
  {https://doi.org/10.1021/ct500079y} {\bibfield  {journal} {\bibinfo
  {journal} {Journal of Chemical Theory and Computation}\ }\textbf {\bibinfo
  {volume} {10}},\ \bibinfo {pages} {1599} (\bibinfo {year} {2014})},\ \bibinfo
  {note} {pMID: 26580372},\ \Eprint
  {https://arxiv.org/abs/https://doi.org/10.1021/ct500079y}
  {https://doi.org/10.1021/ct500079y} \BibitemShut {NoStop}%
\bibitem [{\citenamefont {Medders}\ \emph {et~al.}(2014)\citenamefont
  {Medders}, \citenamefont {Babin},\ and\ \citenamefont
  {Paesani}}]{Medders2014}%
  \BibitemOpen
  \bibfield  {author} {\bibinfo {author} {\bibfnamefont {G.~R.}\ \bibnamefont
  {Medders}}, \bibinfo {author} {\bibfnamefont {V.}~\bibnamefont {Babin}},\
  and\ \bibinfo {author} {\bibfnamefont {F.}~\bibnamefont {Paesani}},\
  }\bibfield  {title} {\bibinfo {title} {Development of a
  “first-principles” water potential with flexible monomers. iii. liquid
  phase properties},\ }\href {https://doi.org/10.1021/ct5004115} {\bibfield
  {journal} {\bibinfo  {journal} {Journal of Chemical Theory and Computation}\
  }\textbf {\bibinfo {volume} {10}},\ \bibinfo {pages} {2906} (\bibinfo {year}
  {2014})},\ \bibinfo {note} {pMID: 26588266},\ \Eprint
  {https://arxiv.org/abs/https://doi.org/10.1021/ct5004115}
  {https://doi.org/10.1021/ct5004115} \BibitemShut {NoStop}%
\bibitem [{\citenamefont {Reddy}\ \emph {et~al.}(2016)\citenamefont {Reddy},
  \citenamefont {Straight}, \citenamefont {Bajaj}, \citenamefont {Huy~Pham},
  \citenamefont {Riera}, \citenamefont {Moberg}, \citenamefont {Morales},
  \citenamefont {Knight}, \citenamefont {Götz},\ and\ \citenamefont
  {Paesani}}]{Reddy2016}%
  \BibitemOpen
  \bibfield  {author} {\bibinfo {author} {\bibfnamefont {S.~K.}\ \bibnamefont
  {Reddy}}, \bibinfo {author} {\bibfnamefont {S.~C.}\ \bibnamefont {Straight}},
  \bibinfo {author} {\bibfnamefont {P.}~\bibnamefont {Bajaj}}, \bibinfo
  {author} {\bibfnamefont {C.}~\bibnamefont {Huy~Pham}}, \bibinfo {author}
  {\bibfnamefont {M.}~\bibnamefont {Riera}}, \bibinfo {author} {\bibfnamefont
  {D.~R.}\ \bibnamefont {Moberg}}, \bibinfo {author} {\bibfnamefont {M.~A.}\
  \bibnamefont {Morales}}, \bibinfo {author} {\bibfnamefont {C.}~\bibnamefont
  {Knight}}, \bibinfo {author} {\bibfnamefont {A.~W.}\ \bibnamefont {Götz}},\
  and\ \bibinfo {author} {\bibfnamefont {F.}~\bibnamefont {Paesani}},\
  }\bibfield  {title} {\bibinfo {title} {On the accuracy of the mb-pol
  many-body potential for water: Interaction energies, vibrational frequencies,
  and classical thermodynamic and dynamical properties from clusters to liquid
  water and ice},\ }\href {https://doi.org/10.1063/1.4967719} {\bibfield
  {journal} {\bibinfo  {journal} {The Journal of Chemical Physics}\ }\textbf
  {\bibinfo {volume} {145}},\ \bibinfo {pages} {194504} (\bibinfo {year}
  {2016})},\ \Eprint {https://arxiv.org/abs/https://doi.org/10.1063/1.4967719}
  {https://doi.org/10.1063/1.4967719} \BibitemShut {NoStop}%
\end{thebibliography}%
	
\end{document}


\title{Supplementary material for: ``Machine learning potentials from transfer learning of periodic correlated electronic structure methods: Application to liquid water with AFQMC, CCSD, and CCSD(T)"}

\author{Michael S. Chen}
\affiliation{Department of Chemistry, Stanford University, Stanford, California, 94305, USA}

\author{Joonho Lee}
\affiliation{Department of Chemistry, Columbia University, New York, New York 10027, USA}

\author{Hong-Zhou Ye}
\affiliation{Department of Chemistry, Columbia University, New York, New York 10027, USA}

\author{Timothy C. Berkelbach}
\email{t.berkelbach@columbia.edu}
\affiliation{Department of Chemistry, Columbia University, New York, New York 10027, USA}
\affiliation{Center for Computational Quantum Physics, Flatiron Institute, New York, New York 10010, USA}

\author{David R. Reichman}
\email{drr2103@columbia.edu}
\affiliation{Department of Chemistry, Columbia University, New York, New York 10027, USA}

\author{Thomas E. Markland}
\email{tmarkland@stanford.edu}
\affiliation{Department of Chemistry, Stanford University, Stanford, California, 94305, USA}

\date{\today}

\maketitle

\tableofcontents

\section{Electronic structure calculation details}
\subsection{DFT, HF, and DFTB calculations}
\label{sec:si-dft-hf-dftb}
We used the CP2K package\cite{Hutter2014} for our training set HF and DFT calculations, employing the BLYP\cite{Becke1988,Lee1988} and revPBE0\cite{Perdew1996,Zhang1998,Adamo1999} functionals. Our SCC-DFTB\cite{Elstner1998} calculations, used to perform an initial MD simulation for sampling the first 50 training set configurations, were also conducted using CP2K. All calculations were run using periodic boundary conditions.

We used Goedecker-Tetter-Hutter (GTH) pseudopotentials\cite{Goedecker1996Jul} for our HF and DFT calculations conducted in CP2K. For our BLYP calculations, we employed a mixed Gaussian and plane wave method\cite{Lippert1997,VandeVondele2005} using the DZVP\cite{VandeVondele2007Sep} basis set and an auxiliary plane wave basis with a 300 Ry cutoff. These HF and revPBE0 calculations made use of the TZV2P\cite{VandeVondele2007Sep} basis set and a 400 Ry cutoff auxiliary plane wave basis. We employed a truncated Coulomb operator\cite{Guidon2009} with a cutoff of 3.9~\AA~and the auxiliary density matrix method\cite{Guidon2010} with the cpFIT3 fitting basis set. The short cutoff we employed for the truncated Coulomb operator was due to the small periodic boxes we used for our training set (side lengths of 7.831~\AA). In checking the effects of employing such a short cutoff, and in general these small periodic boxes for the training set, we found that the effects seemed minimal with regards to the properties we are examining in this work (see SI Sec.~\ref{sec:si-revpbe0-refit-tests} for more details). For the revPBE0 calculations, we applied D3 dispersion correction\cite{Grimme2010}.

Our SCC-DFTB calculations were run using parametrizations provided by CP2K for O and H atoms. We also employed an Ewald sum\cite{Ewald1921}, in the form of a smooth particle mesh using beta-Euler splines\cite{Essman1995}, for electrostatics.

\subsection{CCSD and CCSD(T)}
\label{sec:si-ccsd-ccsdt}

We performed our CCSD and CCSD(T) calculations using the PySCF \cite{sun2020recent} software package and the cc-pVQZ basis set. To reduce the high computational cost of these calculations, we use the frozen natural orbital (FNO) approximation \cite{Taube08JCP} as detailed in previous work \cite{Lange20MP} to systematically compress the virtual space. Particularly, we found that using $200$ virtual orbitals is enough to reach a maximum absolute error of $0.03$ kcal/mol for CCSD and $0.02$ kcal/mol for CCSD(T) or a root-mean-square error of $0.02$ kcal/mol for CCSD and $0.01$ kcal/mol for CCSD(T) in the relative energy (per water molecule) between $10$ randomly drawn $16$-water molecule configurations. We thus performed all CCSD and CCSD(T) calculations in this work with the FNO approximation and $200$ virtual orbitals.

\subsection{AFQMC}
\label{sec:si-afqmc}
We performed our AFQMC calculations sing QMCPACK\cite{kim2018qmcpack,Kent2020May} and ipie\cite{Malone2022Sep}, using the TZV2P basis set\cite{VandeVondele2007Sep} and Goedecker-Tetter-Hutter PBE pseudopotentials\cite{Goedecker1996Jul,Hartwigsen1998Aug}. Integrals were generated from PySCF\cite{sun2020recent} including the point charge probe finite size correction\cite{Fraser1814Jan}. Note that the AFQMC energies we ultimately trained our MLPs on used the HF energies computed as part of our CCSD/CCSD(T) calculations, where we employed a larger basis set (SI Sec.~\ref{sec:si-ccsd-ccsdt}), in lieu of the energies associated with the HF trial wavefunctions employed for the AFQMC calculations.

\subsection{Basis set convergence tests}
\begin{figure*}[h!]
    \begin{center}
        \includegraphics{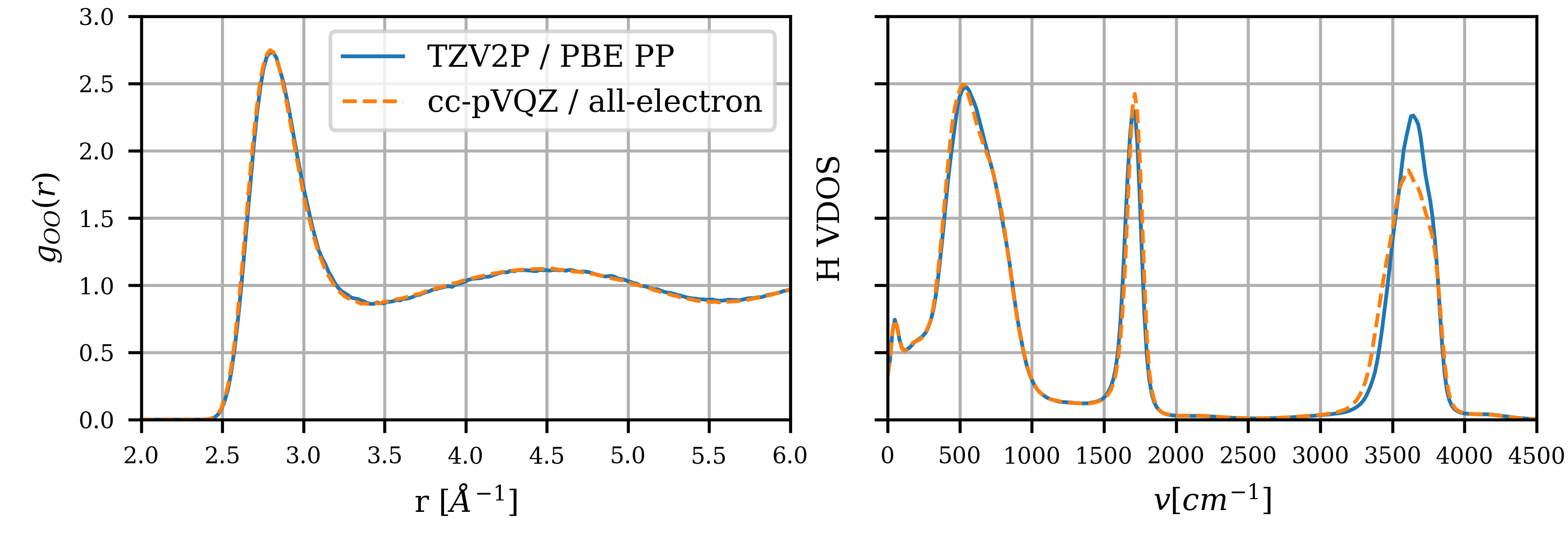}
    \end{center}
    \caption{Oxygen-oxygen RDFs and hydrogen VDOS from two different transfer learning committee MLP models both initialized using weights from a committee of revPBE0-D3 MLPs. One model was trained to N=200 configurations where the CCSD(T) energies were computed with the TZV2P basis set and the GTH pseudopotentials for the PBE functional. The other model was trained on the same N=200 configurations, but with the CCSD(T) energies computed using the cc-pVQZ basis set and all-electron potentials}
	\label{fig:basis_set_test}
\end{figure*}

\subsection{Correlations coefficients between methods}
\begin{figure*}[h!]
    \begin{center}
        \includegraphics[width=0.65\textwidth]{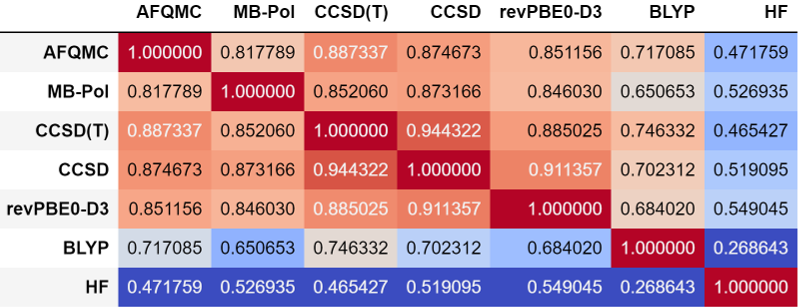}
    \end{center}
    \caption{Table of Kendall rank correlation coefficients between different electronic structure methods, computed over the training set of 200 liquid water configurations (small periodic boxes of 16 waters) we used in this work.}
    \label{tab:si_electronic_structure_correlations}
\end{figure*}

\section{Machine learning details}
\label{sec:si-ml-details}
All of our committee machine learning potentials (MLPs), each consisting of 8 separate MLPs, were fit using the n2p2 package\cite{n2p2} v2.1.0. We employed the same set of input features (symmetry functions) and neural network architectures for all individual MLPs in our committee approach as was used in previous work fitting MLPs for liquid water\cite{Morawietz2016}. In short, for our MLPs we used Behler-Parinello high dimensional neural network potentials employing local atom-centered symmetry function descriptors\cite{Behler2007} that decompose the potential energy of a system as a sum of predicted atomic contributions.

When fitting our MLPs to both configuration energies and atomic forces, as was the case with our HF and BLYP committee MLPs, we employed the same training and optimizer hyperparameters (e.g. a 90-10 training to validation set split, early-stopping, Nguyen-Widrow weight initializiation\cite{Nguyen1990}, the extended Kalman Filter\cite{Kalman1960,Kalman1961,Smith1962}, etc.) as were used and benchmarked in previously published works\cite{Morawietz2016,Singraber2019,Schran2020}. These force-fitted MLPs were fit to 531 configurations of 16 waters and were used to initialize refits targeting revPBE0-D3, which served as preliminary checks on the effectiveness of this transfer learning procedure. For the refits to CCSD, CCSD(T), and AFQMC, in addition to starting from HF and BLYP, we also used the generation 1 revPBE0-D3 fits \cite{Schran2020}  from Schran et. al. to initialize the transfer learning models. All of the refits we conducted were trained using just the energies of 200 configurations and employed early-stopping, using a 90-10 training to validation set split, to mitigate overfitting.

\section{Molecular dynamics simulation details}
\label{sec:si-md-details}

All classical MD and PIMD NVT simulations employing our committee of ML potentials were conducted using the i-PI\cite{Ceriotti2014,Kapil2019} and LAMMPS\cite{LAMMPS} packages in conjunction, the latter of which is interfaced to n2p2\cite{Singraber2019}. Our classical MD simulations were run using a 0.5~fs timestep and the global canonical sampling through velocity rescaling (CSVR) thermostat\cite{Bussi2007} with a time constant of 2~ps in order to sample the T=300K and 370K NVT ensembles. The properties reported in this work were obtained from classical MD simulations that were run for 1~ns and preceded by a 50~ps equilibration run. The PIMD simulations were run using a 0.25~fs timestep and the global version of the path integral Langevin equation thermostat (PILE-G)\cite{Ceriotti2010} to sample the T=300~K and 370~K NVT ensembles using a time constant of 20~ps for the centroid. For the PIMD simulations, 32 replicas were employed. The properties reported in this work were obtained from PIMD simulations run for 500~ps and preceded by a 50~ps equilibration run. Note that the classical and PIMD simulations used to compute reported properties were performed with boxes of 64 water molecules. For our 300~K simulations, we used the experimental density of liquid water at that temperature, 0.996~g/cm$^3$\cite{Vedamuthu1996,Mallamace2007}, to obtain a simulation box length of 12.432~\AA, and at 370~K we used the experimental density of 0.960~g/cm$^3$\cite{Mallamace2007} to obtain a box length of 12.590~\AA.

The MD simulations we ran as part of our active learning scheme for building our training set were run for 50~ps and for a system consisting of 16 water molecules in a simulation box with side lengths of 7.831~\AA, with all other settings kept the same as the simulations used to compute properties. The initial 100~ps MD simulation using DFTB that we used to initialize our active sampling scheme was also run using the same settings.

\section{Computational details for benchmark calculations}

\subsection{Self-diffusion coefficients}
\label{sec:si-diffusion-coefficients}
We calculated the molecular self-diffusion coefficients by performing a linear fit of the long-time slope of the mean square displacement (MSD) of the oxygen atoms vs. time. We calculate our MSDs out to 10~ps and fit the region from 1~ps to 10~ps. The value we report is the mean diffusion coefficient as calculated over separate 40~ps chunks of the sampled trajectory and the error bars correspond to the standard deviation of the mean.

We used the following relation to correct for the systems size scaling of the simulated diffusion coefficient\cite{Yeh2004}:

\begin{align}
    D(\infty) = D(L) + \frac{\xi}{6\pi\beta\eta L}
\end{align}

$L$ is the length of the simulation box, $\xi$ is a geometry specific constant that is 2.837297 for a cubic box, and $\eta$ is the shear viscosity (we use the experimental values at T=300~K and 370~K depending on which NVT ensemble we are sampling\cite{Kestin1978}).

\subsection{Hydrogen vibrational density of states (VDOS)}
The hydrogen VDOS was computed by Fourier transforming in time the velocity autocorrelation function. We calculated the velocity autocorrelation function out to 2~ps.

\section{Transferability of relative training set performance}
\label{sec:si-trainset-relative-performance}
\begin{figure*}[h!]
    \begin{center}
        \includegraphics{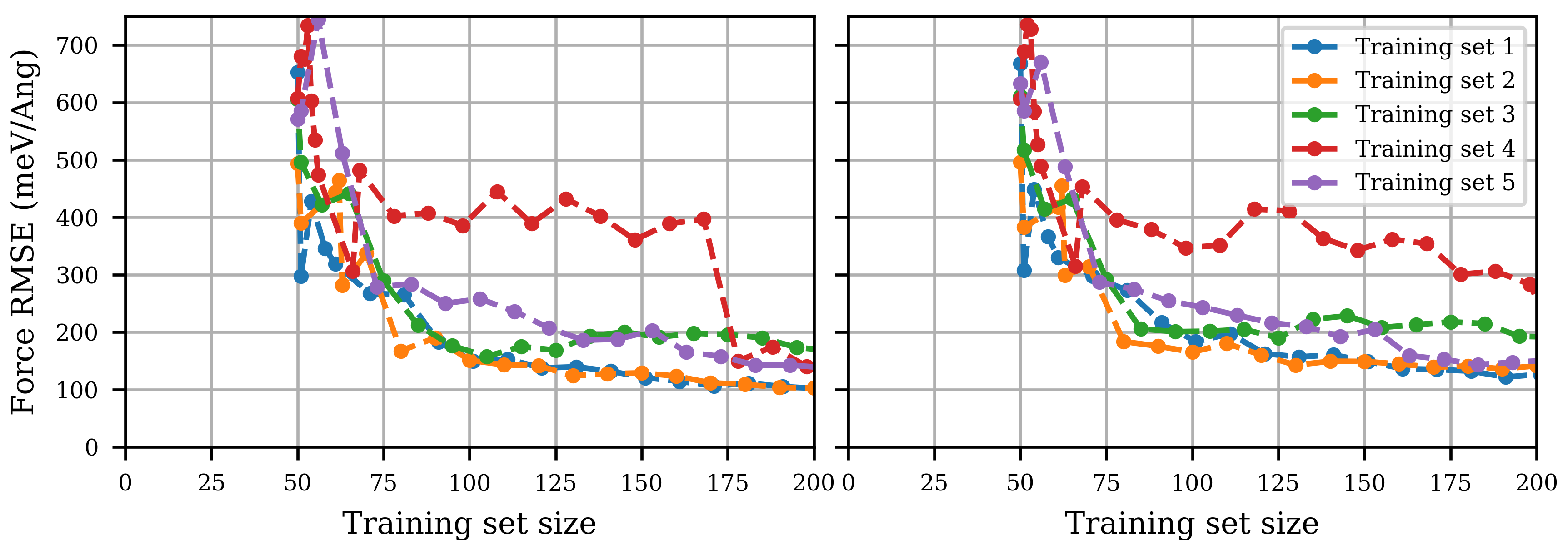}
    \end{center}
    \caption{(Left) Learning curves showing committee MLPs' force prediction errors, on a test set of 1000 configurations (64 waters) sampled randomly from a reference AIMD trajectory\cite{Marsalek2017}, with increasing training set size (16 waters) for 5 separate active learning sampled training sets where the target level of electronic structure theory was DFT using the revPBE0-D3 functional. (Right) The same 5 training set configurations result in similar relative test set force prediction errors when training a model to DFT using the BLYP functional, highlighting the transferability of relative utility, across different electronic structure methods, of the training sets for liquid water.}
\label{fig:trainingset_force_lc}
\end{figure*}

SI Fig.~\ref{fig:trainingset_force_lc} demonstrates that the relative performance of our training sets sampled via the active learning procedure we used is transferable between different electronic structure methods. Specifically, we see that training set 1 (blue), which contains configurations sampled with revPBE0-D3 as the target level of electronic structure theory, results in the lowest test set force error predictions for when revPBE0-D3 is the target (left) and also when BLYP is the target (right). Conversely, training set 4 (red) is the highest error training set for both revPBE0-D3 and BLYP and in general the relative performance of training sets is consistent for both DFT functionals. Given the transferability of the relative performance of these sets of training configurations across different electronic structure methods, we whose to use training set 1 for our CCSD, CCSD(T), and AFQMC transfer learning fits.

\section{Testing transfer learning procedure on DFT/revPBE0-D3}
\label{sec:si-revpbe0-refit-tests}
The results in this section show that the transfer learning procedure we employed starting with a committee of MLPs trained to either HF or BLYP energies and forces that are then refitted using revPBE0-D3 energies works successfully in reproducing certain target AIMD and PIMD properties. These results also show that it is possible to train MLPs on small periodic boxes of 16 water molecules and use them to run simulations for larger box sizes, i.e. the boxes of 64 water molecules used in the reference AIMD and PIMD simulations, that reproduce certain static and dynamical properties of liquid water. 

\begin{figure*}[]
    \begin{center}
        \includegraphics[width=\textwidth]{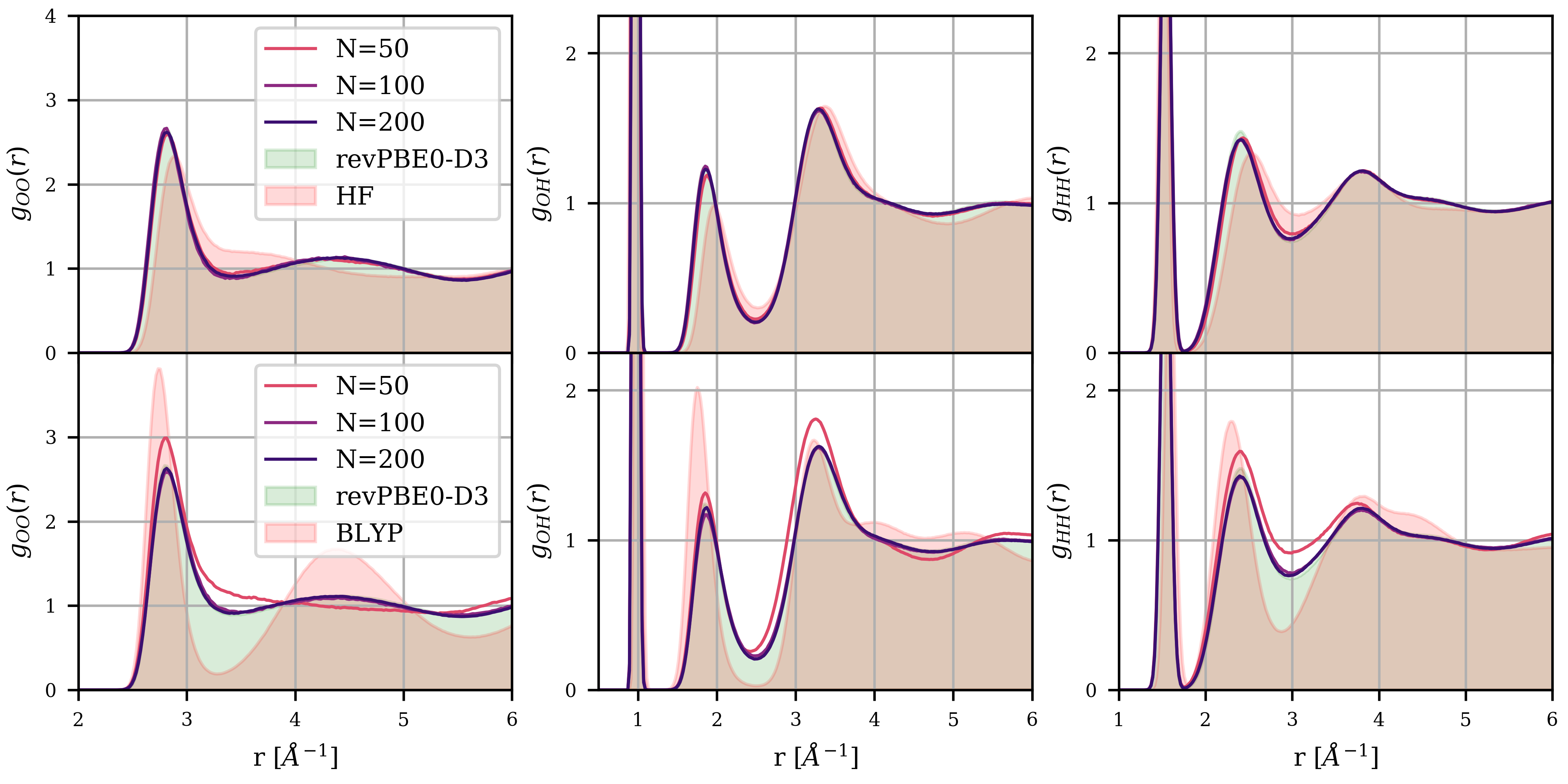}
    \end{center}
    \caption{Transfer learning RDF results (classical NVT MD @300K) targeting DFT using the revPBE0-D3 functional starting with initial weights from fits to HF (top row) and DFT using the BLYP functional (bottom row) showing that the procedure reproduces RDF's of the target method despite starting from an understructured (HF) or overstructured (BLYP) description of the $g_{OO}(r)$. N denotes the number of configuration energies  trained on where each configuration is a periodic simulation boxes of 16 water molecules.}
	\label{fig:revpbe0-D3_refits_rdfs_classical_300K}
\end{figure*}

\begin{figure*}[]
    \begin{center}
        \includegraphics{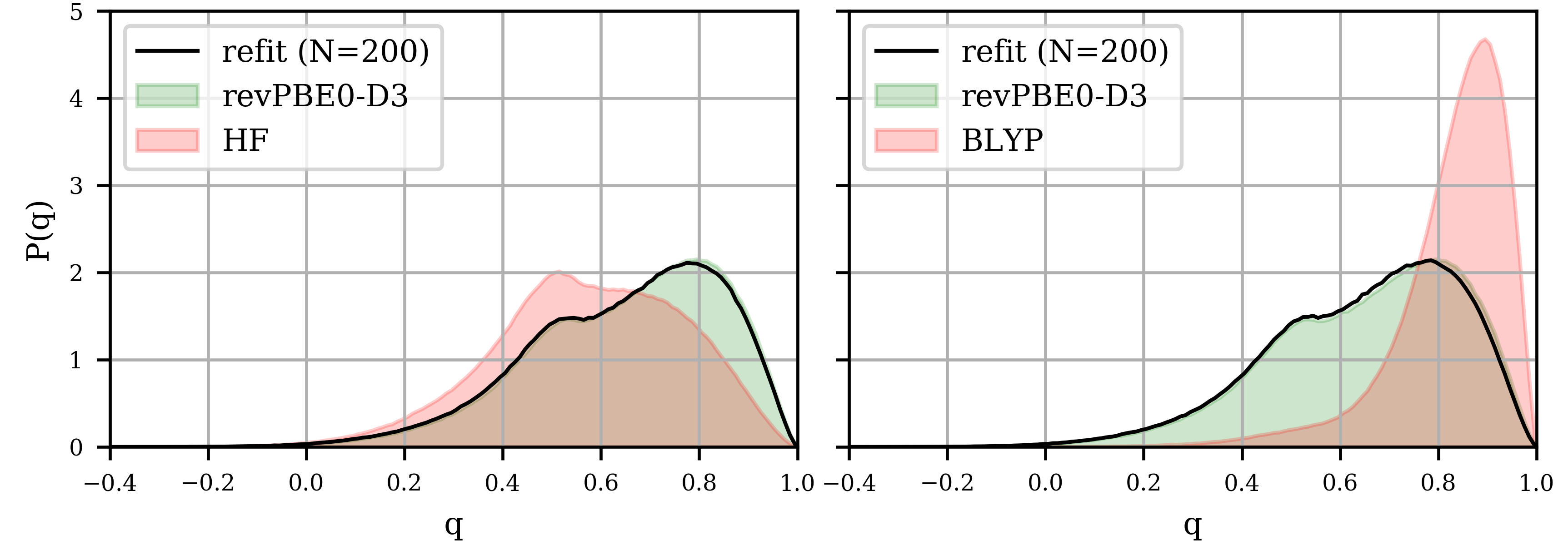}
    \end{center}
    \caption{Transfer learning tetrahedrality parameter, $q$, distribution results (classical NVT MD @300K) targeting DFT using the revPBE0-D3 functional starting with initial weights from fits to HF (left) and DFT using the BLYP functional (right) showing that the procedure, when using a training set of N=200 energy calculations, reproduces the $q$ distribution of the target method despite starting from vastly different initial descriptions.}
	\label{fig:revpbe0-D3_refits_tetra_classical_300K}
\end{figure*}

\begin{figure*}[]
    \begin{center}
        \includegraphics{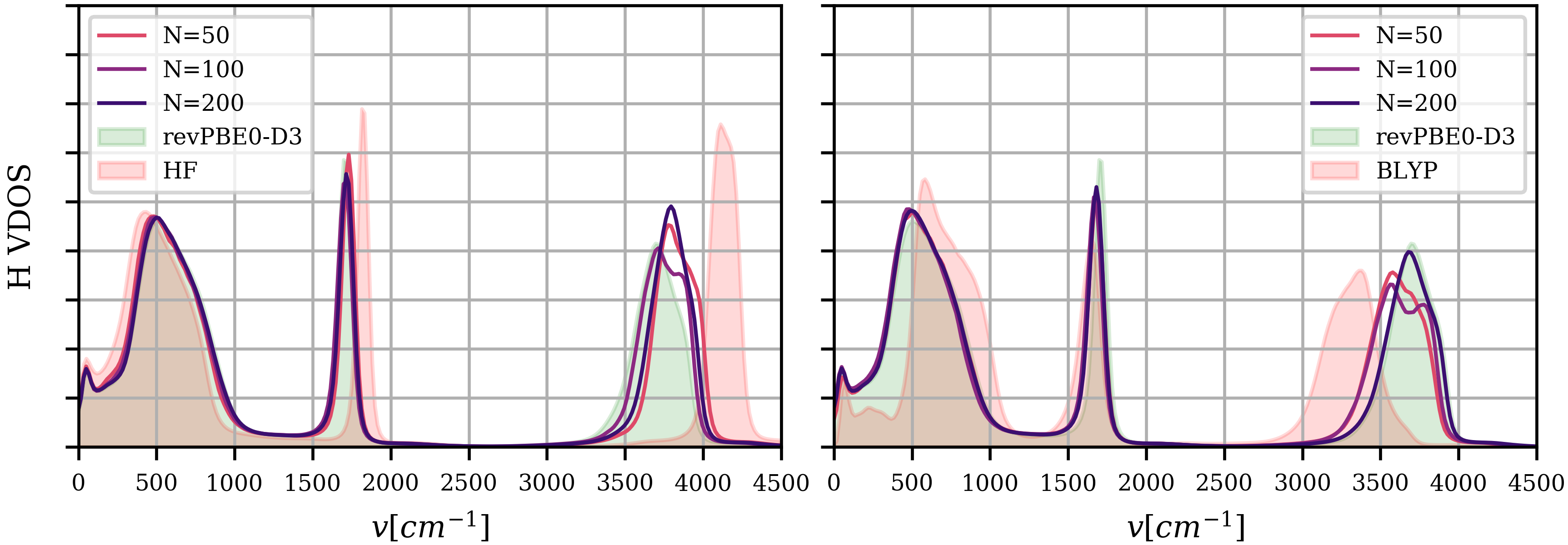}
    \end{center}
    \caption{Transfer learning VDOS results (classical NVT MD @300K) targeting DFT using the revPBE0-D3 functional starting with initial weights from fits to HF (top row) and DFT using the BLYP functional (bottom row) showing that the procedure reproduces the VDOS of the target method despite starting from descriptions that underpredict (BLYP) or overpredict (HF) peaks positions in the VDOS. N denotes the number of configuration energies  trained on where each configuration is a periodic simulation boxes of 16 water molecules.}
	\label{fig:revpbe0-D3_refits_vdos_classical_300K}
\end{figure*}

SI Figs.~\ref{fig:revpbe0-D3_refits_rdfs_classical_300K}, \ref{fig:revpbe0-D3_refits_tetra_classical_300K}, and \ref{fig:revpbe0-D3_refits_vdos_classical_300K} show that the procedure we employed, whether starting from MLPs fitted to HF or BLYP (corresponding MD results shaded in red), gives us a committee of transfer learned MLPs we can use to run classical MD simulations that reproduce the reference AIMD revPBE0-D3 RDFs, distribution of tetrahedral order parameters, and hydrogen VDOS (shaded in green)\cite{Marsalek2017} when trained on as few as N=200 revPBE0-D3 energies. The first row of SI Table~\ref{tab:si-diffusion-coeffs} also shows that the committee of MLPs from this transfer learning procedure also reproduces the AIMD diffusion coefficient.

\begin{figure*}[]
    \begin{center}
        \includegraphics[width=\textwidth]{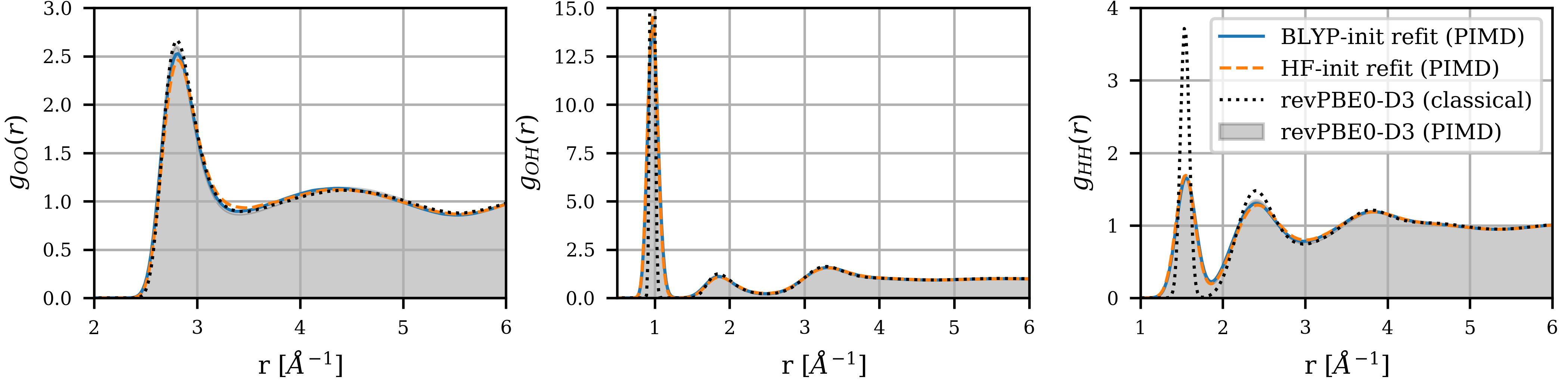}
    \end{center}
    \caption{Transfer learning RDF (NVT PIMD @300K) results targeting DFT using the revPBE0-D3 functional, when training on N=200 configurations energies, starting with initial weights from fits to HF and DFT using the BLYP functional showing that the procedure approximately reproduces RDF's of the target method.}
	\label{fig:revpbe0-D3_refits_rdfs_pimd_300K}
\end{figure*}

\begin{figure*}[]
    \begin{center}
        \includegraphics[width=\textwidth]{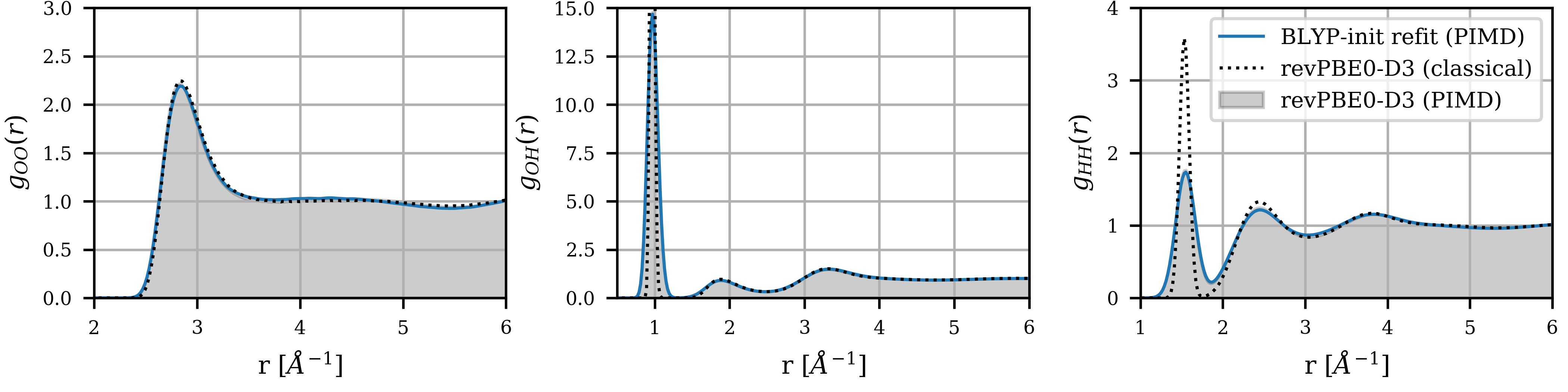}
    \end{center}
    \caption{Transfer learning RDF (NVT PIMD @370K) results targeting DFT using the revPBE0-D3 functional, when training on N=200 configurations energies, starting with initial weights from fits to HF and DFT using the BLYP functional showing that the procedure approximately reproduces RDF's of the target method.}
	\label{fig:revpbe0-D3_refits_rdfs_pimd_370K}
\end{figure*}

\begin{figure*}[]
    \begin{center}
        \includegraphics{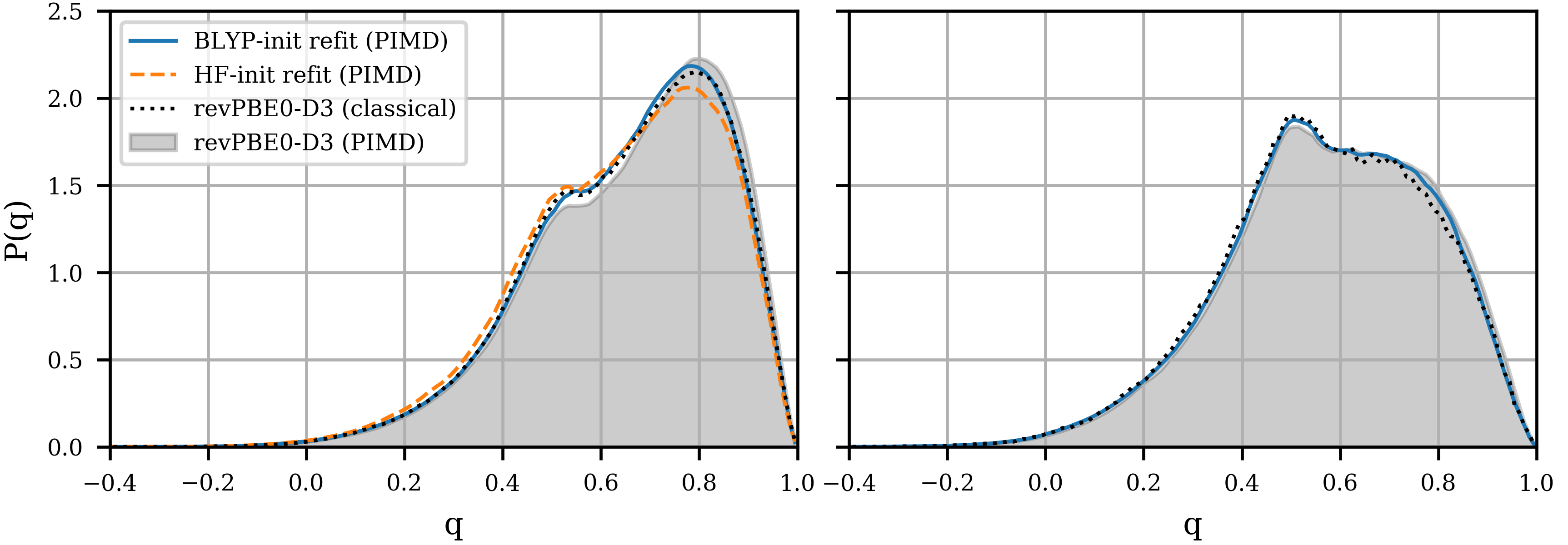}
    \end{center}
    \caption{Transfer learning tetrahedrality parameter, $q$, distribution results from PIMD simulations at T=300K (left) and T=370K (right) targeting DFT using the revPBE0-D3 functional, when training on N=200 configurations energies, starting with initial weights from fits to HF and DFT using the BLYP functional. The procedure, when using a training set of N=200 energy calculations, roughly reproduces the $q$ distribution of the target method when using initial weights from a fit to DFT/BLYP.}
	\label{fig:revpbe0-D3_refits_tetras_rpmd_300K_370K}
\end{figure*}

\begin{figure*}[]
    \begin{center}
        \includegraphics{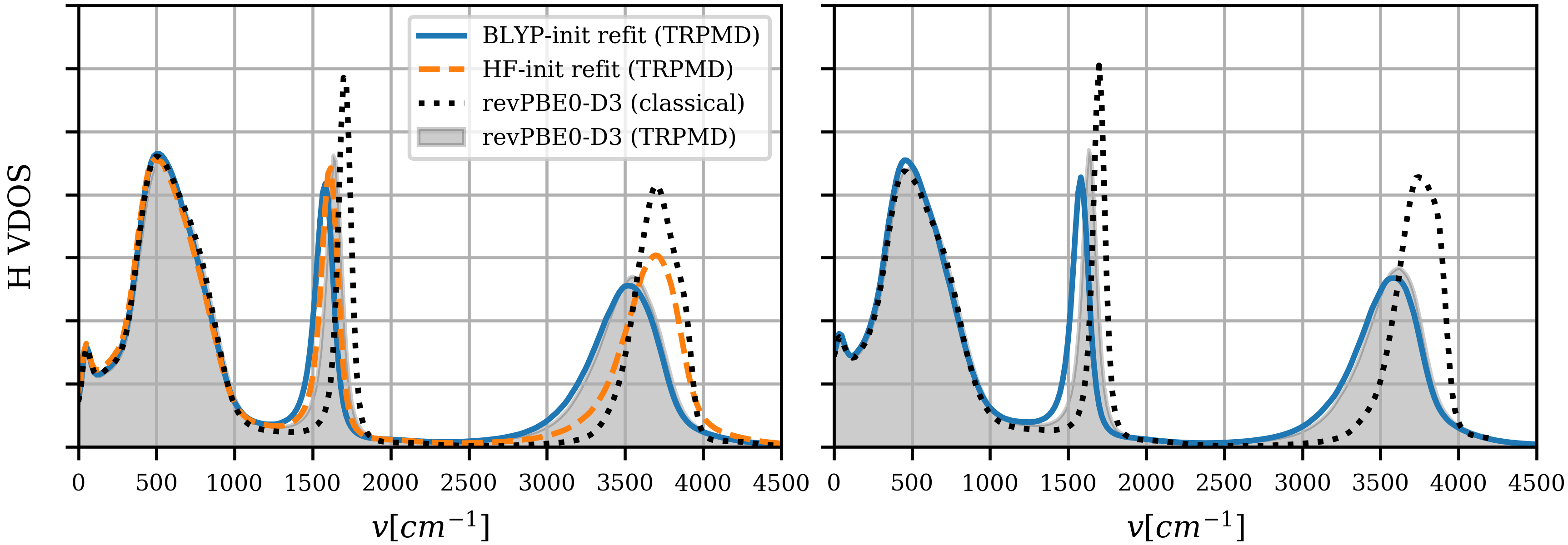}
    \end{center}
    \caption{Transfer learning VDOS results from NVT TRPMD simulations at T=300K (left) and T=370K (right) targeting DFT using the revPBE0-D3 functional, when training on N=200 configurations energies, starting with initial weights from fits to HF and DFT using the BLYP functional showing that the latter approximately reproduces the hydrogen VDOS of the target method.}
	\label{fig:revpbe0-D3_refits_vdos_pimd_300K_370K}
\end{figure*}

SI Figs.~\ref{fig:revpbe0-D3_refits_rdfs_pimd_300K}, \ref{fig:revpbe0-D3_refits_rdfs_pimd_370K}, \ref{fig:revpbe0-D3_refits_tetras_rpmd_300K_370K}, \ref{fig:revpbe0-D3_refits_vdos_pimd_300K_370K} compares the PIMD results using our transfer learned committee of MLPs to reference revPBE0-D3 PIMD results at 300~K\cite{Marsalek2017} as well as reference PIMD results at 370~K, which we obtained using a published committee MLP for liquid water trained to reproduce PIMD results at the level of revPBE0-D3 at ambient and elevated temperatures and pressures\cite{Schran2020}. In general for these PIMD results at 300~K and 370~K, our transfer learned committee MLP reproduces the revPBE0-D3 RDFs, VDOS, and diffusion coefficients within error. On the other hand, the probability distribution of the tetrahedrality parameter ($q$) shown in SI Fig.~\ref{fig:revpbe0-D3_refits_tetras_rpmd_300K_370K} seems to indicate that the transfer learned model PIMD result is closer to the classical MD result for revPBE0-D3 than its PIMD result. However, the differences between the classical MD and PIMD results for the revPBE0-D3 distribution of $q$ are quite small to begin with.



\begin{table}[]
    \vspace{4mm}
    \centering
    \begin{tabular}{ | m{0.15\textwidth} | m{0.25\textwidth} | m{0.25\textwidth}| m{0.25\textwidth} | } 
        \hline
         & BLYP & BLYP$\rightarrow$revPBE0-D3 & revPBE0-D3 \\ \hline
        Classical T=300K ($10^{-9} $ m$^2$/s) & 0.58 (0.004) & 2.67 (0.07) & 2.67 (0.03) \\ \hline
        TRPMD T=300K ($10^{-9} $ m$^2$/s) & X & 2.44 (0.13) & 2.33 (0.09) \\ \hline
        TRPMD T=370K ($10^{-9} $ m$^2$/s) & X & 7.81 (0.15) & 7.64 (0.29) \\ \hline
    \end{tabular}
    \caption{Diffusion coefficients obtained from using revPBE0-D3\cite{Marsalek2017}, a committee of MLPs trained to BLYP enegies and forces, and a transfer learning model that was initialized with the weights from the BLYP MLPs and trained to our N=200 training set of revPBE0-D3 energies. Both classical MD and TRPMD results are shown.}
    \label{tab:si-diffusion-coeffs}
\end{table}


\section{Test evaluating effect of stochastic error in AFQMC energies}
\label{sec:si-energy-noise-tests}
Here we show results from a test we conducted on how the level of stochastic error in the AFQMC energies affects the properties we are benchmarking. Each of our N=200 training set configurations have a calculated AFQMC energy and a corresponding estimate for the standard error that ranges from 1-2~mH depending on the specific configuration. For this test, we sampled new training sets where the same N=200 configuration are used but a random value is added to each AFQMC energy. These random values were drawn from Gaussian distributions with standard deviations set by the estimated standard error for a given configurations AFQMC energy.

\begin{figure*}[h!]
    \begin{center}
        \includegraphics{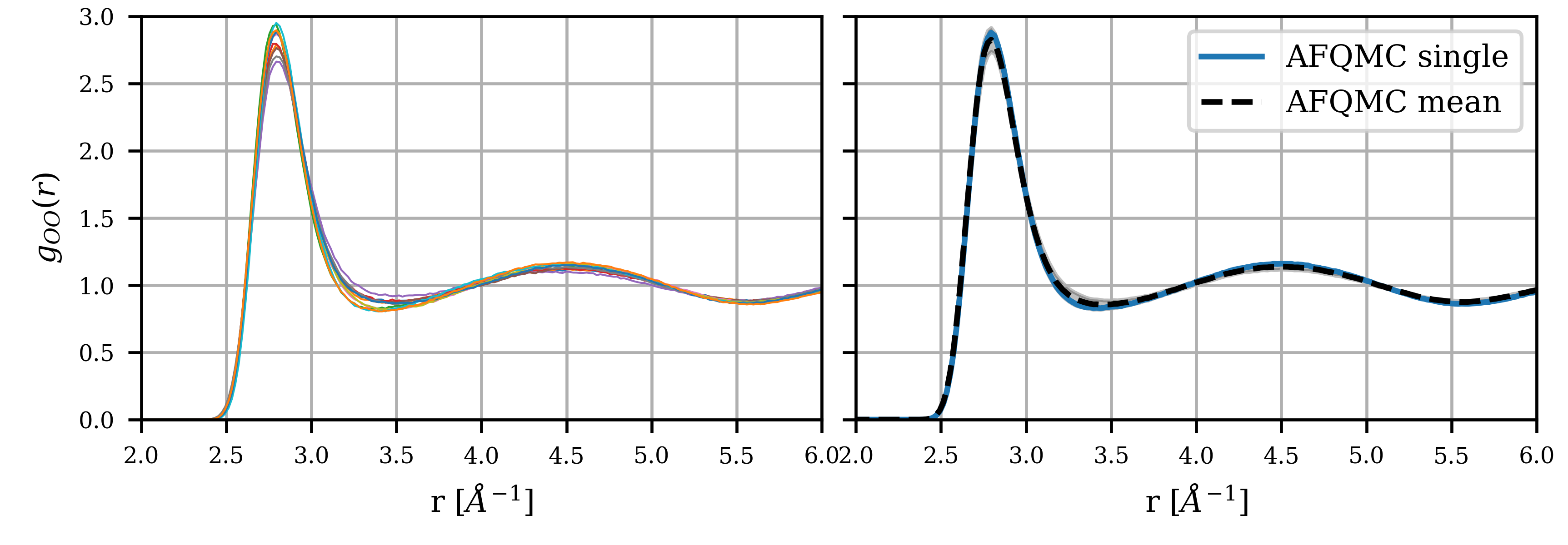}
    \end{center}
    \caption{(left) oxygen-oxygen RDFs obtained from classical MD (T=300~K) using transfer learned committee MLPs fitted to 12 separately training sets of AFQMC energies where random values have been sampled and added to the energies (see SI Sec.~\ref{sec:si-energy-noise-tests} for details). (right) the black dashed line depicts the mean of the 12 RDFs shown on the left panel and the blue line corresponds to the RDF obtained from a single committee MLP fit to the original training set of AFQMC energies. Note that the grey shading around the mean denotes the standard deviations.}
	\label{fig:afqmc_noise_rdf}
\end{figure*}

\begin{figure*}[h!]
    \begin{center}
        \includegraphics{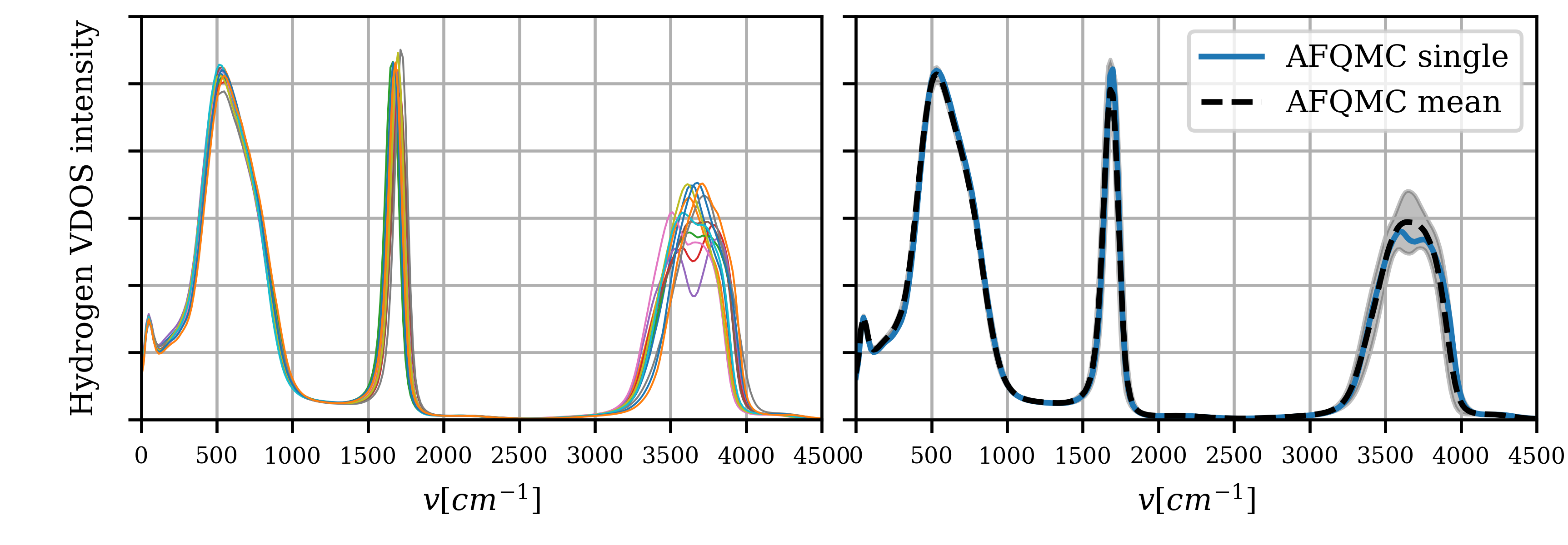}
    \end{center}
    \caption{(left) hydrogen VDOSs obtained from classical MD (T=300~K) using transfer learned committee MLPs fitted to 12 separately training sets of AFQMC energies where random values have been sampled and added to the energies (see SI Sec.~\ref{sec:si-energy-noise-tests} for details). (right) the black dashed line depicts the mean of the 12 VDOSs shown on the left panel and the blue line corresponds to the VDOS obtained from a single committee MLP fit to the original training set of AFQMC energies. Note that the grey shading around the mean denotes the standard deviations.}
	\label{fig:afqmc_noise_vdos}
\end{figure*}

We generated 12 separate training sets in this fashion and the panels on the left side of SI Figs.~\ref{fig:afqmc_noise_rdf} and \ref{fig:afqmc_noise_vdos} show the corresponding oxygen-oxygen RDFs and hydrogen VDOS obtained from transfer learned committee MLPs fitted to each of these 12 training sets, using the revPBE0-D3 MLPs to initialize the weights. We observe that the level of stochastic error in the AFQMC energies results in a spread of predicted RDFs and VDOS, with the variability in the OH stretch peak of the VDOS being particularly pronounced. The right-hand panels of SI Figs.~\ref{fig:afqmc_noise_rdf} and \ref{fig:afqmc_noise_vdos} depict the means of the 12 sets of RDFs and VDOS as a black dashed line with the corresponding standard deviations shaded in grey and barely visible.

\section{Comparison of AFQMC transfer learning results using HF, BLYP, and revPBE0-D3 MLPs to initialize weights}
\label{sec:si-afqmc-different-inits}

\begin{figure*}[h!]
    \begin{center}
        \includegraphics{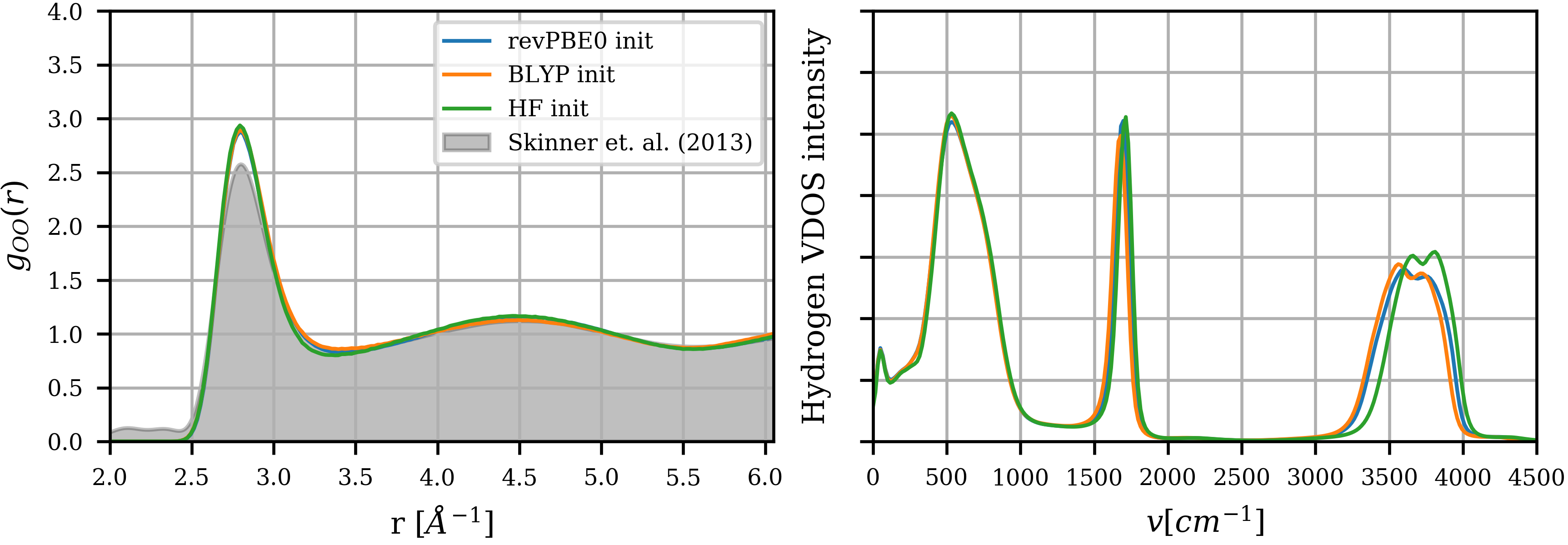}
    \end{center}
    \caption{Transfer learning VDOS results from NVT classical MD simulations targeting AFQMC, when training on N=200 configurations energies, starting with initial weights from fits to HF, DFT/BLYP, and DFT/revPBE0-D3 showing that all 3 models give similar oxygen-oxygen RDFs (left) and VDOS (right).}
	\label{fig:afqmc-rdfs-vdos-classical}
\end{figure*}

\begin{figure*}[h!]
    \begin{center}
        \includegraphics{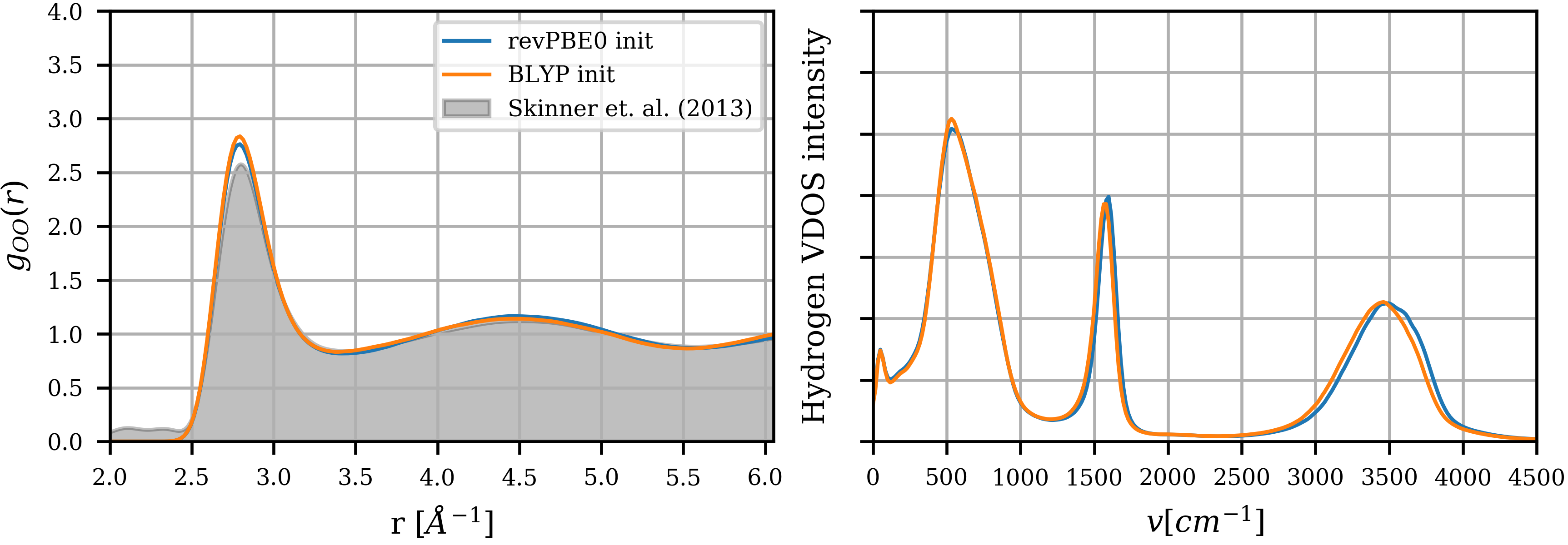}
    \end{center}
    \caption{Transfer learning VDOS results from NVT PIMD/TRPMD simulations targeting AFQMC, when training on N=200 configurations energies, starting with initial weights from fits to DFT/BLYP and DFT/revPBE0-D3 showing that both models give similar oxygen-oxygen RDFs (left) and VDOS (right).}
	\label{fig:afqmc-rdfs-vdos-rpmd}
\end{figure*}

The results in the main text were obtained using committee MLPs that were initially trained to revPBE0-D3 and then refitted to energies from either AFQMC, CCSD, or CCSD(T). SI Figs.~\ref{fig:afqmc-rdfs-vdos-classical} and \ref{fig:afqmc-rdfs-vdos-rpmd} show that the results obtained from different transfer learned models, whether initialized with weights trained to HF, BLYP, or revPBE0-D3 energies and forces, that were refitted to our N=200 AFQMC training set give similar oxygen-oxygen RDFs and hydrogen VDOS. The consistency in these results despite the varying initial conditions gives us additional confidence, on top of our benchmark results presented in SI Sec.~\ref{sec:si-revpbe0-refit-tests}, in the robustness of this transfer learning approach.

\section{Comparisons with MB-Pol}
\label{sec:si-mbpol-comparisons}

\begin{figure*}[h!]
    \begin{center}
        \includegraphics[width=\textwidth]{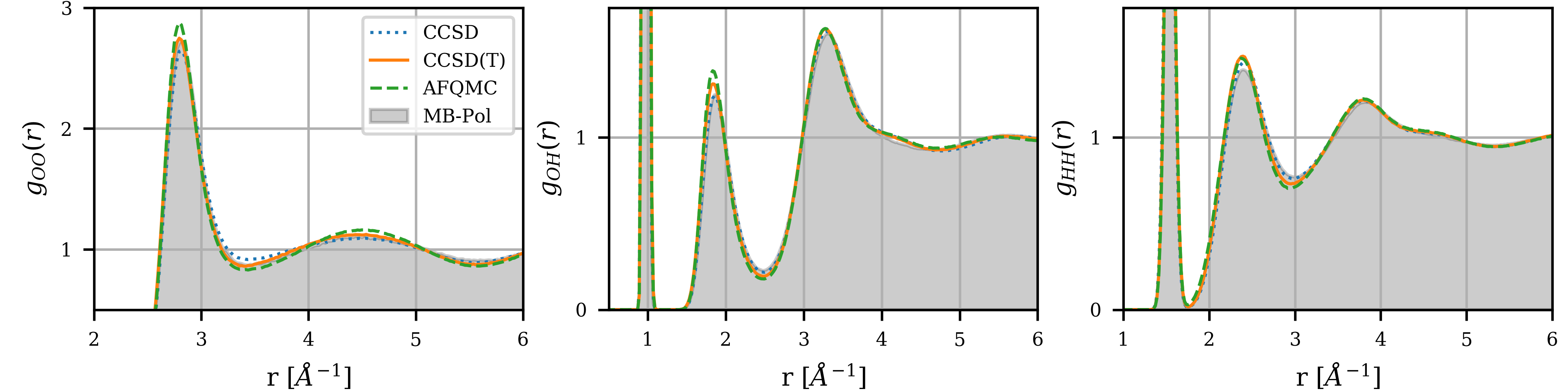}
    \end{center}
    \caption{Comparing RDF results from classical MD simulations (NVT) at T=300K using different electronic structure methods.}
	\label{fig:mbpol_comp_rdfs_classical_300K}
\end{figure*}

\begin{figure*}[h!]
    \begin{center}
        \includegraphics[width=\textwidth]{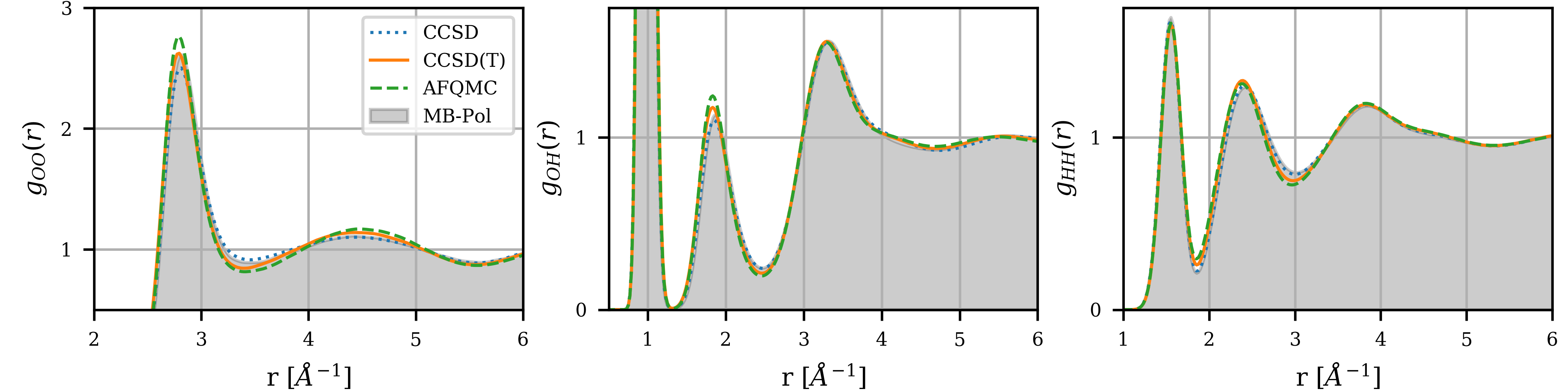}
    \end{center}
    \caption{Comparing RDF results from PIMD simulations (NVT) at T=300K using different electronic structure methods.}
	\label{fig:mbpol_comp_rdfs_pimd_300K}
\end{figure*}

\begin{figure*}[h!]
    \begin{center}
        \includegraphics{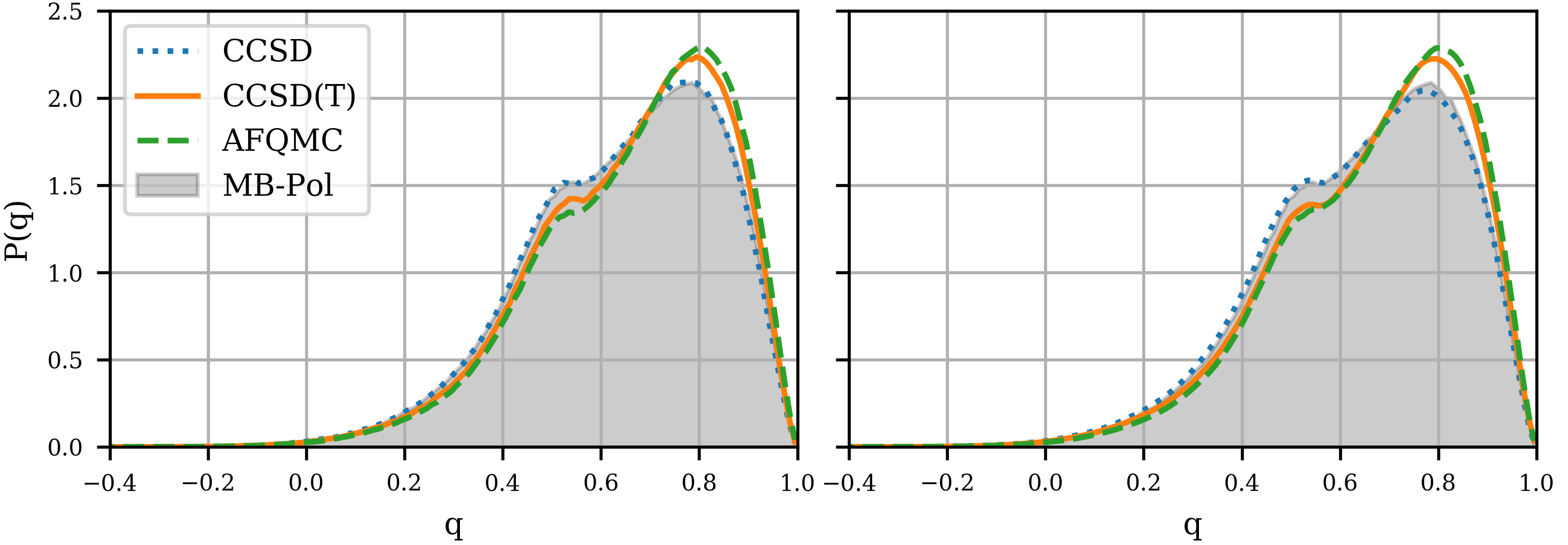}
    \end{center}
    \caption{Comparing $q$ distribution results from classical (left) and PIMD (right) obtained from simulations sampling the NVT ensemble at T=300K using different electronic structure methods.}
	\label{fig:mbpol_comp_tetra_classical_pimd_300K}
\end{figure*}

\begin{figure*}[]
    \begin{center}
        \includegraphics{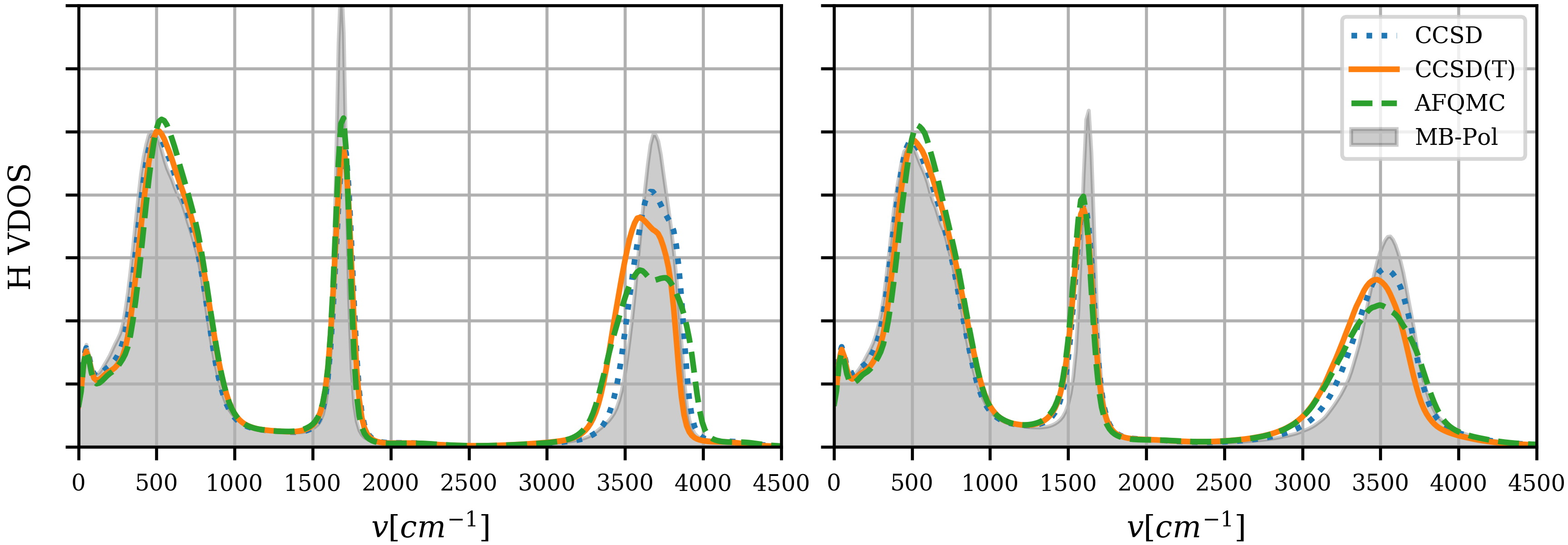}
    \end{center}
    \caption{Comparing VDOS results from classical (left) and PIMD (right) simulations sampling the NVT ensemble at T=300K using different electronic structure methods.}
	\label{fig:mbpol_comp_vdos_classical_pimd_300K}
\end{figure*}

SI Figs.~\ref{fig:mbpol_comp_rdfs_classical_300K}, \ref{fig:mbpol_comp_rdfs_pimd_300K}, \ref{fig:mbpol_comp_tetra_classical_pimd_300K}, and \ref{fig:mbpol_comp_vdos_classical_pimd_300K} depict how the properties obtained from our different transfer learned models fitted to N=200 CCSD, CCSD(T), or AFQMC energies, all initialized with MLPs trained to revPBE0-D3, compare with those obtained from MB-Pol\cite{Babin2013,Babin2014,Medders2014,Reddy2016}. Our classical MD and PIMD simulations using MB-Pol were run for 1~ns and 200~ps, respectively, and otherwise employed the same settings we used for all other MD runs that were used to compute properties (see SI Sec.\ref{sec:si-md-details}).


\pagebreak

\bibliography{references}